\documentclass[reprint,aps,prx,amsmath,amssymb,longbibliography, notitlepage, nofootinbib]{revtex4-2}
\usepackage{amsmath,amssymb, bm,graphicx}
\usepackage{graphics}
\usepackage{float} 
\usepackage[stable]{footmisc}
\usepackage{soul}
\usepackage[colorlinks, linkcolor= blue, citecolor = blue, urlcolor=blue]{hyperref}
\usepackage{makecell}
\usepackage{physics}

\def\be{\begin{equation}}
\def\ee{\end{equation}}
\def\bea{\begin{eqnarray}}
\def\eea{\end{eqnarray}}
\def\nn{\nonumber}
\begin{document}

%\title{ Electrical, thermal and gravitational chiral anomaly in 3D spin-orbit coupled metals}
\title{Chiral Anomalies in 3D Spin-Orbit Coupled Metals: Electrical, Thermal, and Gravitational anomaly}
\author{Sunit Das}
\email{sunitd@iitk.ac.in}
\author{Kamal Das}
\email{kamaldas@iitk.ac.in} 
\author{Amit Agarwal}
\email{amitag@iitk.ac.in}
\affiliation{Department of Physics, Indian Institute of Technology Kanpur, Kanpur-208016, India}

\begin{abstract}
    The discovery of a chiral anomaly in Weyl semimetals, the non-conservation of chiral charge and energy across two opposite chirality Weyl nodes, has sparked immense interest in understanding its impact on various physical phenomena. Here, we demonstrate the existence of electrical, thermal, and gravitational quantum chiral anomalies in 3D spin-orbit coupled systems. 
    Notably, these anomalies involve chiral charge transfer across two Fermi surfaces linked to a single Weyl-like point, rather than across opposite chirality Weyl nodes as in Weyl semimetals. Our findings reveal that the Berry curvature flux piercing the Fermi surface plays a critical role in distinguishing the `chirality' of the carriers and the corresponding chiral charge and energy transfer. Importantly, we demonstrate that these quantum chiral anomalies lead to interesting thermal spin transport such as the spin Nernst effect. Our results suggest that 3D spin-orbit coupled metals offer a promising platform for investigating the interplay between quantum chiral anomalies and charge and spin transport in non-relativistic systems.
\end{abstract}

\maketitle

\section{Introduction}

Chiral anomaly refers to the non-conservation of chiral charges in the presence of collinear electric and magnetic fields. It was first introduced in the context of the relativistic field theory of chiral fermions~\cite{adler_pr69_axial,nielsen_npb81_absence_1, nielsen_npb81_absence_2}. Later it was shown to be achievable in low gap semiconductors~\cite{nielson_plb83_adler}, with signatures in magnetoconductance experiments. Following the discovery of Weyl semimetals (WSMs) in recent years, the physics of chiral anomaly has been widely studied in condensed matter systems, resulting in a variety of non-trivial transport~\cite{son_prb13_chiral,xiong_science15,burkov_prb17_giant,Zhang_nc16,Li_NP16_chiral_magnetic,Nandy_prl17,shekhar_prb18, kim_prb14-boltzmann,kamal_prr20_chiral,kamal_prb19_linear, kamal_prb19_berry,kamal_prb21,kamal_prr20_thermal,debottam_prb22} and optical~\cite{HUTT_SciDirect19,hosur_prb15_tunable,carbotte_prb14_chiral,Ma_prb15,morimoto_prl16_chiral, jadidi_prb20, anmol_prb18,kabya_prb19} effects.
Intriguingly, the presence of a temperature gradient in Weyl systems can also result in an anomaly similar to the axial-gravitational anomaly in flat-space time~\cite{landsteiner_prl11_gravitational, lucas_pnas16_hydrodynamic, Gooth_nature17_experiemnatl,stone_prd18_mixed}. This leads to a range of interesting magneto-thermal transport phenomena~\cite{rex_prb14_thermoelectric, Hirschberger_nm16, Jia_nm16, kim_prb14-boltzmann, spivak_prb16, Stockert_IOP17, zyuzin_prb17_magnetotransport,Vu_NM21_thermal}.

Central to the physics of chiral anomaly is the continuity equation for the chiral charge. The continuity equation for the chiral charges and energy can be derived using semiclassical dynamics in crystalline materials and shows that the Berry curvature monopoles govern the chiral anomaly in Weyl metals~\cite{son_prl12_berry,son_prd13_kinetic, stefanov_prl12}. The concept of chiral anomaly has also been extended to other free fermionic excitations with no high-energy analog, such as multi-Weyl semimetals~\cite{fang_PRL2012_multi, li_PRB2016_weyl, Huang_prb17_topological,dantas_JHEP2018_magne, Sunit_prb22}, which exhibit two band crossings similar to WSMs but with nonlinear momentum dispersion along a particular direction, and semimetals with a higher number of band crossings near the Weyl node~\cite{lepori_JHEP2018_axial}. These systems, while possessing a higher chiral charge, are otherwise similar to Weyl systems in that a theory of chiral anomaly requires the presence of two opposite chirality Weyl nodes. 

\begin{figure}
   \centering
    \includegraphics[width=.96\linewidth]{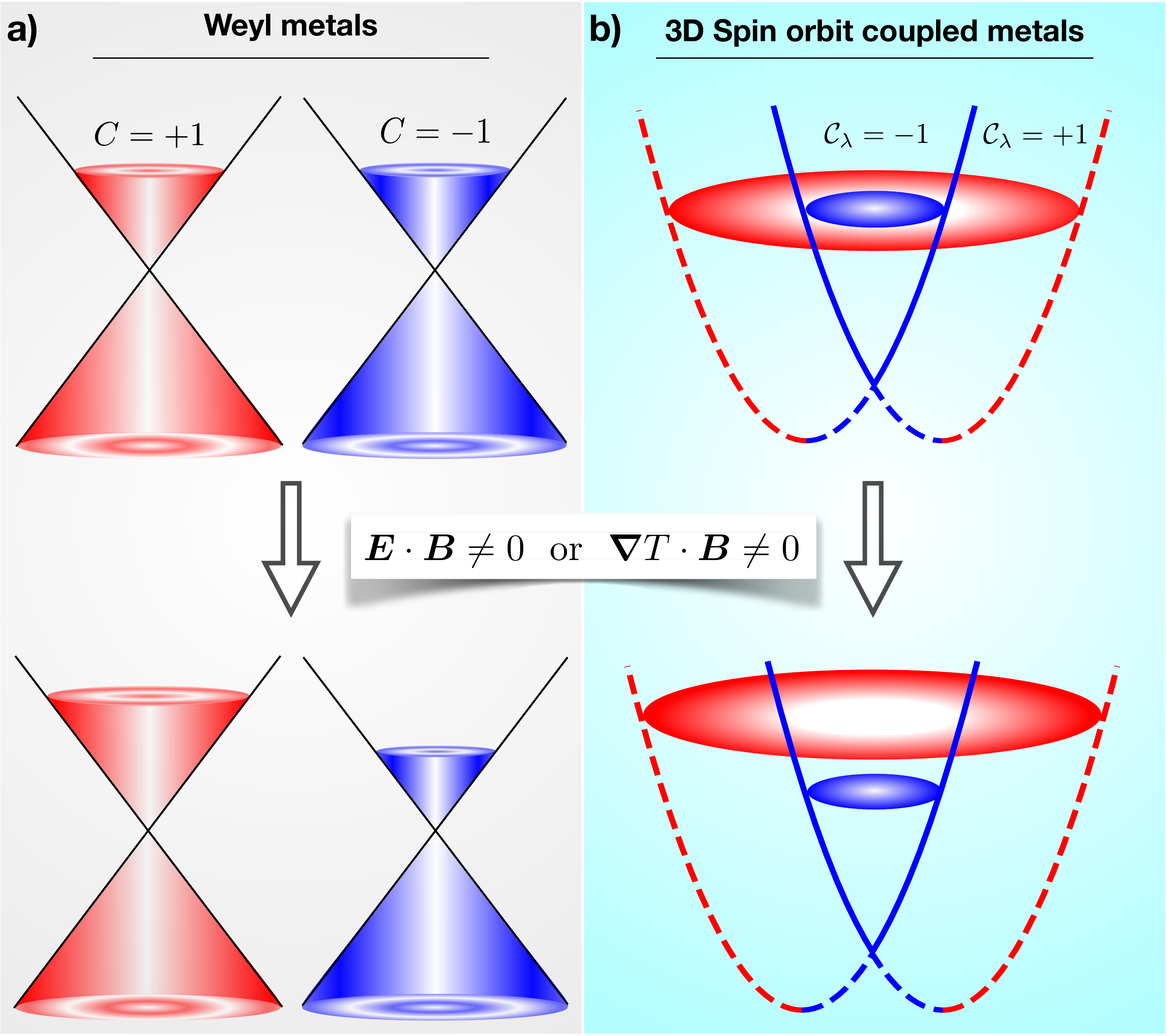}
    \caption{Depiction of the quantum chiral anomalies in (a) Weyl semi-metals and (b) 3D spin-orbit coupled metals or Kramers-Weyl metals. Both systems experience chiral charge and energy pumping, manifesting as electrical, thermal, and gravitational anomalies, when subjected to a magnetic field and collinear electric field (${\bm E}\cdot{\bm B} \neq 0$) or a temperature gradient (${\bm \nabla}T \cdot {\bm B} \neq 0$). In contrast to Weyl semimetals, the chiral charge pumping in 3D spin-orbit coupled metals occurs between two different Fermi surfaces associated with a single `Kramers-Weyl' node, but with opposite Berry curvature flux passing through them.    
    \label{Fig_1}}
\end{figure}

In this paper, we delve into the connection between chiral anomalies and the Berry curvature flux passing through the Fermi surface (FS) \cite{son_prb13_chiral}. This connection was recently explored in Ref.~\cite{cheon_prb22_chiral,Gao_cpl22_chiral}. Motivated by this, we generalize the theory of quantum chiral anomalies to Hamiltonians with non-relativistic terms, specifically ${\bm H} = {\bm h}_k \cdot {\bm \sigma} +  \sigma_0 {\bm k}^{2}$. Here, the ${\bm \sigma}$ represents the real spin of the system, $\sigma_0$ is the identity matrix, and ${\bm h}_k$ is an odd function of ${\bm k}$. The quadratic kinetic energy-like term in the Hamiltonian makes the chiral anomaly in this spin-orbit coupled (SOC) metals to be distinctly different from that in WSM [see Fig.~\ref{Fig_1}]. These types of systems can be found in Kramers-Weyl metals with quadratic corrections to their ${\bm k} \cdot {\bm p}$ Hamiltonians~\cite{Barnevig_science16,Chang18_NM,law21_comm_phys, zhang_prb17,Schroter_NP19, law21_comm_phys,Tan_ADFM22,Rao_N19, Sanchez_N19, debasis_prb22, takane_prl19} or in systems supporting 3D electron gas with SOC. 
%\footnote{Comparing the list of single crystalline point groups which %support 3D spin-orbit coupled metals \cite{SAMOKHIN09} with the list %of Kramers-Weyl metals \cite{Chang18_NM}, we find that these are %identical. However, 3D electron gas with SOC can also arise in %some heterostructures of two different single crystals.} 
While some aspects of the charge, heat, and spin transport in SOC metals have been explored earlier~\cite{Verma_2021,pal_jpcm21_berry,cheon_prb22_chiral}, the physics of quantum chiral anomalies in these systems is largely unexplored and merits further investigation.

%%%%%%%%%%% coarse grain analysis of our detailed result %%%%%%%

In this paper, we demonstrate that Kramers-Weyl and spin-orbit coupled metals can exhibit all three types of quantum chiral anomalies---electrical, thermal, and gravitational. 
%---in the presence of a collinear magnetic field and an electric field or a temperature gradient. 
We investigate the impact of electric field and temperature gradient-induced quantum chiral anomalies on charge, heat, and spin transport phenomena. Similar to the behavior observed in Weyl semi-metals~\cite{kamal_prr20_thermal}, we find that chiral anomalies in 3D SOC systems also result in negative longitudinal magneto-resistance and positive thermal magneto-resistance. However, a distinct feature of 3D SOC systems, as compared to WSMs, is that their low-energy Hamiltonian involves real spins. We show that quantum chiral anomalies in these systems also lead to interesting electrical and thermal spin transport including the spin-Nernst effect. %%%%%%%%%%%%%%%%%%%%%%%%%%

The structure of the rest of this paper is as follows. In Section~\ref{origin}, we discuss the origins of chiral anomalies in three-dimensional (3D) metals with SOC and Kramers-Weyl metals. In Section~\ref{continuity}, we present a mathematical derivation of the continuity equations to demonstrate the existence of these anomalies. The effects of these anomalies on charge and spin transport are examined in Sections~\ref{charge_tran} and~\ref{spin_tran}, respectively. Finally, we summarize our findings in Section~\ref{conclu}.

%The rest of the manuscript is organized as follows. In Sec.~\ref{origin} we highlight the origin of the chiral anomalies in 3D SOC metals and Kramers-Weyl metals. This is followed by a derivation of the continuity equations in Sec.~\ref{continuity} to mathematically establish the existence of chiral anomalies. We study the impact of anomalies on charge and spin transport in Sec.~\ref{charge_tran} and Sec.~\ref{spin_tran}, respectively. Finally, we summarize our results in Sec.~\ref{conclu}.

\section{Origin of Chiral anomalies in spin-orbit coupled metals} 
\label{origin}
To understand the chiral anomaly in 3D spin-orbit coupled metallic systems (or Kramers-Weyl metals), we first revisit the WSM. Specifically, we review the physics of chiral anomaly in WSM from the perspective of semiclassical dynamics. In WSM, the Hamiltonian for a particular Weyl node near the band crossing point can be approximated as ${\mathcal H}_{\rm WSM} = \sum_{a=x,y,z} \hbar ({\bm v}_a \cdot {\bm k}) { \sigma}^a$, where $k$ is measured from the Weyl node. The `chirality' of Weyl node is defined as ${C}={\rm sign}[{\bm v}_x \cdot {\bm v}_y \times {\bm v}_z]$~\cite{armitage_rmp18_weyl}. 
In the semiclassical dynamics picture, the existence of chiral anomaly can be understood by calculating the equilibrium current in the presence of 
an external magnetic field but no electric field. 

The equilibrium charge current for each Weyl node (or the chiral current) arises from the chiral magnetic velocity (see Sec.~\ref{equilibrium_current_Sec} with explicit derivation shown in Appendix~\ref{Appendix_C}). The chiral current for WSM   can be expressed in terms of the Berry curvature flux quantum passing through the FS for the WSM \cite{kamal_prr20_thermal}. This is consistent with the intuitive picture of the Weyl nodes acting as sinks and sources of the Berry curvature. For the pair of Weyl nodes of opposite chirality, their FSs are separated in the momentum space (at least for small energies). In the presence of an external electric field aligned along the magnetic field, 
the chiral charge carriers are pumped across the FSs with distinct Weyl chirality. This flow is stabilized by inter-node scattering. This results in different chiral charge densities on the two Weyl nodes [as shown in Fig.~\ref{Fig_1}(a)], and it manifests in several interesting transport phenomena in WSMs~\cite{son_prb13_chiral, zyuzin_prb17_magnetotransport,kamal_prr20_thermal}. We emphasize two things here: i) A minimum of a pair of Weyl nodes of opposite chirality are needed to produce chiral anomaly in WSM, and ii) the chiral anomaly can be interpreted as an FS phenomenon, where the chiral charges are `pumped' across two FSs enclosing  
opposite quantum of the Berry curvature flux. These two points will be crucial in investigating the chiral anomalies in Kramers-Weyl metals or 3D SOC metals.

3D SOC metals or Kramers-Weyl metals are structurally chiral crystals with broken inversion symmetry. They  
host `Weyl'-like nodal points at all the time-reversal-invariant momentum (TRIM) points in their Brillouin zone. While the form of the SOC can be different, 
a common feature of all such materials is that they have two FSs for each band crossing point (or the Kramers-Weyl node). This is aided by the kinetic energy term of the form $\hbar^2 k^2/(2 m)$ in their dispersion, which is missing in conventional WSM. We have tabulated all crystalline point groups that support Kramers-Weyl points, along with their low energy Hamiltonian in the vicinity of the Kramers-Weyl point in Appendix~\ref{Kramers_Weyl}.

While our discussion applies to all classes of single crystalline systems of 3D SOC metals or Kramers-Weyl metals listed in Table~\ref{table1}, for specific calculations, we consider the Hamiltonian~\cite{kang_prb15_transport,samokhin_prb08_effects, law21_comm_phys},
\be \label{Ham}
{\cal H} = \frac{\hbar^2 {\bm k}^2}{2m} \sigma_0 + \alpha {\bm k} \cdot {\bm \sigma}~. 
\ee
Here, $m$ is the effective electron mass, $\alpha$ is the SOC parameter, $\bm{ \sigma} = (\sigma_x , \sigma_y , \sigma_z )$ denotes the vector of the 
Pauli matrices in spin space and $\bm k$ is the Bloch wave vector. We note that in contrast to conventional WSM, the Pauli matrices here denote the physical spins of the itinerant electrons. The energy dispersion for the Hamiltonian~in Eq.~\eqref{Ham} is, 
\be \label{energy}
\epsilon_\lambda = \frac{\hbar^2 k^2}{2 m} + \lambda \alpha k ~.
\ee
Here, $\lambda= \pm 1$ is the spin-split band index which coincides with the eigenvalues of the operator $ \hat{\cal O} = \hat{\bm k}\cdot {\bm \sigma}$, 
and $k = |{\bm k}|$. 
The corresponding eigenstates are given by $\ket{u}_{+}^{T} = [\cos(\theta_k/2), 
e^{i \phi_k} \sin(\theta_k/2)]$ and $
\ket{u}_{-}^T = [\sin(\theta_k/2), -e^{i \phi_k} \cos(\theta_k/2)]$, with  $\cos\theta_k \equiv k_z/k$ and $\tan\phi_k \equiv k_y/k_x$. In Fig.~\ref{Fig_1}b, the $\lambda=+1$ ($\lambda=-1$) band is represented by the solid (dashed) line. The two bands of the dispersion relation \eqref{energy} have a band-touching point (BTP) at $\epsilon=0$. The $\lambda = +1$ band has a minimum at $\epsilon=0$ and increases monotonically as ${\bm k}$ increases. The $\lambda = -1$ band is non-monotonic, and it has a minimum energy located at 
$\epsilon_{\rm min}  = -\epsilon_\alpha$, with $\epsilon_\alpha = m \alpha^2/2\hbar^2$. %(see Fig.~\ref{Fig_1}). 
The minimum energy point lies on a circular contour specified by $|{\bm k}|^2 = k_\alpha^2$, where $k_\alpha = m \alpha/\hbar^2$.

Clearly, there are two different types of FSs for any value of the Fermi energy greater than the energy of the Kramers-Weyl node. The inner FS resulting from the $\lambda = +1$ band has an electronic character. In contrast, the outer FS can be interpreted to have the hole character. The Berry curvature flux quantum through each of the Fermi surfaces is defined as 
\be \label{BC_flux_ds}
\mathcal{C}_\lambda = \frac{1}{2\pi} \int_{\rm FS} d{\bm S} \cdot {\bm \Omega}_{\lambda}~.
\ee 
Here, $d{\bm S}$ is the elemental surface area of the FS, and ${\bm \Omega}_{\lambda}$ is the Berry curvature.
 More interestingly, the flux quantum associated with the FSs is equal and opposite. We explicitly calculate $\mathcal{C}_\lambda =-\lambda$. See Appendix~\ref{Appendix_B} for details. Hence, the Berry curvature flux quantum piercing the outer (inner) FS is $+1$ ($-1$). We emphasize that this scenario is distinctively different from the usual WSM with chiral symmetry. In WSM, the pair of FS with the opposite sign of the Berry curvature flux quantum corresponds to two distinct Weyl crossing points separated by momentum or energy. However, in this case, a single Weyl crossing is linked to the two FSs with opposite flux quantum. The opposite sign of the Berry curvature flux quantum on the two FSs can be used to define charged fermions of different `flavors' (akin to chirality in the case of WSM) in the two FSs. 
 
The non-zero flux associated with the two FSs in SOC metals gives rise to chiral anomalies. This is captured by the non-conservation of the total 
flavor charge (${\cal N}^\lambda$) and energy (${\cal E}^\lambda$) 
in presence of a magnetic field (${\bm B}$) and an electric field (${\bm E}$) or temperature gradient (${\bm \nabla} T$). In a clean system of 3D SOC metal, we can obtain 
\be \label{collisionless_particle_cont}
\dfrac{\partial {\cal N}^\lambda}{\partial t} \propto 
  - {\cal C}_0^\lambda~{\bm E} \cdot{\bm B}~~~{\rm and}~~~\dfrac{\partial {\cal N}^\lambda}{\partial t} \propto - {\cal C}_1^\lambda~{\bm \nabla}T\cdot{\bm B}~.
\ee 
A similar calculation for the total energy of each flavour of fermions yields, 
\be
\dfrac{\partial {\cal E}^\lambda}{\partial t} \propto  
\begin{cases}
    -(\mu~{\cal C}^\lambda_0 + k_B T ~{\cal C}^\lambda_1) ~{\bm E} \cdot{\bm B}~ \\ 
    -(\mu ~ {\cal C}_1^\lambda +  k_B T~{\cal C}_2^\lambda){\bm \nabla}T\cdot{\bm B}
\end{cases}~.
\ee
Here, $\mu$ is the chemical potential, and $k_B T$ is the energy scale of the temperature. The coefficients ${\cal C}_{\nu}^{\lambda}$ [Eq.~\eqref{Cs}] for $\nu = \{0, 1, 2\}$ are the coefficients of the electrical, thermal, and gravitational chiral anomalies, respectively. See Sec.~\ref{continuity} and Eqs.~\eqref{particle_cont}-\eqref{enrgy_cont} for more details. 
More importantly, these are finite only when the Berry curvature flux quantum ${\cal C}_\lambda$ is finite. Thus, the Berry curvature flux quantum plays an important role in defining the particles' flavor (or chirality) and the associated quantum flavor anomalies (or chiral anomalies). We highlight the chiral charge transfer across the two Fermi surfaces in WSM and in 
3D SOC metals, with opposite Berry curvature flux in Fig~\ref{Fig_1}.
 
%Similar expressions are also obtained to describe the chiral anomaly in WSM~\cite{Gooth_nature17_experiemnatl,kamal_prr20_chiral, kamal_prr20_thermal, Sunit_prb22}. 
%\textcolor{blue}{However, in WSM, ${\mathcal C}_\lambda$ is associated with the FSs of two opposite `chirality' Weyl nodes separated in momentum space (or in energy). In contrast, for 3D SOC metals (or Kramers-Weyl metals), the FSs hosting particles of opposite flavor are associated with a single band crossing node.} \textcolor{cyan}{SD: this is repeatative} 
%This is highlighted schematically in Fig.~\ref{Fig_1}, where the chiral charge transfer between the two FSs has been emphasized. 

In the next section, we explicitly demonstrate the three chiral anomalies in 3D SOC (or Kramers-Weyl) metals using the idea of equilibrium and non-equilibrium chiral charge and energy currents. We specifically focus on the case when the chemical potential is higher in energy than the Kramers-Weyl point ($\mu >0$). The regime when the chemical potential is below the energy of the Kramers-Weyl point is a bit tricky. We find that in this regime there is only one FS. The Fermi surface is associated with the $\lambda=-1$ band, and the total flux through the FS is identically zero. Since the chiral anomaly requires two FSs with opposite Berry curvature flux, there is no chiral anomaly for  $\mu<0$. However, an interesting Brillouin zone partitioning scheme been proposed in Ref.~\cite{cheon_prb22_chiral} to divide the single FS into two parts having opposite Berry curvature flux. We show that such BZ partitioning within a single FS is not physical, and it can lead to chiral anomaly-like physics even in a free electron gas in absence of a magnetic field and Berry curvature. We discuss these subtle issues in detail in Appendix~\ref{Appendix_B}.

%%%%%%%%%%%%%%%%%%%%%%%%%%%%%%%%%%%%%%%%%%%%%%%%%%%%%%
\section{Chiral currents and the chiral anomalies}
\label{continuity}
%\textcolor{red}{Together, the chiral charge and energy pumping specify three chiral anomaly coefficients. These are typically referred to as the coefficients of the electrical, thermal, and gravitational chiral anomaly, respectively \cite{kamal_prr20_chiral, kamal_prr20_thermal}. 

In this section, we first show that the existence of equilibrium currents in the presence of a magnetic field hints at the possible existence of chiral anomalies in the system. Next, we explicitly calculate the continuity equation for the chiral charges and energy current in the presence of a magnetic field and either a collinear electric field or a collinear temperature gradient.

%\textcolor{red}{Physically, they arise from the chiral magnetic velocity term in the velocity of the center of mass of the electron wavepacket.} 
%\textcolor{blue}{KD: Please consider this statement. The magnetic velocity can not be the physical mechanism behind the chiral anomaly. The reason is: magnetic velocity can be induced only by a magnetic field without an electric field. However, for the chiral anomaly, we need both the electric and magnetic fields. Probably what you are trying to say is that if a system possesses a chiral equilibrium current, it gives a hint that the system will host chiral anomaly. Indeed it is the case, and it only happens because the quantity "Berry curvature flux" is associated with both of these phenomena, equilibrium chiral current, and chiral anomaly. But to associate equilibrium current as a physical mechanism of chiral anomaly will be misleading, in my opinion. Let me know what you think. } To show this explicitly, we first calculate the chiral equilibrium current in the presence of only a magnetic field ($\bm B$). 

\subsection{Equilibrium chiral current induced by magnetic field} \label{equilibrium_current_Sec}

The equations of motion of charge carriers in the presence of Berry curvature are described by the following semiclassical equation of motion~\cite{xiao_rmp10_berry, morimoto_prb16_semiclassical}
\begin{subequations}
\bea \label{eom_r}
\dot{\bm r}_\lambda & = &D_\lambda \left[{\bm v}_\lambda +\frac{e}{\hbar}{\bm E}\times {\bm \Omega}_\lambda +\frac{e}{\hbar}({\bm v}_\lambda \cdot {\bm \Omega}_\lambda){\bm B}\right],
\\
\hbar\dot{\bm k }_\lambda &=& D_\lambda \left[-e{\bm E} - e{{\bm v}}_\lambda \times {\bm B}-\frac{e^{2}}{\hbar}({\bm E}\cdot{\bm B}){\bm  \Omega}_\lambda\right]. \label{eom_k}
\eea
\end{subequations}
Here, `$-e$' is the electronic charge, ${\bm v}_{\lambda}$ is the band velocity, and $\bm \Omega_{\lambda}$ is the Berry curvature. 
In Eq.~\eqref{eom_r}, $D_\lambda \equiv 1/(1 + \frac{e}{\hbar}{\bm \Omega}_\lambda \cdot {\bm B})$ is the phase-space factor, which modifies the invariant phase-space volume according to $[d{\bm k}] \to [d{\bm k}] D_\lambda^{-1}$~\cite{xiao_prl05}. The term $\frac{e}{\hbar}({\bm v}_\lambda \cdot {\bm \Omega}_\lambda){\bm B}$ in Eq.~\eqref{eom_r} is known as the chiral magnetic velocity and as will see it plays an important role in anomaly related transport. %This term plays an important role in generating the chiral charge and energy currents~\cite{kamal_prr20_thermal,kamal_prr20_chiral}. 

%The origin of the anomaly-induced current can be understood by calculating the equilibrium (no bias voltage or temperature gradient) charge and energy currents in the presence of an external magnetic field. 
For a given FS, the equilibrium chiral charge and energy currents are calculated to be~\cite{kamal_prr20_chiral}
\be \label{eq:equil}
\{ {\bm j}^\lambda_{e,\rm eq}, {\bm j}^\lambda_{\epsilon,\rm eq} \} =   
\int_{{\rm BZ}_\lambda} [d{\bm k}]\{-e, \epsilon_\lambda \} \frac{e}{\hbar} \left({\bm v}_\lambda \cdot{\bm \Omega}_\lambda \right) f_\lambda~.
\ee 
%
%Here, $\chi$ represents the FS index and $\lambda$ is the band index. 
In Eq.~\eqref{eq:equil}, $f_\lambda$ is the equilibrium Fermi distribution function corresponding to the FS $\lambda$. We emphasize that the chiral magnetic velocity solely determines the chiral currents, and the band gradient velocity does not contribute to it. 
Evaluating Eq.~\eqref{eq:equil} for our model Hamiltonian, we obtain general relations for the charge and the energy current~\cite{ma_PRB2015_chiral,stone_prd18_mixed,kamal_prr20_thermal},
\begin{subequations}
\bea \label{j_eq_e}
{\bm j}_{e,\rm eq}^\lambda &&= -e  \left(\mu \mathcal{C}_0^\lambda + k_B T \mathcal{C}_1^\lambda \right) {\bm B}~, \\
{\bm j}_{\epsilon,\rm eq}^\lambda &&= \left(\frac{\mu}{2} \mathcal{C}_0^\lambda + \mu k_B T \mathcal{C}_1^\lambda + \frac{k_B^2 T^2}{2} \mathcal{C}_2^\lambda \right) {\bm B}~. \label{j_eq_E}
\eea 
\end{subequations} 
Here, we note that all the anomaly coefficients appear in the equilibrium current. In Eqs.~\eqref{j_eq_e}-\eqref{j_eq_E}, the coefficients are specified by, 
\be \label{Cs}
\mathcal{C}^{\lambda}_{\nu} = \frac{e}{4 \pi^2 \hbar^2 } \int d\epsilon \left(\frac{\epsilon -\mu }{k_B T} \right)^{\nu} \left(- \frac{\partial f_\lambda}{\partial \epsilon_\lambda} \right) \mathcal{C}_\lambda~.
\ee 
%
%The $\mathcal{C}_0^\lambda$ generates the electrical chiral anomaly~\cite{son_prb13_chiral}, $\mathcal{C}_1^\lambda$ is the  coefficient of the thermal chiral anomaly~\cite{kamal_prr20_thermal}, and $\mathcal{C}_2^\lambda$ gives rise to the mixed-gravitational chiral anomaly ~\cite{Gooth_nature17_experiemnatl, clemens_prb20_anisotropic,landsteiner_prl11_gravitational}. %Here, we have defined $\mathcal{C}_\lambda$ as follows 
%
% \be 
% {\cal C}_\lambda = \frac{\hbar}{2\pi}  \int d{\bm k}~ {\bm \Omega}_{\lambda}\cdot {\bm v}_\lambda \delta(\mu - \epsilon_\lambda),
% \ee 
%
%which represents the Berry curvature flux quantum through the Fermi surface.
It is evident from Eq.~\eqref{Cs} that for any quantum system with finite $\mathcal{C}_\lambda$, all the chiral anomaly coefficients are non-zero. 
%Consequently, systems that have a pair of FS, with the opposite quantum of Berry curvature flux piercing through them ($\mathcal{C}_\lambda = -\mathcal{C}_{- \lambda}$), will support quantum chiral anomalies. 
We mention here that in defining the anomaly coefficients in Eq.~\eqref{Cs}, we have converted the Fermi sea integration of Eq.~\eqref{eq:equil} into Fermi surface integration using the rule of partial derivative. We provide the details of the calculations in Appendix~\ref{Appendix_C}. 

%%%%% why one should calculate equilibrium current %%%%%
The importance of the equilibrium currents given in Eqs.~\eqref{j_eq_e}-\eqref{j_eq_E} is multifold. First of all, the presence of finite chiral charges and energy currents in equilibrium is an indication of the existence of chiral anomalies. This is because, for both chiral anomaly and non-zero chiral equilibrium current, non-zero Berry curvature flux is a prerequisite. Second, the chiral charge (${\bm j}^+_{e, \rm eq}-{\bm j}^-_{e, \rm eq}$) and energy (${\bm j}^+_{\epsilon, \rm eq}-{\bm j}^-_{\epsilon, \rm eq}$) currents are non-zero. This highlights that in systems hosting a pair of fermions with opposite Berry curvature flux quantum, the chiral magnetic velocity induces a dissipationless chiral charge and energy current along 
$\bm B$~\cite{fukushima_prd08_chiral, kim_prb14-boltzmann, li_npa16_chiral, kharjeev_ppnp14_chiral, Kharzeev_epj18_chiral}. %Third, since any system should not have a total equilibrium current, the finite equilibrium chiral current hints that the Berry curvature flux associated with the two surfaces should be equal and opposite, implying that \textcolor{red}{one flavor must be most} with another flavor. This can be viewed as some sort of fermion doubling theorem. 
Finally, we can expect a finite anomaly-induced current in non-equilibrium. In equilibrium, the total charge (${\bm j}^+_{e, \rm eq}+{\bm j}^-_{e, \rm eq}$) and energy (${\bm j}^+_{\epsilon, \rm eq}+{\bm j}^-_{\epsilon, \rm eq}$) currents from the two opposite chirality  FSs will add up to zero due to same chemical potential and temperature. However, in the presence of chiral chemical potential ($\mu_+ \neq \mu_-$) and chiral temperature ($T_{+}\neq T_{-}$) imbalance induced by the quantum anomalies, these expressions will result in finite charge and energy current. 

%%%%%%%%%% emphasizing that two-dimensional systems do not host chiral anomalies %%%%%%%%
Note that the general expressions of equilibrium charge and energy currents, ${\bm j}^\lambda_{e, \rm eq}$ and ${\bm j}^\lambda_{\epsilon, \rm eq}$, are valid for any 3D systems with band touching point. These currents originate from the chiral magnetic velocity, $e/\hbar ({\bm v}_\lambda \cdot {\bm \Omega}_\lambda) {\bm B}$. As a result, the equilibrium currents are identically zero for any two-dimensional system, for which ${\bm v}_\lambda \cdot {\bm \Omega}_\lambda=0$. The absence of chiral magnetic velocity in 2D systems forbids the existence of quantum chiral anomalies in two-dimensional systems. 
For three-dimensional systems, ${\bm v}_\lambda \cdot {\bm \Omega}_\lambda$ is generally non-zero, which gives rise to finite equilibrium currents. However, to have quantum chiral anomalies in the system, there should be a pair of FS with opposite Berry curvature flux quantum passing through them so that ${\bm j}^\lambda_{e/\epsilon, \rm eq} = -{\bm j}^{-\lambda}_{e/\epsilon, \rm eq}$.

%\textcolor{blue}{Such a chiral charge and energy current hints at the possibility of quantum chiral anomalies in the presence of external perturbations such as an electric field or a temperature gradient ~\cite{nielson_plb83_adler,Bell, adler_pr69_axial, son_prb13_chiral, son_prl12_berry, Landsteiner_app16_notes,ong_nrp21_experimental, kamal_prr20_chiral}. } \textcolor{cyan}{SD: repeatative}

%\textcolor{red}{will tend to induce a chiral charge (or chemical potential) and chiral energy 
%(or temperature) imbalance in the two opposite chirality fermions~\cite{cheon_prb22_chiral}. In WSMs, the chiral magnetic velocity also manifests as a chiral magnetic effect~\cite{fukushima_prd08_chiral, kim_prb14-boltzmann, li_npa16_chiral, kharjeev_ppnp14_chiral, Kharzeev_epj18_chiral}, in addition to the chiral anomaly.~\cite{nielson_plb83_adler,Bell, adler_pr69_axial, son_prb13_chiral, son_prl12_berry, Landsteiner_app16_notes,ong_nrp21_experimental, kamal_prr20_chiral}. A similar effect can also be expected in Kramers-Weyl systems.}

Having discussed the general expressions for the equilibrium charge and energy currents, we now calculate all the anomaly coefficients for a 3D spin-orbit coupled system.  
For the Hamiltonian in Eq.~\eqref{Ham}, the Berry curvature is given by ${\bm \Omega}_{\lambda} = -\lambda {\bm k}/{2k^3}$. The chiral anomaly coefficients are obtained to be
\bea \label{coffcnts}
\{ \mathcal{C}_0^\lambda,~ \mathcal{C}_1^\lambda,~ \mathcal{C}_2^\lambda \} = -\lambda \frac{e}{4 \pi^2 \hbar^2} \left \{ \mathcal{F}_0 , \mathcal{F}_1 , \mathcal{F}_2 \right\}~.
\eea 
We note that the equilibrium currents of Eqs.~\eqref{j_eq_e} and \eqref{j_eq_E}, along with the chiral anomaly coefficients of the above equations, do not get affected by the orbital magnetic moment.
Here, $\mathcal{F}_\nu$'s are the dimensionless functions of i) $x= \beta( \epsilon_\alpha + \mu)$ for $\lambda=-1$ band, and ii) $x= \beta \mu$ for $\lambda=+1$ band with $\beta = 1/k_B T$ being the inverse temperature. Their functional form is given by 
\bea \label{eq:F}
\mathcal{F}_0(x) & \equiv & 1/(1+e^{-x})~, \nn \\ 
\mathcal{F}_1(x) &\equiv & x/(1+e^{x}) + {\rm ln}[1+e^{-x}]~, \\ 
\mathcal{F}_2(x) &\equiv & \frac{\pi^2}{3}- x\left(\frac{x}{1+e^{x}} +2 {\rm ln}[1+e^{-x}]\right) + 2 {\rm Li}_2[-e^{-x}]. \nn
\eea
Here, $\rm Li_2$ is the polylogarithmic function of order two. With the replacement of $(\epsilon_\alpha + \mu) \to \mu$, Eqs.~\eqref{coffcnts} and \eqref{eq:F} become identical to that in the WSMs~\cite{kamal_prr20_thermal}. The temperature dependence of all three chiral anomaly coefficients is similar to Fig. (6) in Ref.~\cite{kamal_prr20_thermal}. In the zero temperature limit, $\mathcal{F}_0 \to 1$, and $\mathcal{F}_2 \to \pi^2/3$. It is worth noting that for $T = 0$, the thermal chiral anomaly coefficient ${\cal C}_1^\lambda \propto {\cal F}_1 \to 0$ becomes finite only for finite $T$. 
% 
%\subsection{Chiral anomaly in non-equilibrium current with finite ${\bf E} \cdot {\bf B}$ or finite ${\bf \nabla} T \cdot {\bf B}$}
\subsection{Steady state in the presence of chiral anomaly}
The presence of external perturbations, such as an electric field ${\bm E}$, or a temperature gradient ${\bm \nabla}T$, drives the system out of equilibrium.  
In the non-equilibrium steady-state, the distribution function ($g_\lambda$) corresponding to the FS $\lambda$ satisfies the following Boltzmann transport equation 
\be\label{bte_1}
\dfrac{\partial g_{\lambda}}{\partial t} + \dot{\bm r}_\lambda \cdot {\bm \nabla}_{\bm r } ~g_{\lambda} +\dot{\bm k }_\lambda \cdot {\bm \nabla}_{\bm k}~g_{\lambda} = {\cal I}_{\rm coll} \{g_{\lambda} \}~.
\ee
Here, ${\cal I}_{\rm coll}\{g_{\lambda} \}$ is the collision integral and $g_{\lambda}$ is the non-equilibrium distribution function for each Fermi function. Similar to that in WSM, the charge and energy pumping between the two FSs dictates that the collision integral should incorporate both the intra- and inter-Fermi surface scattering processes~\cite{yip_arxiv15_kinetic, zyuzin_prb17_magnetotransport, kamal_prr20_thermal}. Within the relaxation time approximation, both the scattering process can be captured by the following form of the collision integral~\cite{kamal_prr20_chiral, deng_prl19_quantum},
\be \label{I_coll}
{\cal I}_{\rm coll}^\lambda= -\dfrac{g_\lambda-\Bar{g}_\lambda}{\tau} - \dfrac{\bar{g}_\lambda-f_\lambda}{\tau_{v}}~.
\ee 
Here, $\bar{g}_\lambda$ represents the `local' steady-state distribution function for each FS with a local chemical potential $\mu_\lambda \equiv \mu + \delta \mu_\lambda$, and local temperature $T_\lambda \equiv T + \delta T_\lambda$~\cite{yip_arxiv15_kinetic}, and $f_\lambda$ specifies the global equilibrium function. The first term in the right-hand side of Eq.~\eqref{I_coll} represents the intra-Fermi surface scattering (with scattering rate $1/\tau$), which establishes the local equilibrium. The inter-Fermi surface scattering has been represented by the 
second term in Eq.~\eqref{I_coll} with scattering rate $1/\tau_v$. The ratio of inter- and intra- Fermi surface scattering time for Hamiltonian~\eqref{Ham} considering screened Coulomb impurity potential has been calculated in Ref.~\cite{cheon_prb22_chiral}. In the small $\mu$ limit, it is given by $\tau_v/\tau \sim (2 m \alpha^2/\hbar^2)^2 / \mu^2$~\cite{cheon_prb22_chiral}. Hence, for small $\mu$, similar to the WSM~\cite{burkov_prb15_negative, jun_science15_evidence}, we have  $\tau_v> \tau$.

 Now, we construct the continuity equation for the particle number and the energy density. Substituting Eq.~\eqref{I_coll} in Eq.~\eqref{bte_1}, and then integrating over all the momentum states for the FS $\lambda$, we obtain 
% %
 \be\label{particle_cont}
 \dfrac{\partial {\cal N}^\lambda}{\partial t}
 +e{\bm E} \cdot{\bm B} {\cal C}_{0}^\lambda+  {\bm \nabla}_{\bm r} \cdot {\bm J}^\lambda  =- \dfrac{{\cal N}^\lambda -{\cal N}_0^{ \lambda}}{\tau_v}.
 \ee
 Here, ${\bm \nabla}_{\bm r} \cdot {\bm J}^\lambda = k_B {\cal C}_{1}^\lambda {\bm{\nabla} T} \cdot{\bm B}$ is the divergence of particle current.
 The quantities $\{ \mathcal{N}_0^\lambda, \mathcal{N}^\lambda  \} = \int [d{\bm k}] D_\lambda^{-1} \{f_\lambda, g_\lambda \}$ represents the total particle number density in each FS before and after applying the perturbing fields. In Eq.~\eqref{particle_cont}, the terms ${\bm E} \cdot{\bm B} {\cal C}_{0}^\lambda$, and $k_B {\cal C}_{1}^\lambda {\bm{\nabla} T} \cdot{\bm B}$ represents the chiral anomaly induced flow of the charge carriers. Similarly, the continuity equation for the energy density, which we construct by multiplying the energy dispersion $\epsilon_\lambda$ in Eq.~\eqref{bte_1} and integrating over all the momentum states,  is obtained to be 
% %
 \be  \label{enrgy_cont}
 \dfrac{\partial {\cal E}^\lambda}{\partial t} + (\mu {\cal C}^\lambda_0 + k_B T {\cal C}^\lambda_1)~e{\bm E} \cdot{\bm B} + {\bm \nabla}_{\bm r} \cdot {\bm J}^\lambda_\mathcal{E} =- \dfrac{{\cal E}^\lambda -{\cal E}_0^{\lambda}}{\tau_v}.
 \ee
% %
The second term on the left hand side is $-{\bm E} \cdot {\bm j}_{e,\rm eq}^\lambda$ that represents the work performed by the electric field and ${\bm \nabla}_{\bm r} \cdot {\bm J}^\lambda_{\mathcal{E}} = (\mu k_B  {\cal C}_1^\lambda +  k_B^2 T{\cal C}_2^\lambda)~{\bm \nabla T} \cdot{\bm B}$ represents the divergence of energy current in presence of ${\bm \nabla}T$. The quantities $\{ \mathcal{E}_0^\lambda, \mathcal{E}^\lambda  \} = \int [d{\bm k}] D_\lambda^{-1} \epsilon_\lambda \{f_\lambda, g_\lambda \}$ is the total energy density in each FS before and after applying external fields, respectively. Here, $\mu \mathcal{C}_0^\lambda$ and $\mu \mathcal{C}_1^\lambda$ specify the energy carried out by the chiral charge transfer, whereas $T \mathcal{C}_2^\lambda$ represents the energy pumped out by the term ${\bm \nabla}T \cdot {\bm B}$~\cite{kamal_prr20_thermal}. In constructing Eq.~\eqref{particle_cont} and \eqref{enrgy_cont}, we have used the fact that the intra-Fermi surface scattering does not change the number of particles and energy in each FS.  

\section{Chiral anomaly and carrier transport}
\label{charge_tran}

To calculate the chiral anomaly-induced charge, heat, and spin currents, we first calculate the non-equilibrium distribution function to linear order in an applied electric field.  
In the linear response regime, we can safely assume that the change in chiral chemical potential and temperature is small, i.e., $\delta \mu_\lambda < \mu$, and $\delta T_\lambda < T$~\cite{yip_arxiv15_kinetic,kamal_prr20_thermal, kamal_prr20_chiral}. Then, to the lowest order in $\delta \mu_\lambda$ and $\delta T_\lambda$, the non-equilibrium distribution function can be calculated to be
\bea \label{distribution_fn}
&& g_\lambda = f_\lambda + 
\left(-\dfrac{\partial f_\lambda}{\partial  \epsilon_\lambda}\right)\bigg[\left( 1- \frac{\tau}{\tau_v} \right) \left(\delta\mu_\lambda +  \dfrac{\epsilon_\lambda -\mu}{T} \delta T_\lambda\right)  \nn \\
&&- \tau D_\lambda \left( {\bm v}_\lambda + \frac{e}{\hbar} \left( {\bm v}_\lambda \cdot {\bm \Omega}_\lambda \right){\bm B} \right)\cdot
\left(e{\bm E} + \left( \epsilon_\lambda - \mu\right)\dfrac{\bm{\nabla} T}{T}\right)\bigg]
~. \nn \\
\eea 
Here, the chiral chemical potential $\delta \mu_\lambda$ and $\delta T_\lambda$ are given by~\cite{kamal_prr20_chiral}
\bea \label{d_mu, d_T}
\delta \mu_\lambda &=& -\frac{\tau_v}{(\mathcal{D}_2^\lambda \mathcal{D}_0^\lambda - {\mathcal{D}^\lambda_1}^2 )} \left[\left(\mathcal{D}_2^\lambda \mathcal{C}^\lambda_0 - \mathcal{D}^\lambda_1 \mathcal{C}^\lambda_1 \right) e {\bm E}\cdot {\bm B} \right. \nn  \\
&& \left.  + \left(\mathcal{D}^\lambda_2 \mathcal{C}_1^\lambda - \mathcal{D}_1^\lambda \mathcal{C}_2^\lambda \right) k_B {\bm \nabla}T \cdot {\bm B} \right], \\
k_B \delta T_\lambda & =& -\frac{\tau_v}{(\mathcal{D}_2^\lambda \mathcal{D}_0^\lambda - {\mathcal{D}^\lambda_1}^2 )} \left[\left(\mathcal{D}_0^\lambda \mathcal{C}^\lambda_1 - \mathcal{D}^\lambda_1 \mathcal{C}^\lambda_0 \right) e {\bm E}\cdot {\bm B} \right. \nn  \\
&& \left.  + \left(\mathcal{D}^\lambda_0 \mathcal{C}_2^\lambda - \mathcal{D}_1^\lambda \mathcal{C}_1^\lambda \right) k_B {\bm \nabla}T \cdot {\bm B} \right]~.
\eea
In the above equation, we have defined the magnetic field-dependent generalized density of states at finite temperature as
\be 
\mathcal{D}^\lambda_\nu = \int d\epsilon \left( \frac{\epsilon_\lambda -\mu }{k_B T} \right)^\nu \left(-\dfrac{\partial f_\lambda}{\partial  \epsilon_\lambda}\right) \mathcal{D}_\lambda~.
\ee
Here, $\nu=\{0,1,2\}$, and $\mathcal{D}_\lambda = \int [d{\bm k}] (1+e/\hbar {\bm \Omega}_\lambda \cdot {\bm B}) \delta(\mu -\epsilon_\lambda)$ being the density of states corresponding to the FS of the band $\lambda$. It is evident that both the electric field and the temperature
gradient components parallel to ${\bm B}$ contribute to generating the system's chiral chemical potential and chiral temperature imbalance. 

Having obtained the non-equilibrium distribution function, we now calculate the charge and heat current in each FS, which are defined as $\{ {\bm j}_e^\lambda,{\bm j}_{Q}^\lambda \} = \int [d {\bm k}] \{ -e, (\epsilon_\lambda -\mu) \} {\bm \dot{r}}_\lambda g_\lambda$. Focusing only on the anomaly induced contribution $\propto \tau_v$, we obtain~\cite{kamal_prr20_chiral}
\bea \label{current_simplified}
\begin{pmatrix} 
{\bm j}_e^\lambda \\
{\bm j}_Q^\lambda
\end{pmatrix}
&&= \tau_v  {\bm B}
\begin{pmatrix}
\frac{1}{{\cal D}_0^\lambda}({e\cal C}_0^\lambda)^2  & e k_B \frac{{\cal D}_1^\lambda}{{\cal D}_0^\lambda {\cal D}_2^\lambda} {\cal C}_0^\lambda {\cal C}_2^\lambda \\
e k_B T \frac{{\cal D}_1^\lambda}{{\cal D}_0^\lambda {\cal D}_2^\lambda} {\cal C}_0^\lambda {\cal C}_2^\lambda & T \frac{1}{{\cal D}_2^\lambda}(k_B{\cal C}_2^\lambda)^2
\end{pmatrix} \nn
\\
&& \times \begin{pmatrix}
{\bm E} \cdot {\bm B} \\
- {\bm \nabla} T \cdot {\bm B} 
\end{pmatrix}.
\eea 
In deriving the above equation, we used the fact that in the $\mu \gg k_B T$ limit (or $ \beta \mu \gg 1$) limit, $\mathcal{C}_1^\lambda \to 0$, and $\mathcal{D}_0^\lambda, \mathcal{D}_2^\lambda >  \mathcal{D}_1^\lambda$.
Now, the transport coefficients can be obtained by comparing the total currents $\left({\bm j}_{e,Q} = \sum_\lambda {\bm j}^\lambda_{e,Q}\right)$ from Eq.~\eqref{current_simplified} and the phenomenological linear response relations~\cite{Ashcroft76}: $j_{e,a} = \sum_b [\sigma_{ab}~E_b - \alpha_{ab}~\nabla_b T$] and $ j_{Q,a} = \sum_b [{\bar \alpha}_{ab}~E_b - {\bar \kappa}_{ab}~\nabla_b T ]$. 
Here, $\sigma$, $\alpha$, $\bar \alpha$, and $\bar \kappa$ denote the electrical, thermo-electric, electro-thermal, and constant voltage thermal conductivity matrix, respectively. Note that the thermo-power matrix is defined as $S_{ab} = [\sigma^{-1} \alpha]_{ab}$, and the open circuit thermal conductivity matrix is expressed as $\kappa_{ab} =[\bar{\kappa} - \bar{\alpha} \sigma^{-1} \alpha]_{ab}$. From Eq.~\eqref{current_simplified}, we see that both the charge and energy currents flow along the direction of the magnetic field. This is consistent with the fact that these originate from the chiral magnetic velocity. 

We calculate the generalized energy density using the Sommerfeld approximation in the limit $\mu \gg k_B T$. Retaining only the leading order term in the Sommerfeld expansion, we obtain
\bea \label{DOS}
\mathcal{D}_{\nu}^{\lambda} && \approx \frac{m^{3/2} \sqrt{\epsilon_\alpha}}{\sqrt{2} \pi^2 \hbar^3} \begin{cases}  \frac{\left( 1 + \lambda \sqrt{1 + \tilde{\mu}} \right)^2}{\sqrt{1 + \tilde{\mu}}} \mathcal{F}_0 ~~ &  \nu=0,\\
 \frac{\tilde{\mu}}{2 \beta \epsilon_\alpha (1 + \tilde{\mu} )^{3/2} }\mathcal{F}_2 ~~& \nu=1,\\ \frac{\left( 1 + \lambda \sqrt{1 + \tilde{\mu}} \right)^2}{\sqrt{1 + \tilde{\mu}}}  \mathcal{F}_2 ~~&\nu=2~.
\end{cases}
\eea 
Here, we have defined the scaled chemical potential, $\tilde{\mu} = \mu/\epsilon_\alpha$.
In calculating the above-generalized energy densities, we have neglected the magnetic field corrections, which are very small.
Note that i) $\mathcal{D}_0^\lambda$ becomes the exact density of states in the zero temperature limit for the corresponding bands~\cite{verma_prb20_dynamical}, and ii) $\mathcal{D}_1^\lambda$ is independent of $\lambda$ {\it i.e.}, it is identical for both the FSs.

The chiral anomaly induced transport coefficients ($\sigma$, $\alpha$, $\bar{\alpha}$, and $\bar{\kappa}$) is obtained from Eq.~\eqref{current_simplified} using the expressions of $\mathcal{C}_\nu^\lambda$, and $\mathcal{D}_\nu^\lambda$. In the 
$\mu \gg k_B T$ limit, for arbitrary orientation of the magnetic field, the anomalies induced transport coefficients are %$\beta \mu \to \infty$ limit, these are given by
\bea \label{conduct_matrix}
\begin{pmatrix}
\sigma_{ab} &  \alpha_{ab} \\
{\bar \alpha}_{ab} & {\bar \kappa}_{ab}
\end{pmatrix} && =   \frac{\tau_v e^3 B^2}{4 \pi^2 m^2 \alpha \Tilde{\mu}^2} {\cal A}_{ab}(\theta,\phi)  \\
&& \times \begin{pmatrix}
e \sqrt{1+\Tilde{\mu}} (2+\Tilde{\mu})  &  \frac{\pi^2 k_B}{6 \beta  \epsilon_\alpha} \frac{\left( \Tilde{\mu}^2 + 8(1+\Tilde{\mu}) \right)}{\tilde{\mu}  \sqrt{1+\Tilde{\mu}}}  \\
\frac{\pi^2}{6 \beta^2 \epsilon_\alpha } \frac{\left( \Tilde{\mu}^2  + 8(1+\Tilde{\mu}) \right)}{ \tilde{\mu} \sqrt{1+\Tilde{\mu}}} & \frac{\pi^2 k_B}{3 e \beta} \sqrt{1+\Tilde{\mu}} (2+\Tilde{\mu}) \nn
\end{pmatrix}~.
\eea 
Here, $\mathcal{A}(\theta, \phi)$ is a $3 \times 3$ matrix, which captures the angular dependence of all the transport coefficients, with $(\theta, \phi)$ denoting the polar, and azimuthal angle of the spherical polar coordinate for the magnetic field. The $\mathcal{A}(\theta, \phi)$ matrix is obtained to be
\be \label{angular_matrix}
{\cal A (\theta,\phi}) =
\begin{pmatrix}
\sin^2 \theta \cos^2 \phi & \frac{1}{2}\sin^2 \theta  \sin 2\phi & \frac{1}{2}\sin 2\theta \cos \phi \\
\frac{1}{2}\sin^2 \theta \sin2\phi & \sin^2 \theta \sin^2 \phi & \frac{1}{2}\sin 2\theta \sin \phi \\
\frac{1}{2}\sin 2\theta \cos \phi & \frac{1}{2}\sin 2\theta \sin \phi & \cos^2 \theta 
\end{pmatrix}. 
\ee 
As a consistency check, we note that the longitudinal electrical conductivity ($\sigma_{aa}$) derived above matches with that obtained recently in Ref.~\cite{cheon_prb22_chiral}. The conductivity matrix of Eq.~\eqref{conduct_matrix} is valid for the arbitrary direction of the applied magnetic field. So, in the planar configuration of the magnetic field ($\theta=\pi/2$), the $xy$-component of the transport coefficients represents various planar Hall effects. For instance, the $\sigma_{xy}$, $\alpha_{xy}$, $\bar \alpha_{xy}$, and $\bar \kappa_{xy}$ represents the usual planar Hall response, planar Nernst effect, planar Ettinghausen effect, and planar Righi-Leduc effects, respectively~\cite{Ashcroft76}. Hence, our work generalizes the chiral anomalies induced transport to the thermo-electric, and thermal conductivity matrices for spin-orbit coupled systems. We emphasize that the chiral anomaly induced responses of Eq.~\eqref{conduct_matrix} become zero for $\epsilon_\alpha=0$. This is expected because the system's inversion symmetry is restored as $\alpha \to 0$, causing the `Weyl' point, related Berry curvature, and chiral magnetic velocity to vanish.
\begin{figure}
    \centering
    \includegraphics[width=0.9\linewidth]{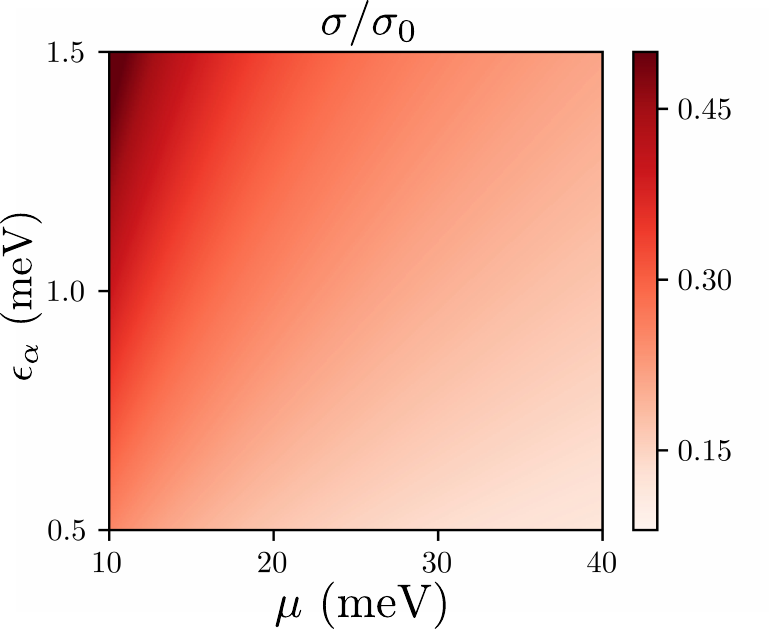}
    \caption{Variation of the chiral anomaly induced electrical conductivity with the chemical potential and the spin-orbit coupling energy strength. The electrical conductivity is expressed in units of $\sigma_0 = \frac{\tau_v e^4 B^2}{4 \sqrt{2} \pi^2 m^{3/2} \hbar}$. The anomaly-induced response is larger for larger SOC strength and smaller chemical potential.
    \label{Fig_2}}
\end{figure}

We present the variation of chiral anomaly-induced electrical conductivity with $\mu$ and $\epsilon_\alpha$  in Fig.~\ref{Fig_2}. We find that the other conductivity components of Eq.~\eqref{conduct_matrix} also follow a similar qualitative trend in $\mu$ and $\alpha$. The anomaly-induced response decreases as $\mu$ increases. This is consistent with the fact that the chiral anomalies originate from the Berry curvature, which peaks in the vicinity of the band touching points. %\textcolor{red}{ KD: Is this a conclusion from the figure? If not then please add it below the analytical result.}

To investigate the impact of the chiral anomaly on various longitudinal transport phenomena, we define the following generalized magneto-resistance, ${\rm MR}_{\cal R} \equiv {\cal R}(B)/{\cal R}(B=0) -1$. Here, $\cal R$ denotes the different transport contributions in Eq.~\eqref{conduct_matrix}. In the $\mu \gg k_B T $ limit, we calculate the Drude conductivities to be 
\bea
\begin{aligned} 
\sigma_{\rm D} &= \frac{e \tau m \alpha}{3 \hbar^4} \times \frac{2 e \epsilon_\alpha}{\pi^2 } (2+\tilde{\mu})\sqrt{1+\tilde{\mu}} ~,
\\
\alpha_{\rm D} &= - \frac{e \tau m \alpha}{3 \hbar^4} \times \frac{k_B }{3 \beta}  \frac{(3\tilde{\mu} + 4)}{\sqrt{1+\tilde{\mu}}}, 
\\
\bar{\kappa}_{\rm D} &= \frac{e \tau m \alpha}{3 \hbar^4} \times 
\frac{2 \epsilon_\alpha}{e \pi^2} \frac{\pi^2 k_B}{3 \beta} (2+\tilde{\mu})\sqrt{1+\tilde{\mu}} ~.
\end{aligned}
\eea
In this limit, the longitudinal MR in resistivity is obtained to be 
\be 
{\rm MR}_\rho  =  -\frac{3 \tau_v \gamma^2 }{3 \tau_v \gamma^2 + 4 \tau}~.
\ee
Here we have defined, $\gamma = \frac{e \hbar^3 B }{m^2 \alpha^2 \tilde{\mu}}$.
The `magneto-resistance' in the Seebeck coefficient can be calculated to be 
\be 
{\rm MR}_S = {\rm MR}_\rho \frac{4 (\tilde{\mu}^2 + 3 \tilde{\mu} +2)}{\tilde{\mu} (3 \tilde{\mu} + 4)}~.
\ee 
We note that both of these, MR$_\rho$ and MR$_S$, show negative `magneto-resistance', similar to the band-inversion WSM~\cite{kamal_prr20_thermal}. 
However, unlike the case of conventional WSM, the relation ${\rm MR}_\rho / {\rm MR}_S = 1/2$ is not satisfied in spin-orbit coupled systems. 
 
%\textcolor{blue}{Note that negative MR in Kramers-Weyl systems (which support 3D SOC-like dispersion at the time-reversal invariant points in the Brillouin zone) has also been predicted in Ref~\cite{Wan_jpcm18}. In their study,  disorder scattering and the formation of discrete Landau levels in a high magnetic field play an important role. In contrast, our results highlight the negative MR in the semiclassical regime considering only one band crossing point (or `node').}

In the case of thermo-electric and constant voltage thermal conductivity, we find 
\bea 
{\rm MR}_{\bar \kappa} &=& \frac{3 \tau_v}{4 \tau} \gamma^2~, \\
{\rm MR}_\alpha &=& -  {\rm MR}_{\bar \kappa}  \frac{  
\tilde{\mu}^2 + 8(1+\tilde{\mu})}{ \tilde{\mu} \sqrt{1+\tilde{\mu}} (3\tilde{\mu} +4)}~.
\eea
Clearly, MR$_\alpha$ is negative while MR$_{\bar \kappa}$ is positive. This is similar to the results obtained for WSM in Refs.~\cite{kamal_prr20_thermal, clemens_prb20_anisotropic}.
%
%%%%%%%%%%%%%%%%%%%%%%%%%%%%%%%%%%%%%%%%%%%%%%%%%%%%%
\section{Chiral anomaly and Spin transport}
\label{spin_tran}
Unlike WSM, where the Pauli matrices in the Hamiltonian represent pseudo-spins, the Pauli matrices in SOC systems described by Eq.~\eqref{Ham} represent physical spins. Consequently, the two bands in SOC systems are spin momentum locked with opposite spin orientations on the inner and outer FSs \cite{kapri_prb21}. 
Thus, it is natural to expect that chiral anomalies can also influence spin transport along with charge transport.
Motivated by this, we explore the chiral anomalies induced linear spin transport ($\propto {\bm E}\cdot{\bm B}$ or ${\bm \nabla}T \cdot {\bm B}$) in this section. 
 Spin transport in a 3D SOC system was recently explored in Ref.~\cite{kapri_prb21} without considering the effect of chiral anomaly. In Ref.~\cite{cheon_prb22_chiral}, the authors studied electrical chiral anomaly induced linear electrical spin current in 3D SOC systems. Here, we include the temperature gradient induced spin currents and study the chiral anomaly induced spin-Nernst effect, in addition to other effects. 

The spin current operator is defined via the anticommutator relation, $\hat{J}_a^{s_b} = \frac{1}{2} \{\hat{v}_a, \hat{s}_b  \}$, where $\hat{v}_a$ is the velocity operator, $\hat{s}_b$ is the spin operator and $a, b$ denote the Cartesian coordinates~\cite{sinova_rmp15}.  
Now, the spin current can be calculated as the expectation value of the spin current operator weighted by the non-equilibrium distribution function, 
\be \label{spin_current_def}
j^{s_b}_{a} = \sum_{\lambda} \int [d{\bm k}] D_\lambda^{-1} \bra{u_{\lambda}({\bm k})} \hat{J}_a^{s_b} \ket{u_{\lambda}({\bm k})} g_\lambda~.
\ee 
The matrix of spin transport coefficients is related to the spin current via the relation $j^{s_b}_{a} =  \sigma_{ac}^{s_b} E_c - \alpha_{ac}^{s_b} \nabla_c T$. Here, $\sigma_{ac}^{s_b}$ is the electrical spin conductivity matrix, and $\alpha_{ac}^{s_b}$ is the thermo-electric spin conductivity matrix. These tensors represent response coefficients for the spin current flowing along the $a$-direction for spin polarization along the $b$-direction, while the electric field or the temperature gradient is applied along the $c$-direction.

The spin current operator for Hamiltonian~\eqref{Ham} is given by 
\be 
\hat{J}_a^{s_b} = 
\frac{\hbar k_a}{m} \sigma_0 +  \delta_{ab} \frac{\alpha}{\hbar} \sigma_b~, 
\ee 
where $\delta_{ab} = 0$ or $1$ depending on $a \neq b$ or $a=b$, respectively. 
Using the eigenstates of Hamiltonian~\eqref{Ham}, we evaluate the expectation value of the above equation to be 
\bea 
&& \bra{u_\lambda} \hat{J}^{s_b}_a \ket{u_\lambda} = \frac{\alpha}{\hbar} \mathbb{I}_{ab} + \lambda \frac{\hbar k}{m} \mathcal{A}_{ab}(\theta_k, \phi_k)~.
\eea
Here, $\mathbb{I}$ denotes the $3\times3$ identity matrix, and $\mathcal{A}(\theta_k, \phi_k)$ is a $3\times3$ matrix defined in Eq.~\eqref{angular_matrix}.
Following the symmetric energy dispersion, the distribution function $g_\lambda$ [see Eq.~\eqref{distribution_fn}] is independent of $\theta_k$ and $\phi_k$. As a consequence, the angular integration over $\phi_k$ makes all the off-diagonal elements of $\bra{u_\lambda} \hat{J}^{s_b}_a \ket{u_\lambda}$ to be zero, and $j_a^{s_b} =0 $ for $a \neq b$. Thus, the spin current is finite only when the spins are aligned along the direction of the velocity of the carriers.  

\begin{figure} 
    \centering
    \includegraphics[width=0.9\linewidth]{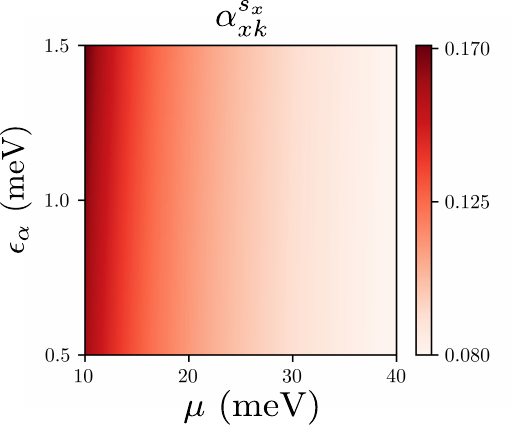}
    \caption{The variation of the longitudinal thermoelectric spin conductivity with the chemical potential ${\mu}$ and the spin-orbit coupling energy strength $\epsilon_\alpha$. The conductivity $\alpha_{xk}^{s_x}$ is scaled by $\frac{\tau_v e k_B B}{9\sqrt{2} \hbar^2 \beta \sqrt{m}}$. Similar to the chiral anomaly induced electrical response, the chiral anomaly induced spin response is also larger for larger spin-orbit coupling and smaller chemical potential. 
    \label{Fig_3}}
\end{figure}
Hence, the chiral anomaly induced spin currents are finite only when the spins are polarized along the respective directions of current, and we have $j_x^{s_x}=j_y^{s_y}=j_z^{s_z}=j^s_{\rm CA}$. 
We calculate the spin current induced by the chiral anomalies to be [see Appendix~\ref{spin_currents_calc} for details]
\bea 
j^s_{\rm CA} =&& \tau_v \sum_\lambda \frac{\mathcal{C}_0^\lambda}{\mathcal{D}_0^\lambda}\left[ \frac{\mathcal{D}_1^\lambda}{\mathcal{D}_2^\lambda } \mathcal{L}_1 - \mathcal{L}_0 \right] e{\bm E}\cdot {\bm B} \nn \\
&& - \frac{\mathcal{C}_2^\lambda}{\mathcal{D}_2^\lambda} \left[ \frac{\mathcal{D}_1^\lambda}{ \mathcal{D}_0^\lambda} \mathcal{L}_0 -  \mathcal{L}_1 \right] k_B {\bm \nabla}T\cdot {\bm B}~.
\eea 
Here, we have defined 
\be \mathcal{L}_\nu = \int [d{\bm k}] \left(\frac{\alpha}{\hbar} + \lambda \frac{\hbar }{m} {\bm k}_a \cdot \hat{\bm k} \right) \left(\frac{\epsilon_\lambda -\mu}{k_B T} \right)^\nu \left(- \frac{\partial f_\lambda}{\partial \epsilon_\lambda} \right)~,
\ee 
with ${\bm k}_a = k_a {\hat a}$ being a vector along $a$-direction with magnitude equal to the component of $\bm k$ along the $a$-direction, and $\hat{\bm k}=\sin\theta_k \cos\phi_k \hat{\bm x}+\sin\theta_k \sin\phi_k \hat{\bm y}+\cos\theta_k \hat{\bm z}$. We now have $j^{s_a}_a \propto {\bm E}\cdot{\bm B}$ for any arbitrary direction of the applied electric field along the $k$-direction.  We calculate the corresponding chiral anomaly induced electrical spin conductivity to be, 
\be \label{electrical_spin_condct}
\sigma_{xc}^{s_x} = \sigma_0^s   \left[ \sqrt{1+\Tilde{\mu}}
 - \frac{\pi^2 }{6 \beta^2 \epsilon_\alpha^2} \frac{(\tilde{\mu}^2 + 9 \tilde{\mu} -20 ) }{\tilde{\mu}^2 (1+\tilde{\mu})^2} \right] \hat{c}\cdot \hat{\bm B}~,  \\
\ee
where we have defined $\sigma^s_0=\frac{\tau_v e^2 B \alpha }{6\pi^2 \hbar^3 }$.
The second term on the right-hand side of Eq.~\eqref{electrical_spin_condct} is the finite temperature correction to the electrical spin conductivity, which vanishes in the $T \to 0$ limit. 

For the thermoelectric part of the spin conductivity, we find that it behaves like the electric spin conductivity. All the thermoelectric spin currents, where the spin is not aligned along the current direction, vanish. We obtain, $j^{s_b}_{a} = 0$ for $b\neq a$, and  $j^{s_a}_{a} \propto {\bm \nabla}T \cdot{\bm B}$.   Our  calculations show that only the conductivity components,  $\alpha_{ac}^{s_a}$ are finite, and $\alpha_{xc}^{s_x} = \alpha_{yc}^{s_y} = \alpha_{zc}^{s_z}$. 
We calculate the thermoelectric spin conductivity for the temperature gradient applied along the $c$-direction to be, 
\be
\alpha_{xc}^{s_x} = \alpha_0^s 
 \left[\frac{2}{\tilde{\mu}^2}+\frac{\tilde{\mu}^2 + 3 \tilde{\mu} -2}{ \tilde{\mu}^2\sqrt{1+\tilde{\mu}}}  - \frac{\tilde{\mu}^2 + 7 \tilde{\mu} +6}{ 2\tilde{\mu} (1+\tilde{\mu})^{3/2}}  \right] \hat{c}\cdot \hat{\bm B}~, \label{thermoelectric_spin_condct} \\
\ee
where $\alpha_0^s = \frac{\tau_v e k_B \alpha B}{18 \hbar^3 \beta \epsilon_\alpha}$.
The above expression represents the chiral anomaly induced spin-Seebeck (for $c=x$) or the spin-Nernst coefficient (for $c \neq x$), with the spins polarized along the $x$-direction. The variation of $\alpha_{xk}^{s_x}$ with $\mu$ and $\epsilon_\alpha$ is presented in Fig.~\ref{Fig_3}. The electrical spin conductivity also follows similar trends in $\mu$ and $\epsilon_\alpha$. The anomaly induced effects in general decrease with increasing $\mu$ and increase with increasing $\alpha$ which is a proxy for the degree of inversion symmetry breaking. 
%
%%%%%%%%%%%%%%%%%%%%%%%%%%%%%%%%%%%%%%%%%%%%%%%%%%%%%
\section{Conclusion}
\label{conclu}
In summary, we have provided evidence that quantum chiral anomalies can be understood as a feature of FSs. Specifically, the chirality of charge carriers can be determined by the sign of the Berry curvature quantum passing through the associated Fermi surface. This has significant implications for 3D SOC metals or Kramers-Weyl metals, where chiral charge pumping can occur across the two Fermi surfaces associated with a single Kramers-Weyl node. To the best of our knowledge, this kind of chiral anomaly has no analog in relativistic field theories of chiral fermions. We have also demonstrated the existence of three distinct types of quantum chiral anomalies -- electrical, thermal, and gravitational -- in 3D SOC metals and Kramers-Weyl metals.

The effect of these quantum chiral anomalies can be observed in electrical and thermo-electric charge and spin transport in 3D SOC metals and Kramers-Weyl metals. While the electrical transport signatures of chiral anomalies in 3D spin-orbit coupled metals are similar to those in Weyl semimetals, the signatures in electrical and thermo-electric spin transport are unique to 3D SOC metals. We have shown that spin conductivities are finite only when spins are polarized along the direction of carrier flow. we found that the chiral anomaly-induced spin conductivities are proportional to the strength of the magnetic field, unlike charge conductivities which scale with the square of the magnetic field. Our findings contribute to the understanding of chiral anomaly induced charge, heat, and spin transport in 3D SOC metals and Kramers-Weyl systems.

 \section*{Acknowledgements}
 We acknowledge the Science and Engineering Research Board (SERB, via project MTR/2019/001520) for financial support. 
% CRG/2018/002400), and the Government of India's Department of Science and Technology (DST, for project DST/NM/TUE/QM-6/2019(G)-IIT
% Kanpur) for financial support. K.D.
% thank IIT Kanpur for the research fellowship. 
S. D. thanks the MHRD, India for funding through the Prime Minister's Research Fellowship (PMRF). We sincerely thank Atasi Chakraborty for the useful discussions.
\appendix

\section{3D non-centrosymmetric SOC metals and Kramers-Weyl metals} \label{Kramers_Weyl}
In this appendix, we discuss the SOC-induced chiral anomaly in other 3D systems with different forms of the SOC, compared to Eq.~\eqref{Ham}. Comparing the list of single crystalline point groups which support 3D spin-orbit coupled metals \cite{SAMOKHIN09} with the list of Kramers Weyl metals \cite{Chang18_NM}, we find that these are identical. However, 3D electron gas with SOC can also arise in some heterostructures of two different single crystals. Both of these systems have doubly degenerate band touching points, which we refer to as `Kramers-Weyl' points. Kramers-Weyl metals are realized in structurally chiral crystals that lack mirror, inversion, or roto-inversion symmetry~\cite{Chang18_NM}. There are 65 Sohncke chiral space groups corresponding to 11 chiral point groups which characterize the structurally chiral crystals~\cite{Schroter_NP19}.

The bands of non-magnetic chiral crystals are at least doubly degenerate at the time-reversal-invariant momenta (TRIM) points due to Kramers theorem~\cite{Chang18_NM}. However, the SOC lifts the Kramer's degeneracy at all other points in the momentum space, leaving behind `Weyl'-like Kramers-Weyl nodes at the TRIM points. All these band degenerate points are topologically non-trivial, carrying finite Chern numbers~\cite{Chang18_NM}. In general, the chiral crystals can host multiple band crossings at the TRIM points in the Brillouin zone along with multi-fold band degeneracy~\cite{Chang18_NM, Schroter_NP19, law21_comm_phys, Barnevig_science16, debasis_prb22}. 

In this paper, we focus on Kramers-Weyl metals that have a two-fold degenerate Kramers-Weyl point at TRIM. In Table~\ref{table1}, we summarize the chiral space groups and point groups which support Kramers-Weyl fermions, along with some material examples \cite{cheon_prb22_chiral, SAMOKHIN09,felsar_materials22, Chang18_NM}. The generic Kramers-Weyl system will have a low energy Hamiltonian of the form: ${\cal H} = \sum_{ab} {\hbar^2 k_a k_b}/{(2m_{ab})} + {\bm h_{\bm k}}\cdot{\bm \sigma}$, in the vicinity of the Kramers-Weyl point for which $|{\bm h_{\bm k}}| = 0$. Here, $a,b=x,y,z$, $m_{ab}$ is the effective mass tensor, and ${\bm k}$ is the momentum with respect to the Kramers-Weyl point. The specific form of symmetry allowed ${\bm h_{\bf k}}$, for each of the chiral point groups is also summarized in Table~\ref{table1}. Each of these Kramers-Weyl points has a chiral charge with value $\pm 1$. For example, the Hamiltonian~\eqref{Ham} with isotropic SOC term $\alpha {\bm k}\cdot{\bm \sigma}$ can be realized in point groups {\bf T} and {\bf O} in K$_2$Sn$_2$O$_3$, $\beta$-RhSi, CoSi crystals~\cite{takane_prl19, Rao_N19, Sanchez_N19, Chang18_NM, debasis_prb22, law21_comm_phys}.

\begin{widetext}

\begin{table}[h!]
 \begin{center}
    \caption{The space groups and the point groups for topologically non-trivial chiral crystals hosting Kramers-Weyl Fermions with chiral charge $\pm 1$. Some material examples, along with the form of the symmetry-allowed SOC terms in the vicinity of the Kramers-Weyl points for each space group are also presented.}
    \label{table1}
    \begin{tabular}{c c c c}
    \hline
    \hline
    \textbf{Space group} ~& \textbf{Point group (Laue class)}  ~&  {\bf Material} ~& \textbf{SOC term}
      \\
     \hline
     \hline 
    1 & ${\bf C}_1(1)$ &   Li$_{6}$CuB$_4$O$_{10}$ & $\thead{(\alpha_1  k_x + \alpha_2 k_y + \alpha_3 k_z) \sigma_x + (\alpha_4 k_x + \alpha_5 k_y + \alpha_6 k_z) \sigma_y \\ + (\alpha_7 k_x + \alpha_8 k_y + \alpha_9 k_z) \sigma_z }$  \\ \hline 
    3-5 & ${\bf C}_2(2)$ & Pb$_3$GeO$_5$  & $(\alpha_1 k_x + \alpha_2 k_y ) \sigma_x   + (\alpha_3 k_x + \alpha_4 k_y ) \sigma_y + \alpha_5    k_z \sigma_z $ \\  \hline 
    16-24 & ${\bf D}_2(222)$ & AlPS$_4$ & $\alpha_1 k_x \sigma_x + \alpha_2 k_y \sigma_y + \alpha_3 k_z \sigma_z$ \\  \hline 
  143-146  & ${\bf C}_3 (3)$  & $\beta$-Ag$_3$IS \\
 75-80   & ${\bf C}_4(4)$ & BaCu$_2$Te$_2$O$_6$Cl$_2$ & $(\alpha_1   k_x + \alpha_2 k_y) \sigma_x   + (   \alpha_1 k_y - \alpha_2 k_x ) \sigma_y   + \alpha_3 k_z \sigma_z $ \\
  168-173  & ${\bf C}_6 (6)$ & $\alpha$-In$_2$Se$_3$\\  \hline
 149-155   & ${\bf D}_3 (32)$ & Ag$_3$BO$_3$ \\
 89-98 & ${\bf D}_4(422)$ & CdAs$_2$ & $\alpha_1 (k_x \sigma_x + k_y           \sigma_y) +  \alpha_2 k_z \sigma_z$ \\
 177-182 & ${\bf D}_6 (622)$ & NbGe$_2$ &
     \\ \hline
  195-199  & ${\bf T}(23)$ & K$_2$Sn$_2$O$_3$, $\beta$-RhSi & $\alpha_1 (k_x \sigma_x + k_y \sigma_y + k_z \sigma_z)$ \\ 
207-214 & ${\bf O}(432)$ & BaSi$_2$, SrSi$_2$  \\ \hline 
    \hline 
    \end{tabular}
 \end{center}
\end{table}

\end{widetext}

\section{Berry curvature flux quantum and chiral anomaly for negative chemical potential}
\label{Appendix_B}
In this Appendix, we calculate the Berry curvature flux quantum for each Fermi surface and discuss the chiral anomaly for Fermi energies below the Kramers-Weyl node, i.e., $\mu<0$. We start by calculating the Berry curvature flux quantum for the FSs. The Berry curvature flux through any FS is defined as $\mathcal{C}_\lambda = \frac{1}{2\pi} \int_{\rm FS} d{\bm S} \cdot {\bm \Omega}_{\lambda}$, where $d{\bm S}$ is the elemental surface area of the FS. Using the divergence theorem, and capturing the Fermi surface via the Heaviside step function [$\Theta(\mu - \epsilon_\lambda)$], we have 
\bea
\mathcal{C}_\lambda &&= \frac{1}{2\pi} \int d{\bm k} {\bm \nabla}_{\bm k}\cdot {\bm \Omega}_\lambda \Theta(\mu-\epsilon_\lambda)  \nn \\
&& = -\frac{1}{2\pi}  \int d{\bm k}~ {\bm \Omega}_{\lambda}\cdot {\bm \nabla}_{\bm k}\Theta(\mu-\epsilon_\lambda)  \nn \\
&&= \frac{\hbar}{2\pi}  \int d{\bm k}~ {\bm \Omega}_{\lambda}\cdot {\bm v}_\lambda \delta(\mu - \epsilon_\lambda)~. \label{BC_flux_velocity_def}
\eea
Note that in the zero-temperature limit, the above expression reduces to the electrical chiral anomaly coefficient defined in Eq.~\eqref{coffcnts}. Below, we explicitly calculate the $\mathcal{C}_\lambda$.

{\bf Case I} ({\bm $\mu>0$}):---   
For $\mu >0$, there are two Fermi wave vectors $k_{\lambda}^F = -\lambda k_\alpha + \sqrt{k_\alpha^2 + 2m \mu/ \hbar^2}$ with $\lambda=\pm$, corresponding to two FSs of the two bands. The $k^F_{+}$ ($k^F_-$) corresponds to the inner (outer) FS. %, and outer ($k_+^F$) FS. 
Now, using the expressions of ${\bm v}_\lambda$, ${\bm \Omega}_\lambda$, and the $\delta$-function property, $\mathcal{C}_\lambda$ for each band $\lambda$ becomes
\bea \label{C_chi_positive energy}
\mathcal{C}_\lambda &&= \frac{\hbar}{2 \pi}\int d{\bm k} \frac{-\lambda}{2 k^2} \left(\frac{\hbar k}{m} + \lambda \frac{\alpha}{\hbar} \right)  \delta(\mu -\epsilon_\lambda)~, \nn \\
&&= -\lambda \int dk \left(\frac{\hbar^2 k}{m} + \lambda \alpha \right) \frac{\delta(k^F_\lambda - k)}{|\epsilon'_\lambda|}~. \label{BC_flux_mu>0}
\eea 
Here, $\epsilon'_\lambda$ is the first derivative of $\epsilon_\lambda$ with respect to $k$. Evaluating this integral yields $\mathcal{C}_\lambda = -\lambda$. 

\begin{figure}
   \centering
     \includegraphics[width=\linewidth]{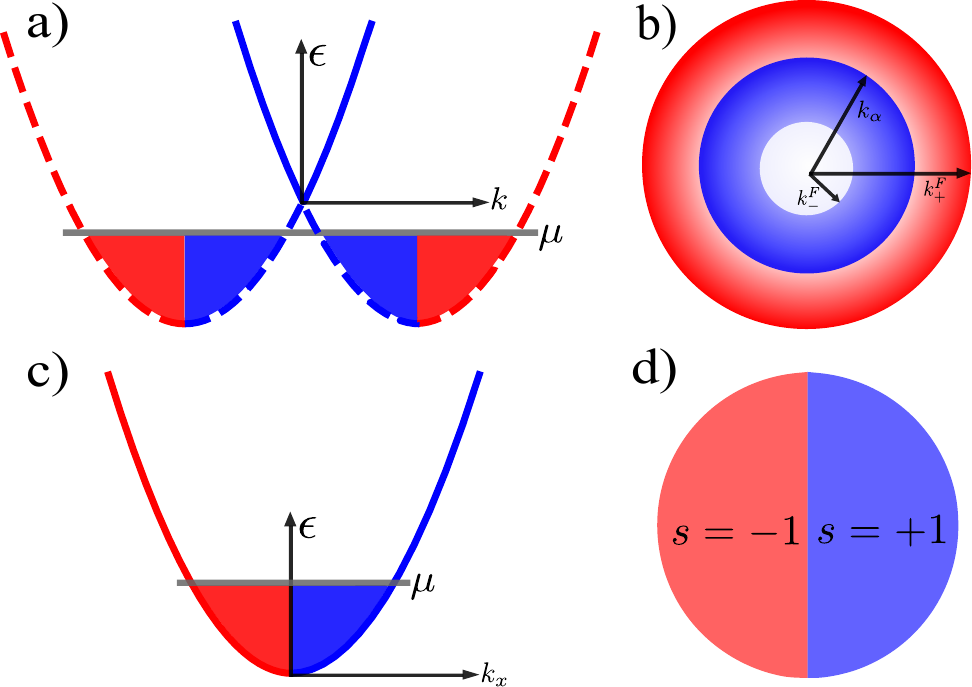}
    \caption{a) The band dispersion and the Brillouin zone partitioning for the $\lambda=-1$ band of a 3D SOC system. %The gray horizontal line represents the position of Fermi energy, $\mu$. Fig. a) highlights the Brillouin zone partitioning for the $\lambda=-1$ band (dashed line). 
    For the blue-shaded region with negative band velocity, the Berry curvature flux is $-1$, while the Berry curvature flux is $+1$ for the red-shaded region with positive band velocity. b) The corresponding cross-section of the Fermi surface for $\mu<0$ for the $\lambda=-1$ band, highlighting the two partitions of the Fermi pocket. c) The band dispersion of 3D electron gas without SOC. This trivial system can also be partitioned into red and blue regions depending on the sign of the $x$ component of the band velocity. d) The cross-section of the spherical FS for a 3D electron gas in the $k_x-k_y$ plane. %For the blue (red)-shaded region, the band velocity along the x-direction is positive (negative). 
     \label{Fig4}}
\end{figure}

{\bf Case II} ({\bm $\mu < 0$}):--- 
For $\mu <0$, there is only one carrier pocket, which looks like an annular sphere. It has two surfaces, the inner and the outer surfaces. 
%Now, if we calculate the Berry curvature flux quantum using the formula $\mathcal{C}_\lambda = \frac{1}{2\pi} \int_{\rm FS} d{\bm S} \cdot {\bm \Omega}_{\lambda}$---assuming the $d{\bm S}$ to point outward, we get $\mathcal{C}_\lambda= +1$ for both the inner and the outer FS (because both $d{\bm S}$ and ${\bm \Omega}_\lambda$ are positive). Thus from the carrier pocket perspective, there is only one kind of `flavored' carrier in the system. Consequently, in contrast to the $\mu>0$ case, a chiral anomaly within a single Kramers-Weyl node is not possible for the $\mu<0$ case. The possibility of having a chiral anomaly between a pair of Kramers Weyl nodes of opposite chirality cannot be ruled out.  
%Interestingly, there is another way to interpret the physics in these systems below $\mu<0$, which allows for the possibility of a chiral anomaly within a single electron pocket. 
Also, the energy dispersion is non-monotonic (see Fig.~\ref{Fig4}). Consequently, in the region near the Kramers-Weyl node, the band velocity is negative, while in other regions, the band velocity is positive. 
As a result, we have regions within the same pocket that have opposite signs of the chiral magnetic velocity ($\propto {\bm v}_\lambda \cdot {\bm \Omega}_\lambda$). This ensures that the Berry curvature flux through the entire FS calculated using Eq.~\eqref{BC_flux_velocity_def} is zero. Hence, we expect that there should not be any chiral anomaly for $\mu<0$. 

However, in Ref.~\cite{cheon_prb22_chiral}, the authors discussed the chiral anomaly for $\mu<0$ with the idea of partitioning the FS into two regions based on the sign of the chiral magnetic velocity. Below, we discuss this in detail.
%This motivates the partitioning of the electron pocket into two regions so that each region supports either positive or negative chiral magnetic velocity. 
We show the partitioning of the FS in Fig~\ref{Fig4}, with the blue and red regions representing the two different partitions. Here, the $\chi$ is used as the index for denoting the inner (outer) region of the FS, with $\chi=-1$ for the blue region ($\chi=+1$ for the red region).
%
%With this Brillouin zone partitioning, we calculate the Berry curvature flux quantum through each region of the Fermi pocket as follows $\mathcal{C}_\lambda^\chi=\frac{1}{2\pi} \int_{\rm FS_{\chi}} d{\bm S}_\lambda \cdot {\bm \Omega}_{\lambda}$. Here, in contrast with the $d{\bm S}$, the $d{\bm S}_\lambda$ has the direction along the group velocity, i.e., inward for the inner FS pocket and outward for the outer FS pocket. Using the above formula, it can be easily seen that $\mathcal{C}_\lambda^\chi  = \chi$ for the two regions. Below, we prove this in another way using Eq.~\eqref{BC_flux_velocity_def}, where it will be more clear how the direction of group velocity plays a vital role in deciding the sign of Berry curvature flux quantum through each region of the Fermi pockets. 
%
To calculate the Berry curvature flux using Eq.~\eqref{BC_flux_velocity_def}, we first compute the Fermi wave vectors corresponding to the two different regions of the Fermi pocket of the $\lambda=-1$ band. The Fermi wave vectors corresponding to the inner ($\chi = -1$) and outer ($\chi = -1$) boundary of the Fermi pocket is given by $k^F_\chi = k_\alpha + \chi \sqrt{k_\alpha^2 + 2m \mu/ \hbar^2}$. Recall that $k_\alpha = m \alpha/\hbar^2$, which corresponds to the minima in the energy of the $\lambda=-1$ band. The $\chi=-$ ($+$) region of the Fermi pocket correspond the $k^F_- < k < k_\alpha$ ($ k_\alpha < k < k^F_+$). These regions are represented by blue and red colors, respectively, in Fig.~\ref{Fig4}. 
For $\lambda=-1$ band, the $\mathcal{C}_\lambda$ is given by  
\be \label{C_chi_negative energy}
\mathcal{C}_\lambda = \frac{\hbar}{2 \pi}\int d{\bm k} \frac{1}{2 k^2} \left(\frac{\hbar k}{m} - \frac{\alpha}{\hbar} \right) \delta(\mu - \epsilon_{-})~.  
\ee
Now, for either of the two regions, the above equation reduces to 
\be \label{C^chi_lamba}
\mathcal{C}_\lambda^\chi =  \int dk \left(\frac{\hbar^2 k}{m} - \alpha \right) \frac{\delta(k^F_\chi - k)}{|\epsilon'_{-}|}~. 
\ee 
As the band velocity, $\epsilon_{-}' = \hbar^2 k/m - \alpha$ is negative (positive) for the region with $k^F_-< k < k_\alpha$ ($ k_\alpha < k < k_+^F$), Eq.~\eqref{C^chi_lamba} yields $\mathcal{C}_\lambda^\chi = \chi$. We note again that the sign of ${\cal C}^\chi_\lambda$ is essentially tied to the sign of the chiral magnetic velocity proportional to the $({\bm \Omega}_{\lambda}\cdot{\bm v}_\lambda)$ term. We emphasize that without partition of the FS, Eq.~\eqref{C_chi_negative energy} itself yields zero due to the two different roots of the $\delta$-function ($k^F_{+}$ and $k^F_{-}$). 
This partitioning of the Brillouin zone, as per the sign of the chiral magnetic velocity, allows one to define two regions of FS with opposite Berry curvature flux quantum. {This had been used in Ref.~\cite{cheon_prb22_chiral} to discuss the continuity equation and the associated electrical chiral anomaly for $\mu<0$, on the same footing as we have discussed for $\mu>0$~\cite{cheon_prb22_chiral} in the main text.} %A similar approach was also presented in Ref.~\cite{cheon_prb22_chiral} for discussing the electrical chiral anomaly in SOC metals. 
While partitioning a single electron pocket to define carriers of different `flavors' is mathematically appealing, we believe that this way of defining the chiral anomaly is superfluous and not physical. 

Here, we present a counter-example to establish the above claim.
Consider a 3D electron gas (without any SOC, without any magnetic field), with an electric field applied along the $x$-direction. We can divide the Fermi sphere of the system into two halves with positive and negative velocities along the $x$-axis and treat the particles with opposite velocities as having different flavors ($s=\pm$). In the bottom panels (c) and (d) of Fig.~\ref{Fig4}, we have schematically shown this partitioning of the FS. The particles in the blue (red) region with $s=+1$ ($s=-1$), have positive (negative) velocity. In the presence of only an electric field along the $x$-direction, the collisionless Boltzmann equation [Eq.~\eqref{bte_1} with $\lambda \to s$ and $\mathcal{I}_{\rm coll}\{g_\lambda\}=0$], upto first order in the electric field strength, becomes 
\be 
\dfrac{\partial g_s}{\partial t} - e E v_x^s \frac{\partial f_s}{\partial \epsilon} = 0~.
\ee 
Here, $g_s$ ($f_s$) is the non-equilibrium (equilibrium Fermi Dirac)   distribution function for the $s$ region of the Brillouin zone. 
Integrating the above equation over all the momentum states within the respective partition of the FS, we obtain
\be
\frac{\partial {\cal N}_s}{\partial t} + s \frac{e E}{2 \pi^3} \left( \frac{2 m \mu}{\hbar^2} \right)^{3/2} = 0~.
\ee  
The above equation resembles Eq.~\eqref{collisionless_particle_cont}, indicating the possibility of a ``chiral anomaly" in a normal 3D electron gas. However, this cannot be physical and is very unlikely to be correct. Due to this, we believe that the partitioning of the BZ to divide one electron/hole pocket into multiple partitions is not physically acceptable. However, the partitioning of the BZ to include full electron/hole pockets is acceptable, and this forms the basis of valley physics in 2D and chiral anomaly related physics in 3D systems.

%It is not clear whether this will manifest in experiments or not. 

%%%%%%%%%%%%%%%%%%%%%%%%%%%%%%%%%%%%%%%%%%%%%%%%%%%%%

Having discussed the chiral anomaly for $\mu<0$, we conclude this Appendix with a small discussion on the chirality of `Weyl'-type nodes and the Berry curvature flux quantum. For the WSM, the Berry curvature flux through the FS of a node represents the `chirality' of that node, irrespective of the conduction or the valence band. This is easily seen because, in the $m \to \infty$ limit, the Hamiltonian in Eq.~\eqref{Ham} reduces to the Hamiltonian for a single Weyl node $\mathcal{H}_{\rm WSM}$. In contrast to the bands of Hamiltonian in Eq.~\eqref{Ham}, both bands of $\mathcal{H}_{\rm WSM}$ are monotonous (around the nodal point) and only one FS exists at any particular energy. Then a straightforward calculation following Eq.~\eqref{BC_flux_mu>0} yields, $\mathcal{C}_\lambda =-{\rm sign}(\alpha)$ for both the conduction and valence band of $\mathcal{H}_{\rm WSM}$. Because the $\mathcal{C }_\lambda$ depends on the sign of $\alpha$, the Berry curvature flux quantum becomes opposite for opposite chirality nodes where $\alpha$ has the opposite sign. This establishes that for WSM, the chirality of each Weyl node can be represented as the Berry curvature flux quantum through the node~\cite{armitage_rmp18_weyl,son_prb13_chiral,kamal_prr20_thermal, Lee_prb22}. However, for the Kramers-Weyl nodes, the Berry curvature flux quantum and the chirality of the node are not identical. The chirality of the Kramers-Weyl nodes depends on the sign of $\alpha$ for Hamiltonian \eqref{Ham}, which is specific to a given TRIM point of the material~\cite{Chang18_NM}. 

\section{Calculation of equilibrium currents} \label{Appendix_C}
In this Appendix, we derive the expressions of the equilibrium currents obtained in Eqs.~\eqref{j_eq_e} and \eqref{j_eq_E}. In the presence of only a magnetic field, the velocity of the center of mass of the wave packets for the carriers in each band is given by ${\bm \dot{r}}_\lambda = D_\lambda \left[{\bm v}_\lambda + \frac{e}{\hbar} ({\bm v}_\lambda \cdot {\bm \Omega}_\lambda) {\bm B} \right]$. The equilibrium charge current for the FS $\lambda$ (corresponding to each band) is given by 
\be
{\bm j}_{e, \rm eq}^\lambda = -e \int [d{\bm k}] {\bm \dot{r}} f_\lambda
 = -e {\bm B}\int [d{\bm k}] \frac{e}{\hbar} ({\bm v}_\lambda \cdot {\bm \Omega}_\lambda) f_\lambda~.
\ee
Here, we have used the fact that the band velocity ${\bm v}_\lambda$ does not contribute to the equilibrium current (due to angular integration being zero).
Now, we use the identity ${\bm \nabla}_k \cdot (\epsilon_\lambda {\bm \Omega}_\lambda) = {\bm \nabla}_k \epsilon_\lambda \cdot {\bm \Omega}_\lambda + \epsilon_\lambda {\bm \nabla}_k \cdot {\bm \Omega}_\lambda$ to express the above equation as,  
\bea
{\bm j}_{e, \rm eq}^\lambda &&=  -\frac{e^2 {\bm B}}{\hbar^2}  \int [d{\bm k}] \left[  {\bm \nabla }_k \cdot (\epsilon_\lambda {\bm \Omega}_\lambda)  - \epsilon_\lambda {\bm \nabla}_k \cdot {\bm \Omega}_\lambda \right]  f_\lambda  \label{eq_current_step1} \\
&& = -\frac{e^2 {\bm B}}{\hbar^2}  \int [d{\bm k}] {\bm \nabla }_k \cdot (\epsilon_\lambda {\bm \Omega}_\lambda) f_\lambda \label{eq_current_step2} \\
&& =  \frac{e^2}{\hbar^2} {\bm B} \int [d{\bm k}]  \epsilon_\lambda {\bm \Omega}_\lambda \cdot \hat{\bm k}  \frac{\partial f_\lambda}{\partial k}, \label{eq_current_step3} \\
&&= -e {\bm B} \int [d{\bm k}] \left( \mu + \epsilon_\lambda -\mu  \right) \frac{e}{\hbar} ({\bm v}_\lambda \cdot {\bm \Omega}_\lambda) \left(- \frac{\partial f_\lambda}{\partial \epsilon_\lambda} \right), \nn \\
&&= -e \left(\mu \mathcal{C}_0^\lambda + k_B T \mathcal{C}_1^\lambda \right) {\bm B}. \label{eq_current_final}
\eea 
To evaluate Eq.~\eqref{eq_current_step2}, we have used the fact that ${\bm \nabla}_k \cdot {\bm \Omega}_\lambda = \pm 2\pi \delta^3({\bm k})$, for a system with doubly degenerate band touching point with linear dispersion. This makes the last integral of Eq.~\eqref{eq_current_step1} to be zero. To obtain Eq.~\eqref{eq_current_step3} from Eq.~\eqref{eq_current_step2}, we have used integrations by parts. 
Here, we have defined $\mathcal{C}_\nu^\lambda$ as,   
\be 
\mathcal{C}_\nu^\lambda= \int [d {\bm k}] \frac{e}{\hbar} {\bm v}_\lambda \cdot {\bm \Omega}_\lambda \left(\frac{\epsilon -\mu }{k_B T} \right)^{\nu} \left(- \frac{\partial f_\lambda}{\partial \epsilon_\lambda} \right)~.
\ee
These can also be rewritten in terms of $\mathcal{C}_\lambda$ given in Eq.~\eqref{coffcnts}.
The energy current, ${\bm j}^\lambda_{\epsilon, \rm eq}$, can be evaluated in a similar manner.

\section{Details of spin current calculations} \label{spin_currents_calc}
To calculate the spin current proportional to the ${\bm E}\cdot {\bm B}$ (or ${\bm \nabla}T \cdot {\bm B}$), we consider the band velocity term of Eq.~\eqref{eom_r} and calculate the spin current operator. The band velocity operator along the $i$-direction is given by $\hat{v}_i = \frac{\hbar k_i}{m} \sigma_0 + \frac{\alpha}{\hbar} \sigma_i$. Without loss of generality, here we show the calculation of spin current in the $x$-direction.
Using the expressions of the eigenstates and the spin current operator given in the main text, we obtain $\bra{u_\lambda} \hat{J}^{s_x}_x \ket{u_\lambda} = \left(\alpha/\hbar + \lambda \hbar k_x \sin\theta_k \cos\phi_k /m \right)$. Now, the chiral anomaly induced spin current is given by 
\begin{widetext}
\bea 
j_x^{s_x} && = \tau_v \sum_\lambda \int [d{\bm k}] \left(\frac{\alpha}{\hbar} + \lambda \frac{\hbar k_x}{m} \sin\theta_k \cos\phi_k \right)  \left(\delta \mu_\lambda + \frac{\epsilon_\lambda -\mu}{T} \delta T_\lambda \right) \left(- \frac{\partial f_\lambda}{\partial \epsilon_\lambda} \right).  
\eea 
In the $\beta \mu \to \infty$ limit, writing the expressions of $\delta \mu_\lambda$ and $\delta T_\lambda$ explicitly, we have
%
%\begin{widetext}
\bea \label{j_xspin_current}
j_x^{s_x} =&& \tau_v \sum_\lambda \left[ \frac{\mathcal{D}_1^\lambda \mathcal{C}_0^\lambda}{\mathcal{D}_2^\lambda \mathcal{D}_0^\lambda} \mathcal{L}_1 - \frac{ \mathcal{C}_0^\lambda}{ \mathcal{D}_0^\lambda} \mathcal{L}_0 \right] e{\bm E}\cdot {\bm B} - \left[ \frac{\mathcal{D}_1^\lambda \mathcal{C}_2^\lambda}{\mathcal{D}_2^\lambda \mathcal{D}_0^\lambda} \mathcal{L}_0 - \frac{\mathcal{C}_2^\lambda}{\mathcal{D}_2^\lambda } \mathcal{L}_1 \right] k_B {\bm \nabla}T\cdot {\bm B}.
\eea 
\end{widetext}
The definition of $\mathcal{L}_\nu$ is given in the main text. We evaluate the $\mathcal{L}_\nu$ using the Sommerfeld approximation in the $\mu \gg k_B T$ limit. We obtain the following expressions
\bea 
\mathcal{L}_0 &&=-\lambda\frac{m^2 \alpha^2}{6 \pi^2 \hbar^5}\frac{[\tilde{\mu} -\tilde{\mu}^2 +2(1+\tilde{\mu})]}{1+\tilde{\mu}},
\\
\mathcal{L}_1 &&=\frac{k_BT}{9\hbar^3}\frac{(-\lambda + \sqrt{1+\tilde{\mu}}) [\lambda(2+\tilde{\mu}) + \sqrt{1+\tilde{\mu}}]}{(1+\tilde{\mu})^{3/2}}.
\eea 
Using these expression along with $\mathcal{C}_\nu^\lambda$ and $\mathcal{D}_\nu^\lambda$ in Eq.~\eqref{j_xspin_current}, we obtain the spin conductivities of Eqs.~\eqref{electrical_spin_condct} and \eqref{thermoelectric_spin_condct}.
Following a similar procedure, we can calculate other spin currents.

We show that due to rotational symmetry $j^{s_j}_i =0$ for $i \neq j$. Without loss of generality, we will explicitly show the calculation for $j^{s_z}_x$. The expectation value of the spin current operator $\hat{J}^{s_z}_x$ is given by $\bra{u_\lambda} \hat{J}^{s_z}_x \ket{u_\lambda} = \lambda\frac{p}{2m} \sin2\theta_k \cos\phi_k$. Now, as the distribution function is independent of $\theta_k$ and $\phi_k$, so the angular integration over $\phi_k$ of the $\bra{u_\lambda} \hat{J}^{s_z}_x \ket{u_\lambda}$ yields $j^{s_z}_x=0$. Similarly, all the spin currents with spin polarization perpendicular to the propagation velocity can be easily shown to be zero due to the vanishing angular integration over $\phi_k$. 

%%%%%%%%%%%%%%%%%%%%%%%%%%%%%%%%%%%%%%%%%%%%%%%%%%%%
\bibliography{Ref}

\begin{thebibliography}{83}%
\makeatletter
\providecommand \@ifxundefined [1]{%
 \@ifx{#1\undefined}
}%
\providecommand \@ifnum [1]{%
 \ifnum #1\expandafter \@firstoftwo
 \else \expandafter \@secondoftwo
 \fi
}%
\providecommand \@ifx [1]{%
 \ifx #1\expandafter \@firstoftwo
 \else \expandafter \@secondoftwo
 \fi
}%
\providecommand \natexlab [1]{#1}%
\providecommand \enquote  [1]{``#1''}%
\providecommand \bibnamefont  [1]{#1}%
\providecommand \bibfnamefont [1]{#1}%
\providecommand \citenamefont [1]{#1}%
\providecommand \href@noop [0]{\@secondoftwo}%
\providecommand \href [0]{\begingroup \@sanitize@url \@href}%
\providecommand \@href[1]{\@@startlink{#1}\@@href}%
\providecommand \@@href[1]{\endgroup#1\@@endlink}%
\providecommand \@sanitize@url [0]{\catcode `\\12\catcode `\$12\catcode
  `\&12\catcode `\#12\catcode `\^12\catcode `\_12\catcode `\%12\relax}%
\providecommand \@@startlink[1]{}%
\providecommand \@@endlink[0]{}%
\providecommand \url  [0]{\begingroup\@sanitize@url \@url }%
\providecommand \@url [1]{\endgroup\@href {#1}{\urlprefix }}%
\providecommand \urlprefix  [0]{URL }%
\providecommand \Eprint [0]{\href }%
\providecommand \doibase [0]{https://doi.org/}%
\providecommand \selectlanguage [0]{\@gobble}%
\providecommand \bibinfo  [0]{\@secondoftwo}%
\providecommand \bibfield  [0]{\@secondoftwo}%
\providecommand \translation [1]{[#1]}%
\providecommand \BibitemOpen [0]{}%
\providecommand \bibitemStop [0]{}%
\providecommand \bibitemNoStop [0]{.\EOS\space}%
\providecommand \EOS [0]{\spacefactor3000\relax}%
\providecommand \BibitemShut  [1]{\csname bibitem#1\endcsname}%
\let\auto@bib@innerbib\@empty
%</preamble>
\bibitem [{\citenamefont {Adler}(1969)}]{adler_pr69_axial}%
  \BibitemOpen
  \bibfield  {author} {\bibinfo {author} {\bibfnamefont {S.~L.}\ \bibnamefont
  {Adler}},\ }\bibfield  {title} {\bibinfo {title} {Axial-vector vertex in
  spinor electrodynamics},\ }\href {https://doi.org/10.1103/PhysRev.177.2426}
  {\bibfield  {journal} {\bibinfo  {journal} {Phys. Rev.}\ }\textbf {\bibinfo
  {volume} {177}},\ \bibinfo {pages} {2426} (\bibinfo {year}
  {1969})}\BibitemShut {NoStop}%
\bibitem [{\citenamefont {Nielsen}\ and\ \citenamefont
  {Ninomiya}(1981{\natexlab{a}})}]{nielsen_npb81_absence_1}%
  \BibitemOpen
  \bibfield  {author} {\bibinfo {author} {\bibfnamefont {H.}~\bibnamefont
  {Nielsen}}\ and\ \bibinfo {author} {\bibfnamefont {M.}~\bibnamefont
  {Ninomiya}},\ }\bibfield  {title} {\bibinfo {title} {Absence of neutrinos on
  a lattice: (i). proof by homotopy theory},\ }\href
  {https://doi.org/https://doi.org/10.1016/0550-3213(81)90361-8} {\bibfield
  {journal} {\bibinfo  {journal} {Nuclear Physics B}\ }\textbf {\bibinfo
  {volume} {185}},\ \bibinfo {pages} {20} (\bibinfo {year}
  {1981}{\natexlab{a}})}\BibitemShut {NoStop}%
\bibitem [{\citenamefont {Nielsen}\ and\ \citenamefont
  {Ninomiya}(1981{\natexlab{b}})}]{nielsen_npb81_absence_2}%
  \BibitemOpen
  \bibfield  {author} {\bibinfo {author} {\bibfnamefont {H.}~\bibnamefont
  {Nielsen}}\ and\ \bibinfo {author} {\bibfnamefont {M.}~\bibnamefont
  {Ninomiya}},\ }\bibfield  {title} {\bibinfo {title} {Absence of neutrinos on
  a lattice: (ii). intuitive topological proof},\ }\href
  {https://doi.org/https://doi.org/10.1016/0550-3213(81)90524-1} {\bibfield
  {journal} {\bibinfo  {journal} {Nuclear Physics B}\ }\textbf {\bibinfo
  {volume} {193}},\ \bibinfo {pages} {173} (\bibinfo {year}
  {1981}{\natexlab{b}})}\BibitemShut {NoStop}%
\bibitem [{\citenamefont {Nielsen}\ and\ \citenamefont
  {Ninomiya}(1983)}]{nielson_plb83_adler}%
  \BibitemOpen
  \bibfield  {author} {\bibinfo {author} {\bibfnamefont {H.}~\bibnamefont
  {Nielsen}}\ and\ \bibinfo {author} {\bibfnamefont {M.}~\bibnamefont
  {Ninomiya}},\ }\bibfield  {title} {\bibinfo {title} {The adler-bell-jackiw
  anomaly and weyl fermions in a crystal},\ }\href
  {https://doi.org/https://doi.org/10.1016/0370-2693(83)91529-0} {\bibfield
  {journal} {\bibinfo  {journal} {Physics Letters B}\ }\textbf {\bibinfo
  {volume} {130}},\ \bibinfo {pages} {389} (\bibinfo {year}
  {1983})}\BibitemShut {NoStop}%
\bibitem [{\citenamefont {Son}\ and\ \citenamefont
  {Spivak}(2013)}]{son_prb13_chiral}%
  \BibitemOpen
  \bibfield  {author} {\bibinfo {author} {\bibfnamefont {D.~T.}\ \bibnamefont
  {Son}}\ and\ \bibinfo {author} {\bibfnamefont {B.~Z.}\ \bibnamefont
  {Spivak}},\ }\bibfield  {title} {\bibinfo {title} {Chiral anomaly and
  classical negative magnetoresistance of weyl metals},\ }\href
  {https://doi.org/10.1103/PhysRevB.88.104412} {\bibfield  {journal} {\bibinfo
  {journal} {Phys. Rev. B}\ }\textbf {\bibinfo {volume} {88}},\ \bibinfo
  {pages} {104412} (\bibinfo {year} {2013})}\BibitemShut {NoStop}%
\bibitem [{\citenamefont {Xiong}\ \emph
  {et~al.}(2015{\natexlab{a}})\citenamefont {Xiong}, \citenamefont {Kushwaha},
  \citenamefont {Liang}, \citenamefont {Krizan}, \citenamefont {Hirschberger},
  \citenamefont {Wang}, \citenamefont {Cava},\ and\ \citenamefont
  {Ong}}]{xiong_science15}%
  \BibitemOpen
  \bibfield  {author} {\bibinfo {author} {\bibfnamefont {J.}~\bibnamefont
  {Xiong}}, \bibinfo {author} {\bibfnamefont {S.~K.}\ \bibnamefont {Kushwaha}},
  \bibinfo {author} {\bibfnamefont {T.}~\bibnamefont {Liang}}, \bibinfo
  {author} {\bibfnamefont {J.~W.}\ \bibnamefont {Krizan}}, \bibinfo {author}
  {\bibfnamefont {M.}~\bibnamefont {Hirschberger}}, \bibinfo {author}
  {\bibfnamefont {W.}~\bibnamefont {Wang}}, \bibinfo {author} {\bibfnamefont
  {R.~J.}\ \bibnamefont {Cava}},\ and\ \bibinfo {author} {\bibfnamefont
  {N.~P.}\ \bibnamefont {Ong}},\ }\bibfield  {title} {\bibinfo {title}
  {Evidence for the chiral anomaly in the dirac semimetal na$_3$bi},\ }\href
  {https://doi.org/10.1126/science.aac6089} {\bibfield  {journal} {\bibinfo
  {journal} {Science}\ }\textbf {\bibinfo {volume} {350}},\ \bibinfo {pages}
  {413} (\bibinfo {year} {2015}{\natexlab{a}})}\BibitemShut {NoStop}%
\bibitem [{\citenamefont {Burkov}(2017)}]{burkov_prb17_giant}%
  \BibitemOpen
  \bibfield  {author} {\bibinfo {author} {\bibfnamefont {A.~A.}\ \bibnamefont
  {Burkov}},\ }\bibfield  {title} {\bibinfo {title} {Giant planar hall effect
  in topological metals},\ }\href {https://doi.org/10.1103/PhysRevB.96.041110}
  {\bibfield  {journal} {\bibinfo  {journal} {Phys. Rev. B}\ }\textbf {\bibinfo
  {volume} {96}},\ \bibinfo {pages} {041110} (\bibinfo {year}
  {2017})}\BibitemShut {NoStop}%
\bibitem [{\citenamefont {Zhang}\ \emph {et~al.}(2016)\citenamefont {Zhang},
  \citenamefont {Xu}, \citenamefont {Belopolski}, \citenamefont {Yuan},
  \citenamefont {Lin}, \citenamefont {Tong}, \citenamefont {Bian},
  \citenamefont {Alidoust}, \citenamefont {Lee}, \citenamefont {Huang},
  \citenamefont {Chang}, \citenamefont {Chang}, \citenamefont {Hsu},
  \citenamefont {Jeng}, \citenamefont {Neupane}, \citenamefont {Sanchez},
  \citenamefont {Zheng}, \citenamefont {Wang}, \citenamefont {Lin},
  \citenamefont {Zhang}, \citenamefont {Lu}, \citenamefont {Shen},
  \citenamefont {Neupert}, \citenamefont {Zahid~Hasan},\ and\ \citenamefont
  {Jia}}]{Zhang_nc16}%
  \BibitemOpen
  \bibfield  {author} {\bibinfo {author} {\bibfnamefont {C.-L.}\ \bibnamefont
  {Zhang}}, \bibinfo {author} {\bibfnamefont {S.-Y.}\ \bibnamefont {Xu}},
  \bibinfo {author} {\bibfnamefont {I.}~\bibnamefont {Belopolski}}, \bibinfo
  {author} {\bibfnamefont {Z.}~\bibnamefont {Yuan}}, \bibinfo {author}
  {\bibfnamefont {Z.}~\bibnamefont {Lin}}, \bibinfo {author} {\bibfnamefont
  {B.}~\bibnamefont {Tong}}, \bibinfo {author} {\bibfnamefont {G.}~\bibnamefont
  {Bian}}, \bibinfo {author} {\bibfnamefont {N.}~\bibnamefont {Alidoust}},
  \bibinfo {author} {\bibfnamefont {C.-C.}\ \bibnamefont {Lee}}, \bibinfo
  {author} {\bibfnamefont {S.-M.}\ \bibnamefont {Huang}}, \bibinfo {author}
  {\bibfnamefont {T.-R.}\ \bibnamefont {Chang}}, \bibinfo {author}
  {\bibfnamefont {G.}~\bibnamefont {Chang}}, \bibinfo {author} {\bibfnamefont
  {C.-H.}\ \bibnamefont {Hsu}}, \bibinfo {author} {\bibfnamefont {H.-T.}\
  \bibnamefont {Jeng}}, \bibinfo {author} {\bibfnamefont {M.}~\bibnamefont
  {Neupane}}, \bibinfo {author} {\bibfnamefont {D.~S.}\ \bibnamefont
  {Sanchez}}, \bibinfo {author} {\bibfnamefont {H.}~\bibnamefont {Zheng}},
  \bibinfo {author} {\bibfnamefont {J.}~\bibnamefont {Wang}}, \bibinfo {author}
  {\bibfnamefont {H.}~\bibnamefont {Lin}}, \bibinfo {author} {\bibfnamefont
  {C.}~\bibnamefont {Zhang}}, \bibinfo {author} {\bibfnamefont {H.-Z.}\
  \bibnamefont {Lu}}, \bibinfo {author} {\bibfnamefont {S.-Q.}\ \bibnamefont
  {Shen}}, \bibinfo {author} {\bibfnamefont {T.}~\bibnamefont {Neupert}},
  \bibinfo {author} {\bibfnamefont {M.}~\bibnamefont {Zahid~Hasan}},\ and\
  \bibinfo {author} {\bibfnamefont {S.}~\bibnamefont {Jia}},\ }\bibfield
  {title} {\bibinfo {title} {Signatures of the adler--bell--jackiw chiral
  anomaly in a weyl fermion semimetal},\ }\href
  {https://doi.org/10.1038/ncomms10735} {\bibfield  {journal} {\bibinfo
  {journal} {Nature Communications}\ }\textbf {\bibinfo {volume} {7}},\
  \bibinfo {pages} {10735} (\bibinfo {year} {2016})}\BibitemShut {NoStop}%
\bibitem [{\citenamefont {Li}\ \emph {et~al.}(2016{\natexlab{a}})\citenamefont
  {Li}, \citenamefont {Kharzeev}, \citenamefont {Zhang}, \citenamefont {Huang},
  \citenamefont {Pletikosi{\'{c}}}, \citenamefont {Fedorov}, \citenamefont
  {Zhong}, \citenamefont {Schneeloch}, \citenamefont {Gu},\ and\ \citenamefont
  {Valla}}]{Li_NP16_chiral_magnetic}%
  \BibitemOpen
  \bibfield  {author} {\bibinfo {author} {\bibfnamefont {Q.}~\bibnamefont
  {Li}}, \bibinfo {author} {\bibfnamefont {D.~E.}\ \bibnamefont {Kharzeev}},
  \bibinfo {author} {\bibfnamefont {C.}~\bibnamefont {Zhang}}, \bibinfo
  {author} {\bibfnamefont {Y.}~\bibnamefont {Huang}}, \bibinfo {author}
  {\bibfnamefont {I.}~\bibnamefont {Pletikosi{\'{c}}}}, \bibinfo {author}
  {\bibfnamefont {A.~V.}\ \bibnamefont {Fedorov}}, \bibinfo {author}
  {\bibfnamefont {R.~D.}\ \bibnamefont {Zhong}}, \bibinfo {author}
  {\bibfnamefont {J.~A.}\ \bibnamefont {Schneeloch}}, \bibinfo {author}
  {\bibfnamefont {G.~D.}\ \bibnamefont {Gu}},\ and\ \bibinfo {author}
  {\bibfnamefont {T.}~\bibnamefont {Valla}},\ }\bibfield  {title} {\bibinfo
  {title} {Chiral magnetic effect in zrte5},\ }\href
  {https://doi.org/10.1038/nphys3648} {\bibfield  {journal} {\bibinfo
  {journal} {Nature Physics}\ }\textbf {\bibinfo {volume} {12}},\ \bibinfo
  {pages} {550} (\bibinfo {year} {2016}{\natexlab{a}})}\BibitemShut {NoStop}%
\bibitem [{\citenamefont {Nandy}\ \emph {et~al.}(2017)\citenamefont {Nandy},
  \citenamefont {Sharma}, \citenamefont {Taraphder},\ and\ \citenamefont
  {Tewari}}]{Nandy_prl17}%
  \BibitemOpen
  \bibfield  {author} {\bibinfo {author} {\bibfnamefont {S.}~\bibnamefont
  {Nandy}}, \bibinfo {author} {\bibfnamefont {G.}~\bibnamefont {Sharma}},
  \bibinfo {author} {\bibfnamefont {A.}~\bibnamefont {Taraphder}},\ and\
  \bibinfo {author} {\bibfnamefont {S.}~\bibnamefont {Tewari}},\ }\bibfield
  {title} {\bibinfo {title} {Chiral anomaly as the origin of the planar hall
  effect in weyl semimetals},\ }\href
  {https://doi.org/10.1103/PhysRevLett.119.176804} {\bibfield  {journal}
  {\bibinfo  {journal} {Phys. Rev. Lett.}\ }\textbf {\bibinfo {volume} {119}},\
  \bibinfo {pages} {176804} (\bibinfo {year} {2017})}\BibitemShut {NoStop}%
\bibitem [{\citenamefont {Kumar}\ \emph {et~al.}(2018)\citenamefont {Kumar},
  \citenamefont {Guin}, \citenamefont {Felser},\ and\ \citenamefont
  {Shekhar}}]{shekhar_prb18}%
  \BibitemOpen
  \bibfield  {author} {\bibinfo {author} {\bibfnamefont {N.}~\bibnamefont
  {Kumar}}, \bibinfo {author} {\bibfnamefont {S.~N.}\ \bibnamefont {Guin}},
  \bibinfo {author} {\bibfnamefont {C.}~\bibnamefont {Felser}},\ and\ \bibinfo
  {author} {\bibfnamefont {C.}~\bibnamefont {Shekhar}},\ }\bibfield  {title}
  {\bibinfo {title} {Planar hall effect in the weyl semimetal gdptbi},\ }\href
  {https://doi.org/10.1103/PhysRevB.98.041103} {\bibfield  {journal} {\bibinfo
  {journal} {Phys. Rev. B}\ }\textbf {\bibinfo {volume} {98}},\ \bibinfo
  {pages} {041103} (\bibinfo {year} {2018})}\BibitemShut {NoStop}%
\bibitem [{\citenamefont {Kim}\ \emph {et~al.}(2014)\citenamefont {Kim},
  \citenamefont {Kim},\ and\ \citenamefont {Sasaki}}]{kim_prb14-boltzmann}%
  \BibitemOpen
  \bibfield  {author} {\bibinfo {author} {\bibfnamefont {K.-S.}\ \bibnamefont
  {Kim}}, \bibinfo {author} {\bibfnamefont {H.-J.}\ \bibnamefont {Kim}},\ and\
  \bibinfo {author} {\bibfnamefont {M.}~\bibnamefont {Sasaki}},\ }\bibfield
  {title} {\bibinfo {title} {Boltzmann equation approach to anomalous transport
  in a weyl metal},\ }\href {https://doi.org/10.1103/PhysRevB.89.195137}
  {\bibfield  {journal} {\bibinfo  {journal} {Phys. Rev. B}\ }\textbf {\bibinfo
  {volume} {89}},\ \bibinfo {pages} {195137} (\bibinfo {year}
  {2014})}\BibitemShut {NoStop}%
\bibitem [{\citenamefont {Das}\ \emph {et~al.}(2020)\citenamefont {Das},
  \citenamefont {Singh},\ and\ \citenamefont {Agarwal}}]{kamal_prr20_chiral}%
  \BibitemOpen
  \bibfield  {author} {\bibinfo {author} {\bibfnamefont {K.}~\bibnamefont
  {Das}}, \bibinfo {author} {\bibfnamefont {S.~K.}\ \bibnamefont {Singh}},\
  and\ \bibinfo {author} {\bibfnamefont {A.}~\bibnamefont {Agarwal}},\
  }\bibfield  {title} {\bibinfo {title} {Chiral anomalies induced transport in
  weyl metals in quantizing magnetic field},\ }\href
  {https://doi.org/10.1103/PhysRevResearch.2.033511} {\bibfield  {journal}
  {\bibinfo  {journal} {Phys. Rev. Research}\ }\textbf {\bibinfo {volume}
  {2}},\ \bibinfo {pages} {033511} (\bibinfo {year} {2020})}\BibitemShut
  {NoStop}%
\bibitem [{\citenamefont {Das}\ and\ \citenamefont
  {Agarwal}(2019{\natexlab{a}})}]{kamal_prb19_linear}%
  \BibitemOpen
  \bibfield  {author} {\bibinfo {author} {\bibfnamefont {K.}~\bibnamefont
  {Das}}\ and\ \bibinfo {author} {\bibfnamefont {A.}~\bibnamefont {Agarwal}},\
  }\bibfield  {title} {\bibinfo {title} {Linear magnetochiral transport in
  tilted type-i and type-ii weyl semimetals},\ }\href
  {https://doi.org/10.1103/PhysRevB.99.085405} {\bibfield  {journal} {\bibinfo
  {journal} {Phys. Rev. B}\ }\textbf {\bibinfo {volume} {99}},\ \bibinfo
  {pages} {085405} (\bibinfo {year} {2019}{\natexlab{a}})}\BibitemShut
  {NoStop}%
\bibitem [{\citenamefont {Das}\ and\ \citenamefont
  {Agarwal}(2019{\natexlab{b}})}]{kamal_prb19_berry}%
  \BibitemOpen
  \bibfield  {author} {\bibinfo {author} {\bibfnamefont {K.}~\bibnamefont
  {Das}}\ and\ \bibinfo {author} {\bibfnamefont {A.}~\bibnamefont {Agarwal}},\
  }\bibfield  {title} {\bibinfo {title} {Berry curvature induced thermopower in
  type-i and type-ii weyl semimetals},\ }\href
  {https://doi.org/10.1103/PhysRevB.100.085406} {\bibfield  {journal} {\bibinfo
   {journal} {Phys. Rev. B}\ }\textbf {\bibinfo {volume} {100}},\ \bibinfo
  {pages} {085406} (\bibinfo {year} {2019}{\natexlab{b}})}\BibitemShut
  {NoStop}%
\bibitem [{\citenamefont {Das}\ and\ \citenamefont
  {Agarwal}(2021)}]{kamal_prb21}%
  \BibitemOpen
  \bibfield  {author} {\bibinfo {author} {\bibfnamefont {K.}~\bibnamefont
  {Das}}\ and\ \bibinfo {author} {\bibfnamefont {A.}~\bibnamefont {Agarwal}},\
  }\bibfield  {title} {\bibinfo {title} {Intrinsic hall conductivities induced
  by the orbital magnetic moment},\ }\href
  {https://doi.org/10.1103/PhysRevB.103.125432} {\bibfield  {journal} {\bibinfo
   {journal} {Phys. Rev. B}\ }\textbf {\bibinfo {volume} {103}},\ \bibinfo
  {pages} {125432} (\bibinfo {year} {2021})}\BibitemShut {NoStop}%
\bibitem [{\citenamefont {Das}\ and\ \citenamefont
  {Agarwal}(2020)}]{kamal_prr20_thermal}%
  \BibitemOpen
  \bibfield  {author} {\bibinfo {author} {\bibfnamefont {K.}~\bibnamefont
  {Das}}\ and\ \bibinfo {author} {\bibfnamefont {A.}~\bibnamefont {Agarwal}},\
  }\bibfield  {title} {\bibinfo {title} {Thermal and gravitational chiral
  anomaly induced magneto-transport in weyl semimetals},\ }\href
  {https://doi.org/10.1103/PhysRevResearch.2.013088} {\bibfield  {journal}
  {\bibinfo  {journal} {Phys. Rev. Research}\ }\textbf {\bibinfo {volume}
  {2}},\ \bibinfo {pages} {013088} (\bibinfo {year} {2020})}\BibitemShut
  {NoStop}%
\bibitem [{\citenamefont {Mandal}\ \emph {et~al.}(2022)\citenamefont {Mandal},
  \citenamefont {Das},\ and\ \citenamefont {Agarwal}}]{debottam_prb22}%
  \BibitemOpen
  \bibfield  {author} {\bibinfo {author} {\bibfnamefont {D.}~\bibnamefont
  {Mandal}}, \bibinfo {author} {\bibfnamefont {K.}~\bibnamefont {Das}},\ and\
  \bibinfo {author} {\bibfnamefont {A.}~\bibnamefont {Agarwal}},\ }\bibfield
  {title} {\bibinfo {title} {Chiral anomaly and nonlinear magnetotransport in
  time reversal symmetric weyl semimetals},\ }\href
  {https://doi.org/10.1103/PhysRevB.106.035423} {\bibfield  {journal} {\bibinfo
   {journal} {Phys. Rev. B}\ }\textbf {\bibinfo {volume} {106}},\ \bibinfo
  {pages} {035423} (\bibinfo {year} {2022})}\BibitemShut {NoStop}%
\bibitem [{\citenamefont {Hütt}\ \emph {et~al.}(2019)\citenamefont {Hütt},
  \citenamefont {Kamenskyi}, \citenamefont {Neubauer}, \citenamefont {Shekhar},
  \citenamefont {Felser}, \citenamefont {Dressel},\ and\ \citenamefont
  {Pronin}}]{HUTT_SciDirect19}%
  \BibitemOpen
  \bibfield  {author} {\bibinfo {author} {\bibfnamefont {F.}~\bibnamefont
  {Hütt}}, \bibinfo {author} {\bibfnamefont {D.}~\bibnamefont {Kamenskyi}},
  \bibinfo {author} {\bibfnamefont {D.}~\bibnamefont {Neubauer}}, \bibinfo
  {author} {\bibfnamefont {C.}~\bibnamefont {Shekhar}}, \bibinfo {author}
  {\bibfnamefont {C.}~\bibnamefont {Felser}}, \bibinfo {author} {\bibfnamefont
  {M.}~\bibnamefont {Dressel}},\ and\ \bibinfo {author} {\bibfnamefont {A.~V.}\
  \bibnamefont {Pronin}},\ }\bibfield  {title} {\bibinfo {title} {Terahertz
  transmission through taas single crystals in simultaneously applied magnetic
  and electric fields: Possible optical signatures of the chiral anomaly in a
  weyl semimetal},\ }\href
  {https://doi.org/https://doi.org/10.1016/j.rinp.2019.102630} {\bibfield
  {journal} {\bibinfo  {journal} {Results in Physics}\ }\textbf {\bibinfo
  {volume} {15}},\ \bibinfo {pages} {102630} (\bibinfo {year}
  {2019})}\BibitemShut {NoStop}%
\bibitem [{\citenamefont {Hosur}\ and\ \citenamefont
  {Qi}(2015)}]{hosur_prb15_tunable}%
  \BibitemOpen
  \bibfield  {author} {\bibinfo {author} {\bibfnamefont {P.}~\bibnamefont
  {Hosur}}\ and\ \bibinfo {author} {\bibfnamefont {X.-L.}\ \bibnamefont {Qi}},\
  }\bibfield  {title} {\bibinfo {title} {Tunable circular dichroism due to the
  chiral anomaly in weyl semimetals},\ }\href
  {https://doi.org/10.1103/PhysRevB.91.081106} {\bibfield  {journal} {\bibinfo
  {journal} {Phys. Rev. B}\ }\textbf {\bibinfo {volume} {91}},\ \bibinfo
  {pages} {081106} (\bibinfo {year} {2015})}\BibitemShut {NoStop}%
\bibitem [{\citenamefont {Ashby}\ and\ \citenamefont
  {Carbotte}(2014)}]{carbotte_prb14_chiral}%
  \BibitemOpen
  \bibfield  {author} {\bibinfo {author} {\bibfnamefont {P.~E.~C.}\
  \bibnamefont {Ashby}}\ and\ \bibinfo {author} {\bibfnamefont {J.~P.}\
  \bibnamefont {Carbotte}},\ }\bibfield  {title} {\bibinfo {title} {Chiral
  anomaly and optical absorption in weyl semimetals},\ }\href
  {https://doi.org/10.1103/PhysRevB.89.245121} {\bibfield  {journal} {\bibinfo
  {journal} {Phys. Rev. B}\ }\textbf {\bibinfo {volume} {89}},\ \bibinfo
  {pages} {245121} (\bibinfo {year} {2014})}\BibitemShut {NoStop}%
\bibitem [{\citenamefont {Ma}\ and\ \citenamefont
  {Pesin}(2015{\natexlab{a}})}]{Ma_prb15}%
  \BibitemOpen
  \bibfield  {author} {\bibinfo {author} {\bibfnamefont {J.}~\bibnamefont
  {Ma}}\ and\ \bibinfo {author} {\bibfnamefont {D.~A.}\ \bibnamefont {Pesin}},\
  }\bibfield  {title} {\bibinfo {title} {Chiral magnetic effect and natural
  optical activity in metals with or without weyl points},\ }\href
  {https://doi.org/10.1103/PhysRevB.92.235205} {\bibfield  {journal} {\bibinfo
  {journal} {Phys. Rev. B}\ }\textbf {\bibinfo {volume} {92}},\ \bibinfo
  {pages} {235205} (\bibinfo {year} {2015}{\natexlab{a}})}\BibitemShut
  {NoStop}%
\bibitem [{\citenamefont {Morimoto}\ and\ \citenamefont
  {Nagaosa}(2016)}]{morimoto_prl16_chiral}%
  \BibitemOpen
  \bibfield  {author} {\bibinfo {author} {\bibfnamefont {T.}~\bibnamefont
  {Morimoto}}\ and\ \bibinfo {author} {\bibfnamefont {N.}~\bibnamefont
  {Nagaosa}},\ }\bibfield  {title} {\bibinfo {title} {Chiral anomaly and giant
  magnetochiral anisotropy in noncentrosymmetric weyl semimetals},\ }\href
  {https://doi.org/10.1103/PhysRevLett.117.146603} {\bibfield  {journal}
  {\bibinfo  {journal} {Phys. Rev. Lett.}\ }\textbf {\bibinfo {volume} {117}},\
  \bibinfo {pages} {146603} (\bibinfo {year} {2016})}\BibitemShut {NoStop}%
\bibitem [{\citenamefont {Jadidi}\ \emph {et~al.}(2020)\citenamefont {Jadidi},
  \citenamefont {Kargarian}, \citenamefont {Mittendorff}, \citenamefont
  {Aytac}, \citenamefont {Shen}, \citenamefont {K\"onig-Otto}, \citenamefont
  {Winnerl}, \citenamefont {Ni}, \citenamefont {Gaeta}, \citenamefont
  {Murphy},\ and\ \citenamefont {Drew}}]{jadidi_prb20}%
  \BibitemOpen
  \bibfield  {author} {\bibinfo {author} {\bibfnamefont {M.~M.}\ \bibnamefont
  {Jadidi}}, \bibinfo {author} {\bibfnamefont {M.}~\bibnamefont {Kargarian}},
  \bibinfo {author} {\bibfnamefont {M.}~\bibnamefont {Mittendorff}}, \bibinfo
  {author} {\bibfnamefont {Y.}~\bibnamefont {Aytac}}, \bibinfo {author}
  {\bibfnamefont {B.}~\bibnamefont {Shen}}, \bibinfo {author} {\bibfnamefont
  {J.~C.}\ \bibnamefont {K\"onig-Otto}}, \bibinfo {author} {\bibfnamefont
  {S.}~\bibnamefont {Winnerl}}, \bibinfo {author} {\bibfnamefont
  {N.}~\bibnamefont {Ni}}, \bibinfo {author} {\bibfnamefont {A.~L.}\
  \bibnamefont {Gaeta}}, \bibinfo {author} {\bibfnamefont {T.~E.}\ \bibnamefont
  {Murphy}},\ and\ \bibinfo {author} {\bibfnamefont {H.~D.}\ \bibnamefont
  {Drew}},\ }\bibfield  {title} {\bibinfo {title} {Nonlinear optical control of
  chiral charge pumping in a topological weyl semimetal},\ }\href
  {https://doi.org/10.1103/PhysRevB.102.245123} {\bibfield  {journal} {\bibinfo
   {journal} {Phys. Rev. B}\ }\textbf {\bibinfo {volume} {102}},\ \bibinfo
  {pages} {245123} (\bibinfo {year} {2020})}\BibitemShut {NoStop}%
\bibitem [{\citenamefont {Thakur}\ \emph {et~al.}(2018)\citenamefont {Thakur},
  \citenamefont {Sadhukhan},\ and\ \citenamefont {Agarwal}}]{anmol_prb18}%
  \BibitemOpen
  \bibfield  {author} {\bibinfo {author} {\bibfnamefont {A.}~\bibnamefont
  {Thakur}}, \bibinfo {author} {\bibfnamefont {K.}~\bibnamefont {Sadhukhan}},\
  and\ \bibinfo {author} {\bibfnamefont {A.}~\bibnamefont {Agarwal}},\
  }\bibfield  {title} {\bibinfo {title} {Dynamic current-current susceptibility
  in three-dimensional dirac and weyl semimetals},\ }\href
  {https://doi.org/10.1103/PhysRevB.97.035403} {\bibfield  {journal} {\bibinfo
  {journal} {Phys. Rev. B}\ }\textbf {\bibinfo {volume} {97}},\ \bibinfo
  {pages} {035403} (\bibinfo {year} {2018})}\BibitemShut {NoStop}%
\bibitem [{\citenamefont {Sonowal}\ \emph {et~al.}(2019)\citenamefont
  {Sonowal}, \citenamefont {Singh},\ and\ \citenamefont
  {Agarwal}}]{kabya_prb19}%
  \BibitemOpen
  \bibfield  {author} {\bibinfo {author} {\bibfnamefont {K.}~\bibnamefont
  {Sonowal}}, \bibinfo {author} {\bibfnamefont {A.}~\bibnamefont {Singh}},\
  and\ \bibinfo {author} {\bibfnamefont {A.}~\bibnamefont {Agarwal}},\
  }\bibfield  {title} {\bibinfo {title} {Giant optical activity and kerr effect
  in type-i and type-ii weyl semimetals},\ }\href
  {https://doi.org/10.1103/PhysRevB.100.085436} {\bibfield  {journal} {\bibinfo
   {journal} {Phys. Rev. B}\ }\textbf {\bibinfo {volume} {100}},\ \bibinfo
  {pages} {085436} (\bibinfo {year} {2019})}\BibitemShut {NoStop}%
\bibitem [{\citenamefont {Landsteiner}\ \emph {et~al.}(2011)\citenamefont
  {Landsteiner}, \citenamefont {Meg\'{\i}as},\ and\ \citenamefont
  {Pena-Benitez}}]{landsteiner_prl11_gravitational}%
  \BibitemOpen
  \bibfield  {author} {\bibinfo {author} {\bibfnamefont {K.}~\bibnamefont
  {Landsteiner}}, \bibinfo {author} {\bibfnamefont {E.}~\bibnamefont
  {Meg\'{\i}as}},\ and\ \bibinfo {author} {\bibfnamefont {F.}~\bibnamefont
  {Pena-Benitez}},\ }\bibfield  {title} {\bibinfo {title} {Gravitational
  anomaly and transport phenomena},\ }\href
  {https://doi.org/10.1103/PhysRevLett.107.021601} {\bibfield  {journal}
  {\bibinfo  {journal} {Phys. Rev. Lett.}\ }\textbf {\bibinfo {volume} {107}},\
  \bibinfo {pages} {021601} (\bibinfo {year} {2011})}\BibitemShut {NoStop}%
\bibitem [{\citenamefont {Lucas}\ \emph {et~al.}(2016)\citenamefont {Lucas},
  \citenamefont {Davison},\ and\ \citenamefont
  {Sachdev}}]{lucas_pnas16_hydrodynamic}%
  \BibitemOpen
  \bibfield  {author} {\bibinfo {author} {\bibfnamefont {A.}~\bibnamefont
  {Lucas}}, \bibinfo {author} {\bibfnamefont {R.~A.}\ \bibnamefont {Davison}},\
  and\ \bibinfo {author} {\bibfnamefont {S.}~\bibnamefont {Sachdev}},\
  }\bibfield  {title} {\bibinfo {title} {Hydrodynamic theory of thermoelectric
  transport and negative magnetoresistance in weyl semimetals},\ }\href
  {https://doi.org/10.1073/pnas.1608881113} {\bibfield  {journal} {\bibinfo
  {journal} {Proceedings of the National Academy of Sciences}\ }\textbf
  {\bibinfo {volume} {113}},\ \bibinfo {pages} {9463} (\bibinfo {year}
  {2016})}\BibitemShut {NoStop}%
\bibitem [{\citenamefont {Gooth}\ \emph {et~al.}(2017)\citenamefont {Gooth},
  \citenamefont {Niemann}, \citenamefont {Meng}, \citenamefont {Grushin},
  \citenamefont {Landsteiner}, \citenamefont {Gotsmann}, \citenamefont
  {Menges}, \citenamefont {Schmidt}, \citenamefont {Shekhar}, \citenamefont
  {S{\"u}{\ss}}, \citenamefont {H{\"u}hne}, \citenamefont {Rellinghaus},
  \citenamefont {Felser}, \citenamefont {Yan},\ and\ \citenamefont
  {Nielsch}}]{Gooth_nature17_experiemnatl}%
  \BibitemOpen
  \bibfield  {author} {\bibinfo {author} {\bibfnamefont {J.}~\bibnamefont
  {Gooth}}, \bibinfo {author} {\bibfnamefont {A.~C.}\ \bibnamefont {Niemann}},
  \bibinfo {author} {\bibfnamefont {T.}~\bibnamefont {Meng}}, \bibinfo {author}
  {\bibfnamefont {A.~G.}\ \bibnamefont {Grushin}}, \bibinfo {author}
  {\bibfnamefont {K.}~\bibnamefont {Landsteiner}}, \bibinfo {author}
  {\bibfnamefont {B.}~\bibnamefont {Gotsmann}}, \bibinfo {author}
  {\bibfnamefont {F.}~\bibnamefont {Menges}}, \bibinfo {author} {\bibfnamefont
  {M.}~\bibnamefont {Schmidt}}, \bibinfo {author} {\bibfnamefont
  {C.}~\bibnamefont {Shekhar}}, \bibinfo {author} {\bibfnamefont
  {V.}~\bibnamefont {S{\"u}{\ss}}}, \bibinfo {author} {\bibfnamefont
  {R.}~\bibnamefont {H{\"u}hne}}, \bibinfo {author} {\bibfnamefont
  {B.}~\bibnamefont {Rellinghaus}}, \bibinfo {author} {\bibfnamefont
  {C.}~\bibnamefont {Felser}}, \bibinfo {author} {\bibfnamefont
  {B.}~\bibnamefont {Yan}},\ and\ \bibinfo {author} {\bibfnamefont
  {K.}~\bibnamefont {Nielsch}},\ }\bibfield  {title} {\bibinfo {title}
  {Experimental signatures of the mixed axial--gravitational anomaly in the
  weyl semimetal nbp},\ }\href {https://doi.org/10.1038/nature23005} {\bibfield
   {journal} {\bibinfo  {journal} {Nature}\ }\textbf {\bibinfo {volume}
  {547}},\ \bibinfo {pages} {324} (\bibinfo {year} {2017})}\BibitemShut
  {NoStop}%
\bibitem [{\citenamefont {Stone}\ and\ \citenamefont
  {Kim}(2018)}]{stone_prd18_mixed}%
  \BibitemOpen
  \bibfield  {author} {\bibinfo {author} {\bibfnamefont {M.}~\bibnamefont
  {Stone}}\ and\ \bibinfo {author} {\bibfnamefont {J.}~\bibnamefont {Kim}},\
  }\bibfield  {title} {\bibinfo {title} {Mixed anomalies: Chiral vortical
  effect and the sommerfeld expansion},\ }\href
  {https://doi.org/10.1103/PhysRevD.98.025012} {\bibfield  {journal} {\bibinfo
  {journal} {Phys. Rev. D}\ }\textbf {\bibinfo {volume} {98}},\ \bibinfo
  {pages} {025012} (\bibinfo {year} {2018})}\BibitemShut {NoStop}%
\bibitem [{\citenamefont {Lundgren}\ \emph {et~al.}(2014)\citenamefont
  {Lundgren}, \citenamefont {Laurell},\ and\ \citenamefont
  {Fiete}}]{rex_prb14_thermoelectric}%
  \BibitemOpen
  \bibfield  {author} {\bibinfo {author} {\bibfnamefont {R.}~\bibnamefont
  {Lundgren}}, \bibinfo {author} {\bibfnamefont {P.}~\bibnamefont {Laurell}},\
  and\ \bibinfo {author} {\bibfnamefont {G.~A.}\ \bibnamefont {Fiete}},\
  }\bibfield  {title} {\bibinfo {title} {Thermoelectric properties of weyl and
  dirac semimetals},\ }\href {https://doi.org/10.1103/PhysRevB.90.165115}
  {\bibfield  {journal} {\bibinfo  {journal} {Phys. Rev. B}\ }\textbf {\bibinfo
  {volume} {90}},\ \bibinfo {pages} {165115} (\bibinfo {year}
  {2014})}\BibitemShut {NoStop}%
\bibitem [{\citenamefont {Hirschberger}\ \emph {et~al.}(2016)\citenamefont
  {Hirschberger}, \citenamefont {Kushwaha}, \citenamefont {Wang}, \citenamefont
  {Gibson}, \citenamefont {Liang}, \citenamefont {Belvin}, \citenamefont
  {Bernevig}, \citenamefont {Cava},\ and\ \citenamefont
  {Ong}}]{Hirschberger_nm16}%
  \BibitemOpen
  \bibfield  {author} {\bibinfo {author} {\bibfnamefont {M.}~\bibnamefont
  {Hirschberger}}, \bibinfo {author} {\bibfnamefont {S.}~\bibnamefont
  {Kushwaha}}, \bibinfo {author} {\bibfnamefont {Z.}~\bibnamefont {Wang}},
  \bibinfo {author} {\bibfnamefont {Q.}~\bibnamefont {Gibson}}, \bibinfo
  {author} {\bibfnamefont {S.}~\bibnamefont {Liang}}, \bibinfo {author}
  {\bibfnamefont {C.}~\bibnamefont {Belvin}}, \bibinfo {author} {\bibfnamefont
  {B.~A.}\ \bibnamefont {Bernevig}}, \bibinfo {author} {\bibfnamefont {R.~J.}\
  \bibnamefont {Cava}},\ and\ \bibinfo {author} {\bibfnamefont {N.~P.}\
  \bibnamefont {Ong}},\ }\bibfield  {title} {\bibinfo {title} {The chiral
  anomaly and thermopower of weyl fermions in the half-heusler gdptbi},\
  }\href {https://doi.org/10.1038/nmat4684} {\bibfield  {journal} {\bibinfo
  {journal} {Nature Materials}\ }\textbf {\bibinfo {volume} {15}},\ \bibinfo
  {pages} {1161} (\bibinfo {year} {2016})}\BibitemShut {NoStop}%
\bibitem [{\citenamefont {Jia}\ \emph {et~al.}(2016)\citenamefont {Jia},
  \citenamefont {Li}, \citenamefont {Li}, \citenamefont {Shi}, \citenamefont
  {Liao}, \citenamefont {Yu},\ and\ \citenamefont {Wu}}]{Jia_nm16}%
  \BibitemOpen
  \bibfield  {author} {\bibinfo {author} {\bibfnamefont {Z.}~\bibnamefont
  {Jia}}, \bibinfo {author} {\bibfnamefont {C.}~\bibnamefont {Li}}, \bibinfo
  {author} {\bibfnamefont {X.}~\bibnamefont {Li}}, \bibinfo {author}
  {\bibfnamefont {J.}~\bibnamefont {Shi}}, \bibinfo {author} {\bibfnamefont
  {Z.}~\bibnamefont {Liao}}, \bibinfo {author} {\bibfnamefont {D.}~\bibnamefont
  {Yu}},\ and\ \bibinfo {author} {\bibfnamefont {X.}~\bibnamefont {Wu}},\
  }\bibfield  {title} {\bibinfo {title} {Thermoelectric signature of the chiral
  anomaly in cd3as2},\ }\href {https://doi.org/10.1038/ncomms13013} {\bibfield
  {journal} {\bibinfo  {journal} {Nature Communications}\ }\textbf {\bibinfo
  {volume} {7}},\ \bibinfo {pages} {13013} (\bibinfo {year}
  {2016})}\BibitemShut {NoStop}%
\bibitem [{\citenamefont {Spivak}\ and\ \citenamefont
  {Andreev}(2016)}]{spivak_prb16}%
  \BibitemOpen
  \bibfield  {author} {\bibinfo {author} {\bibfnamefont {B.~Z.}\ \bibnamefont
  {Spivak}}\ and\ \bibinfo {author} {\bibfnamefont {A.~V.}\ \bibnamefont
  {Andreev}},\ }\bibfield  {title} {\bibinfo {title} {Magnetotransport
  phenomena related to the chiral anomaly in weyl semimetals},\ }\href
  {https://doi.org/10.1103/PhysRevB.93.085107} {\bibfield  {journal} {\bibinfo
  {journal} {Phys. Rev. B}\ }\textbf {\bibinfo {volume} {93}},\ \bibinfo
  {pages} {085107} (\bibinfo {year} {2016})}\BibitemShut {NoStop}%
\bibitem [{\citenamefont {Stockert}\ \emph {et~al.}(2017)\citenamefont
  {Stockert}, \citenamefont {dos Reis}, \citenamefont {Ajeesh}, \citenamefont
  {Watzman}, \citenamefont {Schmidt}, \citenamefont {Shekhar}, \citenamefont
  {Heremans}, \citenamefont {Felser}, \citenamefont {Baenitz},\ and\
  \citenamefont {Nicklas}}]{Stockert_IOP17}%
  \BibitemOpen
  \bibfield  {author} {\bibinfo {author} {\bibfnamefont {U.}~\bibnamefont
  {Stockert}}, \bibinfo {author} {\bibfnamefont {R.~D.}\ \bibnamefont {dos
  Reis}}, \bibinfo {author} {\bibfnamefont {M.~O.}\ \bibnamefont {Ajeesh}},
  \bibinfo {author} {\bibfnamefont {S.~J.}\ \bibnamefont {Watzman}}, \bibinfo
  {author} {\bibfnamefont {M.}~\bibnamefont {Schmidt}}, \bibinfo {author}
  {\bibfnamefont {C.}~\bibnamefont {Shekhar}}, \bibinfo {author} {\bibfnamefont
  {J.~P.}\ \bibnamefont {Heremans}}, \bibinfo {author} {\bibfnamefont
  {C.}~\bibnamefont {Felser}}, \bibinfo {author} {\bibfnamefont
  {M.}~\bibnamefont {Baenitz}},\ and\ \bibinfo {author} {\bibfnamefont
  {M.}~\bibnamefont {Nicklas}},\ }\bibfield  {title} {\bibinfo {title}
  {Thermopower and thermal conductivity in the weyl semimetal nbp},\ }\href
  {https://doi.org/10.1088/1361-648X/aa7a3b} {\bibfield  {journal} {\bibinfo
  {journal} {Journal of Physics: Condensed Matter}\ }\textbf {\bibinfo {volume}
  {29}},\ \bibinfo {pages} {325701} (\bibinfo {year} {2017})}\BibitemShut
  {NoStop}%
\bibitem [{\citenamefont {Zyuzin}(2017)}]{zyuzin_prb17_magnetotransport}%
  \BibitemOpen
  \bibfield  {author} {\bibinfo {author} {\bibfnamefont {V.~A.}\ \bibnamefont
  {Zyuzin}},\ }\bibfield  {title} {\bibinfo {title} {Magnetotransport of weyl
  semimetals due to the chiral anomaly},\ }\href
  {https://doi.org/10.1103/PhysRevB.95.245128} {\bibfield  {journal} {\bibinfo
  {journal} {Phys. Rev. B}\ }\textbf {\bibinfo {volume} {95}},\ \bibinfo
  {pages} {245128} (\bibinfo {year} {2017})}\BibitemShut {NoStop}%
\bibitem [{\citenamefont {Vu}\ \emph {et~al.}(2021)\citenamefont {Vu},
  \citenamefont {Zhang}, \citenamefont {Sahin}, \citenamefont {Flatte},
  \citenamefont {Trivedi},\ and\ \citenamefont {Heremans}}]{Vu_NM21_thermal}%
  \BibitemOpen
  \bibfield  {author} {\bibinfo {author} {\bibfnamefont {D.}~\bibnamefont
  {Vu}}, \bibinfo {author} {\bibfnamefont {W.}~\bibnamefont {Zhang}}, \bibinfo
  {author} {\bibfnamefont {C.}~\bibnamefont {Sahin}}, \bibinfo {author}
  {\bibfnamefont {M.~E.}\ \bibnamefont {Flatte}}, \bibinfo {author}
  {\bibfnamefont {N.}~\bibnamefont {Trivedi}},\ and\ \bibinfo {author}
  {\bibfnamefont {J.~P.}\ \bibnamefont {Heremans}},\ }\bibfield  {title}
  {\bibinfo {title} {Thermal chiral anomaly in the magnetic-field-induced ideal
  weyl phase of bi1-xsbx},\ }\href {https://doi.org/10.1038/s41563-021-00983-8}
  {\bibfield  {journal} {\bibinfo  {journal} {Nature Materials}\ }\textbf
  {\bibinfo {volume} {20}},\ \bibinfo {pages} {1525} (\bibinfo {year}
  {2021})}\BibitemShut {NoStop}%
\bibitem [{\citenamefont {Son}\ and\ \citenamefont
  {Yamamoto}(2012)}]{son_prl12_berry}%
  \BibitemOpen
  \bibfield  {author} {\bibinfo {author} {\bibfnamefont {D.~T.}\ \bibnamefont
  {Son}}\ and\ \bibinfo {author} {\bibfnamefont {N.}~\bibnamefont {Yamamoto}},\
  }\bibfield  {title} {\bibinfo {title} {Berry curvature, triangle anomalies,
  and the chiral magnetic effect in fermi liquids},\ }\href
  {https://doi.org/10.1103/PhysRevLett.109.181602} {\bibfield  {journal}
  {\bibinfo  {journal} {Phys. Rev. Lett.}\ }\textbf {\bibinfo {volume} {109}},\
  \bibinfo {pages} {181602} (\bibinfo {year} {2012})}\BibitemShut {NoStop}%
\bibitem [{\citenamefont {Son}\ and\ \citenamefont
  {Yamamoto}(2013)}]{son_prd13_kinetic}%
  \BibitemOpen
  \bibfield  {author} {\bibinfo {author} {\bibfnamefont {D.~T.}\ \bibnamefont
  {Son}}\ and\ \bibinfo {author} {\bibfnamefont {N.}~\bibnamefont {Yamamoto}},\
  }\bibfield  {title} {\bibinfo {title} {Kinetic theory with berry curvature
  from quantum field theories},\ }\href
  {https://doi.org/10.1103/PhysRevD.87.085016} {\bibfield  {journal} {\bibinfo
  {journal} {Phys. Rev. D}\ }\textbf {\bibinfo {volume} {87}},\ \bibinfo
  {pages} {085016} (\bibinfo {year} {2013})}\BibitemShut {NoStop}%
\bibitem [{\citenamefont {Stephanov}\ and\ \citenamefont
  {Yin}(2012)}]{stefanov_prl12}%
  \BibitemOpen
  \bibfield  {author} {\bibinfo {author} {\bibfnamefont {M.~A.}\ \bibnamefont
  {Stephanov}}\ and\ \bibinfo {author} {\bibfnamefont {Y.}~\bibnamefont
  {Yin}},\ }\bibfield  {title} {\bibinfo {title} {Chiral kinetic theory},\
  }\href {https://doi.org/10.1103/PhysRevLett.109.162001} {\bibfield  {journal}
  {\bibinfo  {journal} {Phys. Rev. Lett.}\ }\textbf {\bibinfo {volume} {109}},\
  \bibinfo {pages} {162001} (\bibinfo {year} {2012})}\BibitemShut {NoStop}%
\bibitem [{\citenamefont {Fang}\ \emph {et~al.}(2012)\citenamefont {Fang},
  \citenamefont {Gilbert}, \citenamefont {Dai},\ and\ \citenamefont
  {Bernevig}}]{fang_PRL2012_multi}%
  \BibitemOpen
  \bibfield  {author} {\bibinfo {author} {\bibfnamefont {C.}~\bibnamefont
  {Fang}}, \bibinfo {author} {\bibfnamefont {M.~J.}\ \bibnamefont {Gilbert}},
  \bibinfo {author} {\bibfnamefont {X.}~\bibnamefont {Dai}},\ and\ \bibinfo
  {author} {\bibfnamefont {B.~A.}\ \bibnamefont {Bernevig}},\ }\bibfield
  {title} {\bibinfo {title} {Multi-weyl topological semimetals stabilized by
  point group symmetry},\ }\href
  {https://doi.org/10.1103/PhysRevLett.108.266802} {\bibfield  {journal}
  {\bibinfo  {journal} {Phys. Rev. Lett.}\ }\textbf {\bibinfo {volume} {108}},\
  \bibinfo {pages} {266802} (\bibinfo {year} {2012})}\BibitemShut {NoStop}%
\bibitem [{\citenamefont {Li}\ \emph {et~al.}(2016{\natexlab{b}})\citenamefont
  {Li}, \citenamefont {Roy},\ and\ \citenamefont
  {Das~Sarma}}]{li_PRB2016_weyl}%
  \BibitemOpen
  \bibfield  {author} {\bibinfo {author} {\bibfnamefont {X.}~\bibnamefont
  {Li}}, \bibinfo {author} {\bibfnamefont {B.}~\bibnamefont {Roy}},\ and\
  \bibinfo {author} {\bibfnamefont {S.}~\bibnamefont {Das~Sarma}},\ }\bibfield
  {title} {\bibinfo {title} {Weyl fermions with arbitrary monopoles in magnetic
  fields: Landau levels, longitudinal magnetotransport, and density-wave
  ordering},\ }\href {https://doi.org/10.1103/PhysRevB.94.195144} {\bibfield
  {journal} {\bibinfo  {journal} {Phys. Rev. B}\ }\textbf {\bibinfo {volume}
  {94}},\ \bibinfo {pages} {195144} (\bibinfo {year}
  {2016}{\natexlab{b}})}\BibitemShut {NoStop}%
\bibitem [{\citenamefont {Huang}\ \emph {et~al.}(2017)\citenamefont {Huang},
  \citenamefont {Zhou},\ and\ \citenamefont {Shen}}]{Huang_prb17_topological}%
  \BibitemOpen
  \bibfield  {author} {\bibinfo {author} {\bibfnamefont {Z.-M.}\ \bibnamefont
  {Huang}}, \bibinfo {author} {\bibfnamefont {J.}~\bibnamefont {Zhou}},\ and\
  \bibinfo {author} {\bibfnamefont {S.-Q.}\ \bibnamefont {Shen}},\ }\bibfield
  {title} {\bibinfo {title} {Topological responses from chiral anomaly in
  multi-weyl semimetals},\ }\href {https://doi.org/10.1103/PhysRevB.96.085201}
  {\bibfield  {journal} {\bibinfo  {journal} {Phys. Rev. B}\ }\textbf {\bibinfo
  {volume} {96}},\ \bibinfo {pages} {085201} (\bibinfo {year}
  {2017})}\BibitemShut {NoStop}%
\bibitem [{\citenamefont {Dantas}\ \emph {et~al.}(2018)\citenamefont {Dantas},
  \citenamefont {Pe{\~{n}}a-Benitez}, \citenamefont {Roy},\ and\ \citenamefont
  {Sur{\'o}wka}}]{dantas_JHEP2018_magne}%
  \BibitemOpen
  \bibfield  {author} {\bibinfo {author} {\bibfnamefont {R.~M.~A.}\
  \bibnamefont {Dantas}}, \bibinfo {author} {\bibfnamefont {F.}~\bibnamefont
  {Pe{\~{n}}a-Benitez}}, \bibinfo {author} {\bibfnamefont {B.}~\bibnamefont
  {Roy}},\ and\ \bibinfo {author} {\bibfnamefont {P.}~\bibnamefont
  {Sur{\'o}wka}},\ }\bibfield  {title} {\bibinfo {title} {Magnetotransport in
  multi-weyl semimetals: a kinetic theory approach},\ }\href
  {https://doi.org/10.1007/JHEP12(2018)069} {\bibfield  {journal} {\bibinfo
  {journal} {Journal of High Energy Physics}\ }\textbf {\bibinfo {volume}
  {2018}},\ \bibinfo {pages} {69} (\bibinfo {year} {2018})}\BibitemShut
  {NoStop}%
\bibitem [{\citenamefont {Das}\ \emph {et~al.}(2022)\citenamefont {Das},
  \citenamefont {Das},\ and\ \citenamefont {Agarwal}}]{Sunit_prb22}%
  \BibitemOpen
  \bibfield  {author} {\bibinfo {author} {\bibfnamefont {S.}~\bibnamefont
  {Das}}, \bibinfo {author} {\bibfnamefont {K.}~\bibnamefont {Das}},\ and\
  \bibinfo {author} {\bibfnamefont {A.}~\bibnamefont {Agarwal}},\ }\bibfield
  {title} {\bibinfo {title} {Nonlinear magnetoconductivity in weyl and
  multi-weyl semimetals in quantizing magnetic field},\ }\href
  {https://doi.org/10.1103/PhysRevB.105.235408} {\bibfield  {journal} {\bibinfo
   {journal} {Phys. Rev. B}\ }\textbf {\bibinfo {volume} {105}},\ \bibinfo
  {pages} {235408} (\bibinfo {year} {2022})}\BibitemShut {NoStop}%
\bibitem [{\citenamefont {Lepori}\ \emph {et~al.}(2018)\citenamefont {Lepori},
  \citenamefont {Burrello},\ and\ \citenamefont
  {Guadagnini}}]{lepori_JHEP2018_axial}%
  \BibitemOpen
  \bibfield  {author} {\bibinfo {author} {\bibfnamefont {L.}~\bibnamefont
  {Lepori}}, \bibinfo {author} {\bibfnamefont {M.}~\bibnamefont {Burrello}},\
  and\ \bibinfo {author} {\bibfnamefont {E.}~\bibnamefont {Guadagnini}},\
  }\bibfield  {title} {\bibinfo {title} {Axial anomaly in multi-weyl and
  triple-point semimetals},\ }\href {https://doi.org/10.1007/JHEP06(2018)110}
  {\bibfield  {journal} {\bibinfo  {journal} {Journal of High Energy Physics}\
  }\textbf {\bibinfo {volume} {2018}},\ \bibinfo {pages} {110} (\bibinfo {year}
  {2018})}\BibitemShut {NoStop}%
\bibitem [{\citenamefont {Cheon}\ \emph {et~al.}(2022)\citenamefont {Cheon},
  \citenamefont {Cho}, \citenamefont {Kim},\ and\ \citenamefont
  {Lee}}]{cheon_prb22_chiral}%
  \BibitemOpen
  \bibfield  {author} {\bibinfo {author} {\bibfnamefont {S.}~\bibnamefont
  {Cheon}}, \bibinfo {author} {\bibfnamefont {G.~Y.}\ \bibnamefont {Cho}},
  \bibinfo {author} {\bibfnamefont {K.-S.}\ \bibnamefont {Kim}},\ and\ \bibinfo
  {author} {\bibfnamefont {H.-W.}\ \bibnamefont {Lee}},\ }\bibfield  {title}
  {\bibinfo {title} {Chiral anomaly in noncentrosymmetric systems induced by
  spin-orbit coupling},\ }\href {https://doi.org/10.1103/PhysRevB.105.L180303}
  {\bibfield  {journal} {\bibinfo  {journal} {Phys. Rev. B}\ }\textbf {\bibinfo
  {volume} {105}},\ \bibinfo {pages} {L180303} (\bibinfo {year}
  {2022})}\BibitemShut {NoStop}%
\bibitem [{\citenamefont {Gao}\ and\ \citenamefont
  {Huang}(2022)}]{Gao_cpl22_chiral}%
  \BibitemOpen
  \bibfield  {author} {\bibinfo {author} {\bibfnamefont {L.-L.}\ \bibnamefont
  {Gao}}\ and\ \bibinfo {author} {\bibfnamefont {X.-G.}\ \bibnamefont
  {Huang}},\ }\bibfield  {title} {\bibinfo {title} {Chiral anomaly in
  non-relativistic systems: Berry curvature and chiral kinetic theory},\ }\href
  {https://doi.org/10.1088/0256-307x/39/2/021101} {\bibfield  {journal}
  {\bibinfo  {journal} {Chinese Physics Letters}\ }\textbf {\bibinfo {volume}
  {39}},\ \bibinfo {pages} {021101} (\bibinfo {year} {2022})}\BibitemShut
  {NoStop}%
\bibitem [{\citenamefont {Bradlyn}\ \emph {et~al.}(2016)\citenamefont
  {Bradlyn}, \citenamefont {Cano}, \citenamefont {Wang}, \citenamefont
  {Vergniory}, \citenamefont {Felser}, \citenamefont {Cava},\ and\
  \citenamefont {Bernevig}}]{Barnevig_science16}%
  \BibitemOpen
  \bibfield  {author} {\bibinfo {author} {\bibfnamefont {B.}~\bibnamefont
  {Bradlyn}}, \bibinfo {author} {\bibfnamefont {J.}~\bibnamefont {Cano}},
  \bibinfo {author} {\bibfnamefont {Z.}~\bibnamefont {Wang}}, \bibinfo {author}
  {\bibfnamefont {M.~G.}\ \bibnamefont {Vergniory}}, \bibinfo {author}
  {\bibfnamefont {C.}~\bibnamefont {Felser}}, \bibinfo {author} {\bibfnamefont
  {R.~J.}\ \bibnamefont {Cava}},\ and\ \bibinfo {author} {\bibfnamefont
  {B.~A.}\ \bibnamefont {Bernevig}},\ }\bibfield  {title} {\bibinfo {title}
  {Beyond dirac and weyl fermions: Unconventional quasiparticles in
  conventional crystals},\ }\href {https://doi.org/10.1126/science.aaf5037}
  {\bibfield  {journal} {\bibinfo  {journal} {Science}\ }\textbf {\bibinfo
  {volume} {353}},\ \bibinfo {pages} {aaf5037} (\bibinfo {year}
  {2016})}\BibitemShut {NoStop}%
\bibitem [{\citenamefont {Chang}\ \emph {et~al.}(2018)\citenamefont {Chang},
  \citenamefont {Wieder}, \citenamefont {Schindler}, \citenamefont {Sanchez},
  \citenamefont {Belopolski}, \citenamefont {Huang}, \citenamefont {Singh},
  \citenamefont {Wu}, \citenamefont {Chang}, \citenamefont {Neupert},
  \citenamefont {Xu}, \citenamefont {Lin},\ and\ \citenamefont
  {Hasan}}]{Chang18_NM}%
  \BibitemOpen
  \bibfield  {author} {\bibinfo {author} {\bibfnamefont {G.}~\bibnamefont
  {Chang}}, \bibinfo {author} {\bibfnamefont {B.~J.}\ \bibnamefont {Wieder}},
  \bibinfo {author} {\bibfnamefont {F.}~\bibnamefont {Schindler}}, \bibinfo
  {author} {\bibfnamefont {D.~S.}\ \bibnamefont {Sanchez}}, \bibinfo {author}
  {\bibfnamefont {I.}~\bibnamefont {Belopolski}}, \bibinfo {author}
  {\bibfnamefont {S.-M.}\ \bibnamefont {Huang}}, \bibinfo {author}
  {\bibfnamefont {B.}~\bibnamefont {Singh}}, \bibinfo {author} {\bibfnamefont
  {D.}~\bibnamefont {Wu}}, \bibinfo {author} {\bibfnamefont {T.-R.}\
  \bibnamefont {Chang}}, \bibinfo {author} {\bibfnamefont {T.}~\bibnamefont
  {Neupert}}, \bibinfo {author} {\bibfnamefont {S.-Y.}\ \bibnamefont {Xu}},
  \bibinfo {author} {\bibfnamefont {H.}~\bibnamefont {Lin}},\ and\ \bibinfo
  {author} {\bibfnamefont {M.~Z.}\ \bibnamefont {Hasan}},\ }\bibfield  {title}
  {\bibinfo {title} {Topological quantum properties of chiral crystals},\
  }\href {https://doi.org/10.1038/s41563-018-0169-3} {\bibfield  {journal}
  {\bibinfo  {journal} {Nature Materials}\ }\textbf {\bibinfo {volume} {17}},\
  \bibinfo {pages} {978} (\bibinfo {year} {2018})}\BibitemShut {NoStop}%
\bibitem [{\citenamefont {He}\ \emph {et~al.}(2021)\citenamefont {He},
  \citenamefont {Xu},\ and\ \citenamefont {Law}}]{law21_comm_phys}%
  \BibitemOpen
  \bibfield  {author} {\bibinfo {author} {\bibfnamefont {W.-Y.}\ \bibnamefont
  {He}}, \bibinfo {author} {\bibfnamefont {X.~Y.}\ \bibnamefont {Xu}},\ and\
  \bibinfo {author} {\bibfnamefont {K.~T.}\ \bibnamefont {Law}},\ }\bibfield
  {title} {\bibinfo {title} {Kramers weyl semimetals as quantum solenoids and
  their applications in spin-orbit torque devices},\ }\href
  {https://doi.org/10.1038/s42005-021-00564-w} {\bibfield  {journal} {\bibinfo
  {journal} {Communications Physics}\ }\textbf {\bibinfo {volume} {4}},\
  \bibinfo {pages} {66} (\bibinfo {year} {2021})}\BibitemShut {NoStop}%
\bibitem [{\citenamefont {Zhang}\ \emph {et~al.}(2017)\citenamefont {Zhang},
  \citenamefont {Schindler}, \citenamefont {Liu}, \citenamefont {Chang},
  \citenamefont {Xu}, \citenamefont {Chang}, \citenamefont {Hua}, \citenamefont
  {Jiang}, \citenamefont {Yuan}, \citenamefont {Sun}, \citenamefont {Jeng},
  \citenamefont {Lu}, \citenamefont {Lin}, \citenamefont {Hasan}, \citenamefont
  {Xie}, \citenamefont {Neupert},\ and\ \citenamefont {Jia}}]{zhang_prb17}%
  \BibitemOpen
  \bibfield  {author} {\bibinfo {author} {\bibfnamefont {C.-L.}\ \bibnamefont
  {Zhang}}, \bibinfo {author} {\bibfnamefont {F.}~\bibnamefont {Schindler}},
  \bibinfo {author} {\bibfnamefont {H.}~\bibnamefont {Liu}}, \bibinfo {author}
  {\bibfnamefont {T.-R.}\ \bibnamefont {Chang}}, \bibinfo {author}
  {\bibfnamefont {S.-Y.}\ \bibnamefont {Xu}}, \bibinfo {author} {\bibfnamefont
  {G.}~\bibnamefont {Chang}}, \bibinfo {author} {\bibfnamefont
  {W.}~\bibnamefont {Hua}}, \bibinfo {author} {\bibfnamefont {H.}~\bibnamefont
  {Jiang}}, \bibinfo {author} {\bibfnamefont {Z.}~\bibnamefont {Yuan}},
  \bibinfo {author} {\bibfnamefont {J.}~\bibnamefont {Sun}}, \bibinfo {author}
  {\bibfnamefont {H.-T.}\ \bibnamefont {Jeng}}, \bibinfo {author}
  {\bibfnamefont {H.-Z.}\ \bibnamefont {Lu}}, \bibinfo {author} {\bibfnamefont
  {H.}~\bibnamefont {Lin}}, \bibinfo {author} {\bibfnamefont {M.~Z.}\
  \bibnamefont {Hasan}}, \bibinfo {author} {\bibfnamefont {X.~C.}\ \bibnamefont
  {Xie}}, \bibinfo {author} {\bibfnamefont {T.}~\bibnamefont {Neupert}},\ and\
  \bibinfo {author} {\bibfnamefont {S.}~\bibnamefont {Jia}},\ }\bibfield
  {title} {\bibinfo {title} {Ultraquantum magnetoresistance in the kramers-weyl
  semimetal candidate
  $\ensuremath{\beta}{\text{\ensuremath{-}}\mathrm{ag}}_{2}\mathrm{Se}$},\
  }\href {https://doi.org/10.1103/PhysRevB.96.165148} {\bibfield  {journal}
  {\bibinfo  {journal} {Phys. Rev. B}\ }\textbf {\bibinfo {volume} {96}},\
  \bibinfo {pages} {165148} (\bibinfo {year} {2017})}\BibitemShut {NoStop}%
\bibitem [{\citenamefont {Schr{\"o}ter}\ \emph {et~al.}(2019)\citenamefont
  {Schr{\"o}ter}, \citenamefont {Pei}, \citenamefont {Vergniory}, \citenamefont
  {Sun}, \citenamefont {Manna}, \citenamefont {de~Juan}, \citenamefont
  {Krieger}, \citenamefont {S{\"u}ss}, \citenamefont {Schmidt}, \citenamefont
  {Dudin}, \citenamefont {Bradlyn}, \citenamefont {Kim}, \citenamefont
  {Schmitt}, \citenamefont {Cacho}, \citenamefont {Felser}, \citenamefont
  {Strocov},\ and\ \citenamefont {Chen}}]{Schroter_NP19}%
  \BibitemOpen
  \bibfield  {author} {\bibinfo {author} {\bibfnamefont {N.~B.~M.}\
  \bibnamefont {Schr{\"o}ter}}, \bibinfo {author} {\bibfnamefont
  {D.}~\bibnamefont {Pei}}, \bibinfo {author} {\bibfnamefont {M.~G.}\
  \bibnamefont {Vergniory}}, \bibinfo {author} {\bibfnamefont {Y.}~\bibnamefont
  {Sun}}, \bibinfo {author} {\bibfnamefont {K.}~\bibnamefont {Manna}}, \bibinfo
  {author} {\bibfnamefont {F.}~\bibnamefont {de~Juan}}, \bibinfo {author}
  {\bibfnamefont {J.~A.}\ \bibnamefont {Krieger}}, \bibinfo {author}
  {\bibfnamefont {V.}~\bibnamefont {S{\"u}ss}}, \bibinfo {author}
  {\bibfnamefont {M.}~\bibnamefont {Schmidt}}, \bibinfo {author} {\bibfnamefont
  {P.}~\bibnamefont {Dudin}}, \bibinfo {author} {\bibfnamefont
  {B.}~\bibnamefont {Bradlyn}}, \bibinfo {author} {\bibfnamefont {T.~K.}\
  \bibnamefont {Kim}}, \bibinfo {author} {\bibfnamefont {T.}~\bibnamefont
  {Schmitt}}, \bibinfo {author} {\bibfnamefont {C.}~\bibnamefont {Cacho}},
  \bibinfo {author} {\bibfnamefont {C.}~\bibnamefont {Felser}}, \bibinfo
  {author} {\bibfnamefont {V.~N.}\ \bibnamefont {Strocov}},\ and\ \bibinfo
  {author} {\bibfnamefont {Y.}~\bibnamefont {Chen}},\ }\bibfield  {title}
  {\bibinfo {title} {Chiral topological semimetal with multifold band crossings
  and long fermi arcs},\ }\href {https://doi.org/10.1038/s41567-019-0511-y}
  {\bibfield  {journal} {\bibinfo  {journal} {Nature Physics}\ }\textbf
  {\bibinfo {volume} {15}},\ \bibinfo {pages} {759} (\bibinfo {year}
  {2019})}\BibitemShut {NoStop}%
\bibitem [{\citenamefont {Tan}\ \emph {et~al.}(2022)\citenamefont {Tan},
  \citenamefont {Jiang}, \citenamefont {Li}, \citenamefont {Wu}, \citenamefont
  {Wang},\ and\ \citenamefont {Huang}}]{Tan_ADFM22}%
  \BibitemOpen
  \bibfield  {author} {\bibinfo {author} {\bibfnamefont {W.}~\bibnamefont
  {Tan}}, \bibinfo {author} {\bibfnamefont {X.}~\bibnamefont {Jiang}}, \bibinfo
  {author} {\bibfnamefont {Y.}~\bibnamefont {Li}}, \bibinfo {author}
  {\bibfnamefont {X.}~\bibnamefont {Wu}}, \bibinfo {author} {\bibfnamefont
  {J.}~\bibnamefont {Wang}},\ and\ \bibinfo {author} {\bibfnamefont
  {B.}~\bibnamefont {Huang}},\ }\bibfield  {title} {\bibinfo {title} {A unified
  understanding of diverse spin textures of kramers–weyl fermions in
  nonmagnetic chiral crystals},\ }\href
  {https://doi.org/https://doi.org/10.1002/adfm.202208023} {\bibfield
  {journal} {\bibinfo  {journal} {Advanced Functional Materials}\ }\textbf
  {\bibinfo {volume} {32}},\ \bibinfo {pages} {2208023} (\bibinfo {year}
  {2022})}\BibitemShut {NoStop}%
\bibitem [{\citenamefont {Rao}\ \emph {et~al.}(2019)\citenamefont {Rao},
  \citenamefont {Li}, \citenamefont {Zhang}, \citenamefont {Tian},
  \citenamefont {Li}, \citenamefont {Fu}, \citenamefont {Tang}, \citenamefont
  {Wang}, \citenamefont {Li}, \citenamefont {Fan}, \citenamefont {Li},
  \citenamefont {Huang}, \citenamefont {Liu}, \citenamefont {Long},
  \citenamefont {Fang}, \citenamefont {Weng}, \citenamefont {Shi},
  \citenamefont {Lei}, \citenamefont {Sun}, \citenamefont {Qian},\ and\
  \citenamefont {Ding}}]{Rao_N19}%
  \BibitemOpen
  \bibfield  {author} {\bibinfo {author} {\bibfnamefont {Z.}~\bibnamefont
  {Rao}}, \bibinfo {author} {\bibfnamefont {H.}~\bibnamefont {Li}}, \bibinfo
  {author} {\bibfnamefont {T.}~\bibnamefont {Zhang}}, \bibinfo {author}
  {\bibfnamefont {S.}~\bibnamefont {Tian}}, \bibinfo {author} {\bibfnamefont
  {C.}~\bibnamefont {Li}}, \bibinfo {author} {\bibfnamefont {B.}~\bibnamefont
  {Fu}}, \bibinfo {author} {\bibfnamefont {C.}~\bibnamefont {Tang}}, \bibinfo
  {author} {\bibfnamefont {L.}~\bibnamefont {Wang}}, \bibinfo {author}
  {\bibfnamefont {Z.}~\bibnamefont {Li}}, \bibinfo {author} {\bibfnamefont
  {W.}~\bibnamefont {Fan}}, \bibinfo {author} {\bibfnamefont {J.}~\bibnamefont
  {Li}}, \bibinfo {author} {\bibfnamefont {Y.}~\bibnamefont {Huang}}, \bibinfo
  {author} {\bibfnamefont {Z.}~\bibnamefont {Liu}}, \bibinfo {author}
  {\bibfnamefont {Y.}~\bibnamefont {Long}}, \bibinfo {author} {\bibfnamefont
  {C.}~\bibnamefont {Fang}}, \bibinfo {author} {\bibfnamefont {H.}~\bibnamefont
  {Weng}}, \bibinfo {author} {\bibfnamefont {Y.}~\bibnamefont {Shi}}, \bibinfo
  {author} {\bibfnamefont {H.}~\bibnamefont {Lei}}, \bibinfo {author}
  {\bibfnamefont {Y.}~\bibnamefont {Sun}}, \bibinfo {author} {\bibfnamefont
  {T.}~\bibnamefont {Qian}},\ and\ \bibinfo {author} {\bibfnamefont
  {H.}~\bibnamefont {Ding}},\ }\bibfield  {title} {\bibinfo {title}
  {Observation of unconventional chiral fermions with long fermi arcs in
  cosi},\ }\href {https://doi.org/10.1038/s41586-019-1031-8} {\bibfield
  {journal} {\bibinfo  {journal} {Nature}\ }\textbf {\bibinfo {volume} {567}},\
  \bibinfo {pages} {496} (\bibinfo {year} {2019})}\BibitemShut {NoStop}%
\bibitem [{\citenamefont {Sanchez}\ \emph {et~al.}(2019)\citenamefont
  {Sanchez}, \citenamefont {Belopolski}, \citenamefont {Cochran}, \citenamefont
  {Xu}, \citenamefont {Yin}, \citenamefont {Chang}, \citenamefont {Xie},
  \citenamefont {Manna}, \citenamefont {S{\"u}{\ss}}, \citenamefont {Huang},
  \citenamefont {Alidoust}, \citenamefont {Multer}, \citenamefont {Zhang},
  \citenamefont {Shumiya}, \citenamefont {Wang}, \citenamefont {Wang},
  \citenamefont {Chang}, \citenamefont {Felser}, \citenamefont {Xu},
  \citenamefont {Jia}, \citenamefont {Lin},\ and\ \citenamefont
  {Hasan}}]{Sanchez_N19}%
  \BibitemOpen
  \bibfield  {author} {\bibinfo {author} {\bibfnamefont {D.~S.}\ \bibnamefont
  {Sanchez}}, \bibinfo {author} {\bibfnamefont {I.}~\bibnamefont {Belopolski}},
  \bibinfo {author} {\bibfnamefont {T.~A.}\ \bibnamefont {Cochran}}, \bibinfo
  {author} {\bibfnamefont {X.}~\bibnamefont {Xu}}, \bibinfo {author}
  {\bibfnamefont {J.-X.}\ \bibnamefont {Yin}}, \bibinfo {author} {\bibfnamefont
  {G.}~\bibnamefont {Chang}}, \bibinfo {author} {\bibfnamefont
  {W.}~\bibnamefont {Xie}}, \bibinfo {author} {\bibfnamefont {K.}~\bibnamefont
  {Manna}}, \bibinfo {author} {\bibfnamefont {V.}~\bibnamefont {S{\"u}{\ss}}},
  \bibinfo {author} {\bibfnamefont {C.-Y.}\ \bibnamefont {Huang}}, \bibinfo
  {author} {\bibfnamefont {N.}~\bibnamefont {Alidoust}}, \bibinfo {author}
  {\bibfnamefont {D.}~\bibnamefont {Multer}}, \bibinfo {author} {\bibfnamefont
  {S.~S.}\ \bibnamefont {Zhang}}, \bibinfo {author} {\bibfnamefont
  {N.}~\bibnamefont {Shumiya}}, \bibinfo {author} {\bibfnamefont
  {X.}~\bibnamefont {Wang}}, \bibinfo {author} {\bibfnamefont {G.-Q.}\
  \bibnamefont {Wang}}, \bibinfo {author} {\bibfnamefont {T.-R.}\ \bibnamefont
  {Chang}}, \bibinfo {author} {\bibfnamefont {C.}~\bibnamefont {Felser}},
  \bibinfo {author} {\bibfnamefont {S.-Y.}\ \bibnamefont {Xu}}, \bibinfo
  {author} {\bibfnamefont {S.}~\bibnamefont {Jia}}, \bibinfo {author}
  {\bibfnamefont {H.}~\bibnamefont {Lin}},\ and\ \bibinfo {author}
  {\bibfnamefont {M.~Z.}\ \bibnamefont {Hasan}},\ }\bibfield  {title} {\bibinfo
  {title} {Topological chiral crystals with helicoid-arc quantum states},\
  }\href {https://doi.org/10.1038/s41586-019-1037-2} {\bibfield  {journal}
  {\bibinfo  {journal} {Nature}\ }\textbf {\bibinfo {volume} {567}},\ \bibinfo
  {pages} {500} (\bibinfo {year} {2019})}\BibitemShut {NoStop}%
\bibitem [{\citenamefont {Dutta}\ \emph {et~al.}(2022)\citenamefont {Dutta},
  \citenamefont {Ghosh}, \citenamefont {Singh}, \citenamefont {Lin},
  \citenamefont {Politano}, \citenamefont {Bansil},\ and\ \citenamefont
  {Agarwal}}]{debasis_prb22}%
  \BibitemOpen
  \bibfield  {author} {\bibinfo {author} {\bibfnamefont {D.}~\bibnamefont
  {Dutta}}, \bibinfo {author} {\bibfnamefont {B.}~\bibnamefont {Ghosh}},
  \bibinfo {author} {\bibfnamefont {B.}~\bibnamefont {Singh}}, \bibinfo
  {author} {\bibfnamefont {H.}~\bibnamefont {Lin}}, \bibinfo {author}
  {\bibfnamefont {A.}~\bibnamefont {Politano}}, \bibinfo {author}
  {\bibfnamefont {A.}~\bibnamefont {Bansil}},\ and\ \bibinfo {author}
  {\bibfnamefont {A.}~\bibnamefont {Agarwal}},\ }\bibfield  {title} {\bibinfo
  {title} {Collective plasmonic modes in the chiral multifold fermionic
  material cosi},\ }\href {https://doi.org/10.1103/PhysRevB.105.165104}
  {\bibfield  {journal} {\bibinfo  {journal} {Phys. Rev. B}\ }\textbf {\bibinfo
  {volume} {105}},\ \bibinfo {pages} {165104} (\bibinfo {year}
  {2022})}\BibitemShut {NoStop}%
\bibitem [{\citenamefont {Takane}\ \emph {et~al.}(2019)\citenamefont {Takane},
  \citenamefont {Wang}, \citenamefont {Souma}, \citenamefont {Nakayama},
  \citenamefont {Nakamura}, \citenamefont {Oinuma}, \citenamefont {Nakata},
  \citenamefont {Iwasawa}, \citenamefont {Cacho}, \citenamefont {Kim},
  \citenamefont {Horiba}, \citenamefont {Kumigashira}, \citenamefont
  {Takahashi}, \citenamefont {Ando},\ and\ \citenamefont
  {Sato}}]{takane_prl19}%
  \BibitemOpen
  \bibfield  {author} {\bibinfo {author} {\bibfnamefont {D.}~\bibnamefont
  {Takane}}, \bibinfo {author} {\bibfnamefont {Z.}~\bibnamefont {Wang}},
  \bibinfo {author} {\bibfnamefont {S.}~\bibnamefont {Souma}}, \bibinfo
  {author} {\bibfnamefont {K.}~\bibnamefont {Nakayama}}, \bibinfo {author}
  {\bibfnamefont {T.}~\bibnamefont {Nakamura}}, \bibinfo {author}
  {\bibfnamefont {H.}~\bibnamefont {Oinuma}}, \bibinfo {author} {\bibfnamefont
  {Y.}~\bibnamefont {Nakata}}, \bibinfo {author} {\bibfnamefont
  {H.}~\bibnamefont {Iwasawa}}, \bibinfo {author} {\bibfnamefont
  {C.}~\bibnamefont {Cacho}}, \bibinfo {author} {\bibfnamefont
  {T.}~\bibnamefont {Kim}}, \bibinfo {author} {\bibfnamefont {K.}~\bibnamefont
  {Horiba}}, \bibinfo {author} {\bibfnamefont {H.}~\bibnamefont {Kumigashira}},
  \bibinfo {author} {\bibfnamefont {T.}~\bibnamefont {Takahashi}}, \bibinfo
  {author} {\bibfnamefont {Y.}~\bibnamefont {Ando}},\ and\ \bibinfo {author}
  {\bibfnamefont {T.}~\bibnamefont {Sato}},\ }\bibfield  {title} {\bibinfo
  {title} {Observation of chiral fermions with a large topological charge and
  associated fermi-arc surface states in cosi},\ }\href
  {https://doi.org/10.1103/PhysRevLett.122.076402} {\bibfield  {journal}
  {\bibinfo  {journal} {Phys. Rev. Lett.}\ }\textbf {\bibinfo {volume} {122}},\
  \bibinfo {pages} {076402} (\bibinfo {year} {2019})}\BibitemShut {NoStop}%
\bibitem [{\citenamefont {Verma}\ \emph {et~al.}(2019)\citenamefont {Verma},
  \citenamefont {Biswas},\ and\ \citenamefont {Ghosh}}]{Verma_2021}%
  \BibitemOpen
  \bibfield  {author} {\bibinfo {author} {\bibfnamefont {S.}~\bibnamefont
  {Verma}}, \bibinfo {author} {\bibfnamefont {T.}~\bibnamefont {Biswas}},\ and\
  \bibinfo {author} {\bibfnamefont {T.~K.}\ \bibnamefont {Ghosh}},\ }\bibfield
  {title} {\bibinfo {title} {Thermoelectric and optical probes for a fermi
  surface topology change in noncentrosymmetric metals},\ }\href
  {https://doi.org/10.1103/PhysRevB.100.045201} {\bibfield  {journal} {\bibinfo
   {journal} {Phys. Rev. B}\ }\textbf {\bibinfo {volume} {100}},\ \bibinfo
  {pages} {045201} (\bibinfo {year} {2019})}\BibitemShut {NoStop}%
\bibitem [{\citenamefont {Pal}\ \emph {et~al.}(2021)\citenamefont {Pal},
  \citenamefont {Dey},\ and\ \citenamefont {Ghosh}}]{pal_jpcm21_berry}%
  \BibitemOpen
  \bibfield  {author} {\bibinfo {author} {\bibfnamefont {O.}~\bibnamefont
  {Pal}}, \bibinfo {author} {\bibfnamefont {B.}~\bibnamefont {Dey}},\ and\
  \bibinfo {author} {\bibfnamefont {T.~K.}\ \bibnamefont {Ghosh}},\ }\bibfield
  {title} {\bibinfo {title} {Berry curvature induced magnetotransport in 3d
  noncentrosymmetric metals},\ }\href
  {https://doi.org/10.1088/1361-648x/ac2fd4} {\bibfield  {journal} {\bibinfo
  {journal} {Journal of Physics: Condensed Matter}\ }\textbf {\bibinfo {volume}
  {34}},\ \bibinfo {pages} {025702} (\bibinfo {year} {2021})}\BibitemShut
  {NoStop}%
\bibitem [{\citenamefont {Armitage}\ \emph {et~al.}(2018)\citenamefont
  {Armitage}, \citenamefont {Mele},\ and\ \citenamefont
  {Vishwanath}}]{armitage_rmp18_weyl}%
  \BibitemOpen
  \bibfield  {author} {\bibinfo {author} {\bibfnamefont {N.~P.}\ \bibnamefont
  {Armitage}}, \bibinfo {author} {\bibfnamefont {E.~J.}\ \bibnamefont {Mele}},\
  and\ \bibinfo {author} {\bibfnamefont {A.}~\bibnamefont {Vishwanath}},\
  }\bibfield  {title} {\bibinfo {title} {Weyl and dirac semimetals in
  three-dimensional solids},\ }\href
  {https://doi.org/10.1103/RevModPhys.90.015001} {\bibfield  {journal}
  {\bibinfo  {journal} {Rev. Mod. Phys.}\ }\textbf {\bibinfo {volume} {90}},\
  \bibinfo {pages} {015001} (\bibinfo {year} {2018})}\BibitemShut {NoStop}%
\bibitem [{\citenamefont {Kang}\ and\ \citenamefont
  {Zang}(2015)}]{kang_prb15_transport}%
  \BibitemOpen
  \bibfield  {author} {\bibinfo {author} {\bibfnamefont {J.}~\bibnamefont
  {Kang}}\ and\ \bibinfo {author} {\bibfnamefont {J.}~\bibnamefont {Zang}},\
  }\bibfield  {title} {\bibinfo {title} {Transport theory of metallic $b20$
  helimagnets},\ }\href {https://doi.org/10.1103/PhysRevB.91.134401} {\bibfield
   {journal} {\bibinfo  {journal} {Phys. Rev. B}\ }\textbf {\bibinfo {volume}
  {91}},\ \bibinfo {pages} {134401} (\bibinfo {year} {2015})}\BibitemShut
  {NoStop}%
\bibitem [{\citenamefont {Samokhin}(2008)}]{samokhin_prb08_effects}%
  \BibitemOpen
  \bibfield  {author} {\bibinfo {author} {\bibfnamefont {K.~V.}\ \bibnamefont
  {Samokhin}},\ }\bibfield  {title} {\bibinfo {title} {Effects of impurities on
  the upper critical field ${H}_{c2}$ in superconductors without inversion
  symmetry},\ }\href {https://doi.org/10.1103/PhysRevB.78.144511} {\bibfield
  {journal} {\bibinfo  {journal} {Phys. Rev. B}\ }\textbf {\bibinfo {volume}
  {78}},\ \bibinfo {pages} {144511} (\bibinfo {year} {2008})}\BibitemShut
  {NoStop}%
\bibitem [{\citenamefont {Xiao}\ \emph {et~al.}(2010)\citenamefont {Xiao},
  \citenamefont {Chang},\ and\ \citenamefont {Niu}}]{xiao_rmp10_berry}%
  \BibitemOpen
  \bibfield  {author} {\bibinfo {author} {\bibfnamefont {D.}~\bibnamefont
  {Xiao}}, \bibinfo {author} {\bibfnamefont {M.-C.}\ \bibnamefont {Chang}},\
  and\ \bibinfo {author} {\bibfnamefont {Q.}~\bibnamefont {Niu}},\ }\bibfield
  {title} {\bibinfo {title} {Berry phase effects on electronic properties},\
  }\href {https://doi.org/10.1103/RevModPhys.82.1959} {\bibfield  {journal}
  {\bibinfo  {journal} {Rev. Mod. Phys.}\ }\textbf {\bibinfo {volume} {82}},\
  \bibinfo {pages} {1959} (\bibinfo {year} {2010})}\BibitemShut {NoStop}%
\bibitem [{\citenamefont {Morimoto}\ \emph {et~al.}(2016)\citenamefont
  {Morimoto}, \citenamefont {Zhong}, \citenamefont {Orenstein},\ and\
  \citenamefont {Moore}}]{morimoto_prb16_semiclassical}%
  \BibitemOpen
  \bibfield  {author} {\bibinfo {author} {\bibfnamefont {T.}~\bibnamefont
  {Morimoto}}, \bibinfo {author} {\bibfnamefont {S.}~\bibnamefont {Zhong}},
  \bibinfo {author} {\bibfnamefont {J.}~\bibnamefont {Orenstein}},\ and\
  \bibinfo {author} {\bibfnamefont {J.~E.}\ \bibnamefont {Moore}},\ }\bibfield
  {title} {\bibinfo {title} {Semiclassical theory of nonlinear magneto-optical
  responses with applications to topological dirac/weyl semimetals},\ }\href
  {https://doi.org/10.1103/PhysRevB.94.245121} {\bibfield  {journal} {\bibinfo
  {journal} {Phys. Rev. B}\ }\textbf {\bibinfo {volume} {94}},\ \bibinfo
  {pages} {245121} (\bibinfo {year} {2016})}\BibitemShut {NoStop}%
\bibitem [{\citenamefont {Xiao}\ \emph {et~al.}(2005)\citenamefont {Xiao},
  \citenamefont {Shi},\ and\ \citenamefont {Niu}}]{xiao_prl05}%
  \BibitemOpen
  \bibfield  {author} {\bibinfo {author} {\bibfnamefont {D.}~\bibnamefont
  {Xiao}}, \bibinfo {author} {\bibfnamefont {J.}~\bibnamefont {Shi}},\ and\
  \bibinfo {author} {\bibfnamefont {Q.}~\bibnamefont {Niu}},\ }\bibfield
  {title} {\bibinfo {title} {Berry phase correction to electron density of
  states in solids},\ }\href {https://doi.org/10.1103/PhysRevLett.95.137204}
  {\bibfield  {journal} {\bibinfo  {journal} {Phys. Rev. Lett.}\ }\textbf
  {\bibinfo {volume} {95}},\ \bibinfo {pages} {137204} (\bibinfo {year}
  {2005})}\BibitemShut {NoStop}%
\bibitem [{\citenamefont {Ma}\ and\ \citenamefont
  {Pesin}(2015{\natexlab{b}})}]{ma_PRB2015_chiral}%
  \BibitemOpen
  \bibfield  {author} {\bibinfo {author} {\bibfnamefont {J.}~\bibnamefont
  {Ma}}\ and\ \bibinfo {author} {\bibfnamefont {D.~A.}\ \bibnamefont {Pesin}},\
  }\bibfield  {title} {\bibinfo {title} {Chiral magnetic effect and natural
  optical activity in metals with or without weyl points},\ }\href
  {https://doi.org/10.1103/PhysRevB.92.235205} {\bibfield  {journal} {\bibinfo
  {journal} {Phys. Rev. B}\ }\textbf {\bibinfo {volume} {92}},\ \bibinfo
  {pages} {235205} (\bibinfo {year} {2015}{\natexlab{b}})}\BibitemShut
  {NoStop}%
\bibitem [{\citenamefont {Fukushima}\ \emph {et~al.}(2008)\citenamefont
  {Fukushima}, \citenamefont {Kharzeev},\ and\ \citenamefont
  {Warringa}}]{fukushima_prd08_chiral}%
  \BibitemOpen
  \bibfield  {author} {\bibinfo {author} {\bibfnamefont {K.}~\bibnamefont
  {Fukushima}}, \bibinfo {author} {\bibfnamefont {D.~E.}\ \bibnamefont
  {Kharzeev}},\ and\ \bibinfo {author} {\bibfnamefont {H.~J.}\ \bibnamefont
  {Warringa}},\ }\bibfield  {title} {\bibinfo {title} {Chiral magnetic
  effect},\ }\href {https://doi.org/10.1103/PhysRevD.78.074033} {\bibfield
  {journal} {\bibinfo  {journal} {Phys. Rev. D}\ }\textbf {\bibinfo {volume}
  {78}},\ \bibinfo {pages} {074033} (\bibinfo {year} {2008})}\BibitemShut
  {NoStop}%
\bibitem [{\citenamefont {Li}\ and\ \citenamefont
  {Kharzeev}(2016)}]{li_npa16_chiral}%
  \BibitemOpen
  \bibfield  {author} {\bibinfo {author} {\bibfnamefont {Q.}~\bibnamefont
  {Li}}\ and\ \bibinfo {author} {\bibfnamefont {D.~E.}\ \bibnamefont
  {Kharzeev}},\ }\bibfield  {title} {\bibinfo {title} {Chiral magnetic effect
  in condensed matter systems},\ }\href
  {https://doi.org/https://doi.org/10.1016/j.nuclphysa.2016.03.055} {\bibfield
  {journal} {\bibinfo  {journal} {Nuclear Physics A}\ }\textbf {\bibinfo
  {volume} {956}},\ \bibinfo {pages} {107} (\bibinfo {year}
  {2016})}\BibitemShut {NoStop}%
\bibitem [{\citenamefont {Kharzeev}(2014)}]{kharjeev_ppnp14_chiral}%
  \BibitemOpen
  \bibfield  {author} {\bibinfo {author} {\bibfnamefont {D.~E.}\ \bibnamefont
  {Kharzeev}},\ }\bibfield  {title} {\bibinfo {title} {The chiral magnetic
  effect and anomaly-induced transport},\ }\href
  {https://doi.org/https://doi.org/10.1016/j.ppnp.2014.01.002} {\bibfield
  {journal} {\bibinfo  {journal} {Progress in Particle and Nuclear Physics}\
  }\textbf {\bibinfo {volume} {75}},\ \bibinfo {pages} {133} (\bibinfo {year}
  {2014})}\BibitemShut {NoStop}%
\bibitem [{\citenamefont {Kharzeev}\ \emph {et~al.}(2018)\citenamefont
  {Kharzeev}, \citenamefont {Kikuchi},\ and\ \citenamefont
  {Meyer}}]{Kharzeev_epj18_chiral}%
  \BibitemOpen
  \bibfield  {author} {\bibinfo {author} {\bibfnamefont {D.~E.}\ \bibnamefont
  {Kharzeev}}, \bibinfo {author} {\bibfnamefont {Y.}~\bibnamefont {Kikuchi}},\
  and\ \bibinfo {author} {\bibfnamefont {R.}~\bibnamefont {Meyer}},\ }\bibfield
   {title} {\bibinfo {title} {Chiral magnetic effect without chirality source
  in asymmetric weyl semimetals},\ }\href
  {https://doi.org/10.1140/epjb/e2018-80418-1} {\bibfield  {journal} {\bibinfo
  {journal} {The European Physical Journal B}\ }\textbf {\bibinfo {volume}
  {91}},\ \bibinfo {pages} {83} (\bibinfo {year} {2018})}\BibitemShut {NoStop}%
\bibitem [{\citenamefont {Yip}(2015)}]{yip_arxiv15_kinetic}%
  \BibitemOpen
  \bibfield  {author} {\bibinfo {author} {\bibfnamefont {S.~K.}\ \bibnamefont
  {Yip}},\ }\href {https://doi.org/10.48550/ARXIV.1508.01010} {\bibinfo {title}
  {Kinetic equation and magneto-conductance for weyl metal in the clean limit}}
  (\bibinfo {year} {2015})\BibitemShut {NoStop}%
\bibitem [{\citenamefont {Deng}\ \emph {et~al.}(2019)\citenamefont {Deng},
  \citenamefont {Qi}, \citenamefont {Ma}, \citenamefont {Shen}, \citenamefont
  {Wang}, \citenamefont {Sheng},\ and\ \citenamefont
  {Xing}}]{deng_prl19_quantum}%
  \BibitemOpen
  \bibfield  {author} {\bibinfo {author} {\bibfnamefont {M.-X.}\ \bibnamefont
  {Deng}}, \bibinfo {author} {\bibfnamefont {G.~Y.}\ \bibnamefont {Qi}},
  \bibinfo {author} {\bibfnamefont {R.}~\bibnamefont {Ma}}, \bibinfo {author}
  {\bibfnamefont {R.}~\bibnamefont {Shen}}, \bibinfo {author} {\bibfnamefont
  {R.-Q.}\ \bibnamefont {Wang}}, \bibinfo {author} {\bibfnamefont
  {L.}~\bibnamefont {Sheng}},\ and\ \bibinfo {author} {\bibfnamefont {D.~Y.}\
  \bibnamefont {Xing}},\ }\bibfield  {title} {\bibinfo {title} {Quantum
  oscillations of the positive longitudinal magnetoconductivity: A fingerprint
  for identifying weyl semimetals},\ }\href
  {https://doi.org/10.1103/PhysRevLett.122.036601} {\bibfield  {journal}
  {\bibinfo  {journal} {Phys. Rev. Lett.}\ }\textbf {\bibinfo {volume} {122}},\
  \bibinfo {pages} {036601} (\bibinfo {year} {2019})}\BibitemShut {NoStop}%
\bibitem [{\citenamefont {Burkov}(2015)}]{burkov_prb15_negative}%
  \BibitemOpen
  \bibfield  {author} {\bibinfo {author} {\bibfnamefont {A.~A.}\ \bibnamefont
  {Burkov}},\ }\bibfield  {title} {\bibinfo {title} {Negative longitudinal
  magnetoresistance in dirac and weyl metals},\ }\href
  {https://doi.org/10.1103/PhysRevB.91.245157} {\bibfield  {journal} {\bibinfo
  {journal} {Phys. Rev. B}\ }\textbf {\bibinfo {volume} {91}},\ \bibinfo
  {pages} {245157} (\bibinfo {year} {2015})}\BibitemShut {NoStop}%
\bibitem [{\citenamefont {Xiong}\ \emph
  {et~al.}(2015{\natexlab{b}})\citenamefont {Xiong}, \citenamefont {Kushwaha},
  \citenamefont {Liang}, \citenamefont {Krizan}, \citenamefont {Hirschberger},
  \citenamefont {Wang}, \citenamefont {Cava},\ and\ \citenamefont
  {Ong}}]{jun_science15_evidence}%
  \BibitemOpen
  \bibfield  {author} {\bibinfo {author} {\bibfnamefont {J.}~\bibnamefont
  {Xiong}}, \bibinfo {author} {\bibfnamefont {S.~K.}\ \bibnamefont {Kushwaha}},
  \bibinfo {author} {\bibfnamefont {T.}~\bibnamefont {Liang}}, \bibinfo
  {author} {\bibfnamefont {J.~W.}\ \bibnamefont {Krizan}}, \bibinfo {author}
  {\bibfnamefont {M.}~\bibnamefont {Hirschberger}}, \bibinfo {author}
  {\bibfnamefont {W.}~\bibnamefont {Wang}}, \bibinfo {author} {\bibfnamefont
  {R.~J.}\ \bibnamefont {Cava}},\ and\ \bibinfo {author} {\bibfnamefont
  {N.~P.}\ \bibnamefont {Ong}},\ }\bibfield  {title} {\bibinfo {title}
  {Evidence for the chiral anomaly in the dirac semimetal na$_3$bi},\ }\href
  {https://doi.org/10.1126/science.aac6089} {\bibfield  {journal} {\bibinfo
  {journal} {Science}\ }\textbf {\bibinfo {volume} {350}},\ \bibinfo {pages}
  {413} (\bibinfo {year} {2015}{\natexlab{b}})}\BibitemShut {NoStop}%
\bibitem [{\citenamefont {Ashcroft}\ and\ \citenamefont
  {Mermin}(1976)}]{Ashcroft76}%
  \BibitemOpen
  \bibfield  {author} {\bibinfo {author} {\bibfnamefont {N.}~\bibnamefont
  {Ashcroft}}\ and\ \bibinfo {author} {\bibfnamefont {N.}~\bibnamefont
  {Mermin}},\ }\href {https://books.google.co.in/books?id=1C9HAQAAIAAJ} {\emph
  {\bibinfo {title} {Solid State Physics}}},\ HRW international editions\
  (\bibinfo  {publisher} {Holt, Rinehart and Winston},\ \bibinfo {year}
  {1976})\BibitemShut {NoStop}%
\bibitem [{\citenamefont {Verma}\ \emph {et~al.}(2020)\citenamefont {Verma},
  \citenamefont {Kundu},\ and\ \citenamefont {Ghosh}}]{verma_prb20_dynamical}%
  \BibitemOpen
  \bibfield  {author} {\bibinfo {author} {\bibfnamefont {S.}~\bibnamefont
  {Verma}}, \bibinfo {author} {\bibfnamefont {A.}~\bibnamefont {Kundu}},\ and\
  \bibinfo {author} {\bibfnamefont {T.~K.}\ \bibnamefont {Ghosh}},\ }\bibfield
  {title} {\bibinfo {title} {Dynamical polarization and plasmons in
  noncentrosymmetric metals},\ }\href
  {https://doi.org/10.1103/PhysRevB.102.195208} {\bibfield  {journal} {\bibinfo
   {journal} {Phys. Rev. B}\ }\textbf {\bibinfo {volume} {102}},\ \bibinfo
  {pages} {195208} (\bibinfo {year} {2020})}\BibitemShut {NoStop}%
\bibitem [{\citenamefont {Schindler}\ \emph {et~al.}(2020)\citenamefont
  {Schindler}, \citenamefont {Galeski}, \citenamefont {Schnelle}, \citenamefont
  {Wawrzy\ifmmode~\acute{n}\else \'{n}\fi{}czak}, \citenamefont {Abdel-Haq},
  \citenamefont {Guin}, \citenamefont {Kroder}, \citenamefont {Kumar},
  \citenamefont {Fu}, \citenamefont {Borrmann}, \citenamefont {Shekhar},
  \citenamefont {Felser}, \citenamefont {Meng}, \citenamefont {Grushin},
  \citenamefont {Zhang}, \citenamefont {Sun},\ and\ \citenamefont
  {Gooth}}]{clemens_prb20_anisotropic}%
  \BibitemOpen
  \bibfield  {author} {\bibinfo {author} {\bibfnamefont {C.}~\bibnamefont
  {Schindler}}, \bibinfo {author} {\bibfnamefont {S.}~\bibnamefont {Galeski}},
  \bibinfo {author} {\bibfnamefont {W.}~\bibnamefont {Schnelle}}, \bibinfo
  {author} {\bibfnamefont {R.}~\bibnamefont {Wawrzy\ifmmode~\acute{n}\else
  \'{n}\fi{}czak}}, \bibinfo {author} {\bibfnamefont {W.}~\bibnamefont
  {Abdel-Haq}}, \bibinfo {author} {\bibfnamefont {S.~N.}\ \bibnamefont {Guin}},
  \bibinfo {author} {\bibfnamefont {J.}~\bibnamefont {Kroder}}, \bibinfo
  {author} {\bibfnamefont {N.}~\bibnamefont {Kumar}}, \bibinfo {author}
  {\bibfnamefont {C.}~\bibnamefont {Fu}}, \bibinfo {author} {\bibfnamefont
  {H.}~\bibnamefont {Borrmann}}, \bibinfo {author} {\bibfnamefont
  {C.}~\bibnamefont {Shekhar}}, \bibinfo {author} {\bibfnamefont
  {C.}~\bibnamefont {Felser}}, \bibinfo {author} {\bibfnamefont
  {T.}~\bibnamefont {Meng}}, \bibinfo {author} {\bibfnamefont {A.~G.}\
  \bibnamefont {Grushin}}, \bibinfo {author} {\bibfnamefont {Y.}~\bibnamefont
  {Zhang}}, \bibinfo {author} {\bibfnamefont {Y.}~\bibnamefont {Sun}},\ and\
  \bibinfo {author} {\bibfnamefont {J.}~\bibnamefont {Gooth}},\ }\bibfield
  {title} {\bibinfo {title} {Anisotropic electrical and thermal
  magnetotransport in the magnetic semimetal gdptbi},\ }\href
  {https://doi.org/10.1103/PhysRevB.101.125119} {\bibfield  {journal} {\bibinfo
   {journal} {Phys. Rev. B}\ }\textbf {\bibinfo {volume} {101}},\ \bibinfo
  {pages} {125119} (\bibinfo {year} {2020})}\BibitemShut {NoStop}%
\bibitem [{\citenamefont {Kapri}\ \emph {et~al.}(2021)\citenamefont {Kapri},
  \citenamefont {Dey},\ and\ \citenamefont {Ghosh}}]{kapri_prb21}%
  \BibitemOpen
  \bibfield  {author} {\bibinfo {author} {\bibfnamefont {P.}~\bibnamefont
  {Kapri}}, \bibinfo {author} {\bibfnamefont {B.}~\bibnamefont {Dey}},\ and\
  \bibinfo {author} {\bibfnamefont {T.~K.}\ \bibnamefont {Ghosh}},\ }\bibfield
  {title} {\bibinfo {title} {Role of berry curvature in the generation of spin
  currents in rashba systems},\ }\href
  {https://doi.org/10.1103/PhysRevB.103.165401} {\bibfield  {journal} {\bibinfo
   {journal} {Phys. Rev. B}\ }\textbf {\bibinfo {volume} {103}},\ \bibinfo
  {pages} {165401} (\bibinfo {year} {2021})}\BibitemShut {NoStop}%
\bibitem [{\citenamefont {Sinova}\ \emph {et~al.}(2015)\citenamefont {Sinova},
  \citenamefont {Valenzuela}, \citenamefont {Wunderlich}, \citenamefont
  {Back},\ and\ \citenamefont {Jungwirth}}]{sinova_rmp15}%
  \BibitemOpen
  \bibfield  {author} {\bibinfo {author} {\bibfnamefont {J.}~\bibnamefont
  {Sinova}}, \bibinfo {author} {\bibfnamefont {S.~O.}\ \bibnamefont
  {Valenzuela}}, \bibinfo {author} {\bibfnamefont {J.}~\bibnamefont
  {Wunderlich}}, \bibinfo {author} {\bibfnamefont {C.~H.}\ \bibnamefont
  {Back}},\ and\ \bibinfo {author} {\bibfnamefont {T.}~\bibnamefont
  {Jungwirth}},\ }\bibfield  {title} {\bibinfo {title} {Spin hall effects},\
  }\href {https://doi.org/10.1103/RevModPhys.87.1213} {\bibfield  {journal}
  {\bibinfo  {journal} {Rev. Mod. Phys.}\ }\textbf {\bibinfo {volume} {87}},\
  \bibinfo {pages} {1213} (\bibinfo {year} {2015})}\BibitemShut {NoStop}%
\bibitem [{\citenamefont {Samokhin}(2009)}]{SAMOKHIN09}%
  \BibitemOpen
  \bibfield  {author} {\bibinfo {author} {\bibfnamefont {K.}~\bibnamefont
  {Samokhin}},\ }\bibfield  {title} {\bibinfo {title} {Spin–orbit coupling
  and semiclassical electron dynamics in noncentrosymmetric metals},\ }\href
  {https://doi.org/https://doi.org/10.1016/j.aop.2009.08.008} {\bibfield
  {journal} {\bibinfo  {journal} {Annals of Physics}\ }\textbf {\bibinfo
  {volume} {324}},\ \bibinfo {pages} {2385} (\bibinfo {year}
  {2009})}\BibitemShut {NoStop}%
\bibitem [{\citenamefont {Fecher}\ \emph {et~al.}(2022)\citenamefont {Fecher},
  \citenamefont {Kübler},\ and\ \citenamefont {Felser}}]{felsar_materials22}%
  \BibitemOpen
  \bibfield  {author} {\bibinfo {author} {\bibfnamefont {G.~H.}\ \bibnamefont
  {Fecher}}, \bibinfo {author} {\bibfnamefont {J.}~\bibnamefont {Kübler}},\
  and\ \bibinfo {author} {\bibfnamefont {C.}~\bibnamefont {Felser}},\
  }\bibfield  {title} {\bibinfo {title} {Chirality in the solid state: Chiral
  crystal structures in chiral and achiral space groups},\ }\href
  {https://doi.org/https://doi.org/110.3390/ma15175812} {\bibfield  {journal}
  {\bibinfo  {journal} {Materials}\ }\textbf {\bibinfo {volume} {15}},\
  \bibinfo {pages} {5812} (\bibinfo {year} {2022})}\BibitemShut {NoStop}%
\bibitem [{\citenamefont {Park}\ \emph {et~al.}(2022)\citenamefont {Park},
  \citenamefont {Cheon},\ and\ \citenamefont {Lee}}]{Lee_prb22}%
  \BibitemOpen
  \bibfield  {author} {\bibinfo {author} {\bibfnamefont {M.~J.}\ \bibnamefont
  {Park}}, \bibinfo {author} {\bibfnamefont {S.}~\bibnamefont {Cheon}},\ and\
  \bibinfo {author} {\bibfnamefont {H.-W.}\ \bibnamefont {Lee}},\ }\bibfield
  {title} {\bibinfo {title} {Nondivergent chiral charge pumping in weyl
  semimetals},\ }\href {https://doi.org/10.1103/PhysRevB.106.075140} {\bibfield
   {journal} {\bibinfo  {journal} {Phys. Rev. B}\ }\textbf {\bibinfo {volume}
  {106}},\ \bibinfo {pages} {075140} (\bibinfo {year} {2022})}\BibitemShut
  {NoStop}%
\end{thebibliography}%
\end{document}